\documentclass[twocolumn]{aastex63}

\usepackage{amsmath}
\usepackage{overpic}
\usepackage{CJK}

\newcommand{\dm}{\ensuremath{\mathrm{DM}}}
\newcommand{\dmhalo}{\ensuremath{\mathrm{DM}_\mathrm{halos}}}
\newcommand{\dmigm}{\ensuremath{\mathrm{DM}_\mathrm{igm}}}
\newcommand{\dmmw}{\ensuremath{\mathrm{DM}_\mathrm{MW}}}
\newcommand{\dmhost}{\ensuremath{\mathrm{DM}_\mathrm{host}}}
\newcommand{\bardmhost}{\ensuremath{\overline{\mathrm{DM}}_\mathrm{host}}}
\newcommand{\dmargo}{\ensuremath{\mathrm{DM_{argo}}}}
\newcommand{\pccmcube}{\ensuremath{\mathrm{pc}\,\mathrm{cm}^{-3}}}

\newcommand{\sigmw}{\ensuremath{\sigma_\mathrm{MW}}}
\newcommand{\sigigm}{\ensuremath{\sigma_\mathrm{igm}}}
\newcommand{\sighalo}{\ensuremath{\sigma_\mathrm{halos}}}
\newcommand{\sighost}{\ensuremath{\sigma_\mathrm{host}}}

\newcommand{\figm}{\ensuremath{f_\mathrm{igm}}}
\newcommand{\fhot}{\ensuremath{f_\mathrm{hot}}}

\newcommand{\zgal}{\ensuremath{z_\mathrm{gal}}}
\newcommand{\zfrb}{\ensuremath{z_\mathrm{frb}}}

\newcommand{\persqdeg}{\ensuremath{\mathrm{deg}^{-2}}}
\newcommand{\hmpc}{\ensuremath{h^{-1}\,\mathrm{Mpc}}}

\newcommand{\mhe}{\ensuremath{m_\mathrm{He}}}
\newcommand{\mh}{\ensuremath{m_\mathrm{H}}}

\newcommand{\nbarigm}{\ensuremath{\bar{n}_e^\mathrm{igm}}}

\newcommand{\rmax}{\ensuremath{r_\mathrm{max}}}
\newcommand{\rvir}{\ensuremath{r_{200}}}

\newcommand{\mhalo}{\ensuremath{M_\mathrm{halo}}}

\newcommand{\nfrb}{\ensuremath{N_\mathrm{frb}}}

\newcommand{\argo}{\texttt{ARGO}}

\newcommand{\mbi}[1]{\hbox{\boldmath{$#1$}}}
\newcommand{\mat}[1]{\ensuremath{\rm\bf #1}}

\newcommand{\be}{\begin{equation}}
\newcommand{\ee}{\end{equation}}
\newcommand{\ba}{\begin{eqnarray}}
\newcommand{\ea}{\end{eqnarray}}

\accepted{Jan 25, 2022}
\shorttitle{FRB Foreground Mapping}
\shortauthors{Lee et al.}
\graphicspath{{./}{figures/}}
\begin{document}
\begin{CJK*}{UTF8}{gbsn}

\title{Constraining the Cosmic Baryon Distribution with Fast Radio Burst Foreground Mapping}

\correspondingauthor{Khee-Gan Lee}
\email{kglee@ipmu.jp}

\author[0000-0001-9299-5719]{Khee-Gan Lee}
\affiliation{Kavli IPMU (WPI), UTIAS, The University of Tokyo, Kashiwa, Chiba 277-8583, Japan}

\author[0000-0002-5934-9018]{Metin Ata}
\affiliation{Kavli IPMU (WPI), UTIAS, The University of Tokyo, Kashiwa, Chiba 277-8583, Japan}

\author[0000-0003-0574-7421]{Ilya S. Khrykin}
\affiliation{Kavli IPMU (WPI), UTIAS, The University of Tokyo, Kashiwa, Chiba 277-8583, Japan}

\author[0000-0002-0298-8898]{Yuxin Huang}
\affiliation{Kavli IPMU (WPI), UTIAS, The University of Tokyo, Kashiwa, Chiba 277-8583, Japan}

\author[0000-0002-7738-6875]{J. Xavier Prochaska}
\affiliation{Kavli IPMU (WPI), UTIAS, The University of Tokyo, Kashiwa, Chiba 277-8583, Japan}
\affiliation{University of California, Santa Cruz, 1156 High St., Santa Cruz, CA 95064, USA}

\author[0000-0001-5703-2108]{Jeff Cooke}
\affiliation{Centre for Astrophysics and Supercomputing, Swinburne University of Technology, \\
Mail Number H29, PO Box 218, 31122, Hawthorn, VIC, Australia}
\affiliation{ARC Centre of Excellence for All Sky Astrophysics in 3 Dimensions (ASTRO 3D), Australia}

\author[0000-0001-5310-4186]{Jielai Zhang (张洁莱)}
\affiliation{Centre for Astrophysics and Supercomputing, Swinburne University of Technology, \\
Mail Number H29, PO Box 218, 31122, Hawthorn, VIC, Australia}
\affiliation{ARC Centre of Excellence for Gravitational Wave Discovery (OzGrav), Australia}

\author[0000-0001-7599-6488]{Adam Batten}
\affiliation{Centre for Astrophysics and Supercomputing, Swinburne University of Technology, \\
Mail Number H29, PO Box 218, 31122, Hawthorn, VIC, Australia}
\affiliation{ARC Centre of Excellence for All Sky Astrophysics in 3 Dimensions (ASTRO 3D), Australia}



\begin{abstract}

The dispersion measures (DM) of fast radio bursts (FRBs) encode the integrated
electron density along the line-of-sight, which 
is typically dominated by the intergalactic medium (IGM) contribution in the case of extragalactic FRBs. In
this paper, we show that incorporating wide-field spectroscopic galaxy
survey data in the foreground of localized FRBs can
significantly improve constraints on the partition of diffuse cosmic baryons. Using mock DMs and realistic  
lightcone galaxy catalogs derived from the Millennium simulation, we define
spectroscopic surveys that can be carried out with 4m and 8m-class wide field spectroscopic facilities.
On these simulated surveys, we carry out Bayesian density reconstructions in order to 
estimate the foreground matter density field. In comparison with the `true' matter density field, we show that these can help reduce
the uncertainties in the foreground structures by $\sim 2-3\times$ compared to cosmic variance.
We calculate the Fisher matrix to forecast that $N=30\: (96)$ localized FRBs should be able to constrain the diffuse cosmic
baryon fraction to $\sim 10\%\: (\sim 5\%) $, and parameters governing the size and baryon fraction of galaxy circumgalactic halos to within $\sim 20-25\%\: (\sim 8-12\%)$.
From the Fisher analysis, we show that the foreground data increases the sensitivity
of localized FRBs toward our parameters of interest by $\sim 25\times$.
We briefly introduce FLIMFLAM, an ongoing galaxy redshift survey that aims to obtain foreground data on $\sim 30$
localized FRB fields.

\end{abstract}



\section{Introduction} \label{sec:intro}
The ``missing baryon problem" has been, since the turn of the millennium, one of the major unsolved problems in astrophysics (see, e.g., 
\citealt{fukugita:2004}; \citealt{cen:2006}; \citealt{bregman:2007}).
Despite strong constraints on $\Omega_\mathrm{b}$, the cosmic baryon fraction, from Big Bang nucleosynthesis and cosmic microwave
background experiments \citep{planck-collaboration:2020}, observations of the low-redshift Universe ($z \lesssim 1$) have persistently failed to account for the same amount
of baryonic matter. Stars, interstellar medium (ISM) gas in galaxies, and hot X-ray-emitting gas in galaxy clusters only account for $\sim 10\%$ of the
primordial baryon fraction \citep[e.g.,][]{persic:1992,fukugita:1998}, indicating the rest must reside in the intergalactic medium (IGM) or circum-galactic medium (CGM). 

The baryons in the CGM and IGM at $z\gtrsim 2$ are largely accounted for in the observed population of \ion{H}{1} Lyman-$\alpha$ absorbers, 
 thanks to the relatively simple astrophysics of photoionization equilibrium embedded in quasi-linear structure formation. 
However, the IGM at $z<1$ is strongly affected by gravitational shock-heating and feedback from galaxy formation.
%However, while the IGM and CGM baryons are largely accounted for at $z\gtrsim 2$ in the observed population of \ion{H}{1} Lyman-$\alpha$ absorbers
%thanks to the relatively simple astrophysics of photoionization equilibrium embedded in quasi-linear structure formation, 
%the $z<1$ IGM is strongly affected by gravitational shock-heating andfeedback from galaxy formation. 
This leads to a complex 
multi-phase medium at low redshifts \citep{cen:2006, smith:2011}, 
much of which evade easy detection by either X-ray emission or absorption line tracers.
Heroic efforts in recent years have uncovered part of the missing baryon budget via a plethora of techniques, ranging from \ion{H}{1}
Lyman-$\alpha$ absorption \citep{danforth:2008};
high-ionization metal absorption \citep{prochaska:2011,nicastro:2018}, stacked X-ray emission in filaments \citep{tanimura:2020}, 
to the Sunyaev-Zel'dovich effect \citep{tanimura:2019a, tanimura:2019,lim:2020}. 
Nevertheless, as of 2020, as much as $\sim 20\%$ of the cosmic baryons remained unaccounted for, and it was
also the case that no single method could account for more than $\sim 20-30\%$ of the baryon budget.

The burgeoning field of fast radio burst (FRB) research over the past decade has opened a promising new avenue 
towards addressing the missing baryon problem. Detected as millisecond transient radio pulses (see \citealt{cordes:2019} for 
a comprehensive review of the FRB phenomenon), 
FRB signals exhibit {a time arrival difference between photons of different frequencies} --- usually termed the `dispersion measure' (DM) --- caused by  
 free electrons along the line-of-sight to the FRB, $ \mathrm{DM} = \int n_\mathrm{e}/(1+z) \,\mathrm{d}s$, {where $s$ is the proper line element}. 
 The overall DM signal from an FRB at redshift \zfrb{} is thought to arise from several contributions: 
 \begin{equation}\label{eq:dm_all}
 \dm_\mathrm{FRB} = \dmmw + \dmigm + \dmhalo + \frac{\dmhost}{1+\zfrb},
 \end{equation}
 where \dmmw\ is the contribution from the Milky Way's interstellar medium and halo gas, \dmigm\ arises from the diffuse intergalactic
 medium gas tracing the cosmic large-scale structure on $\geq$Mpc-scales, \dmhalo\ is from the aggregate intervening galaxy halo gas within several arcmin (tens of transverse kpc) of the sightline, and \dmhost\ is from the host galaxy and progenitor source. 
 
 The large DM values observed in FRBs compared 
 to galactic pulsars \citep{manchester:2005} were a key piece of evidence pointing at their extragalactic origin, 
 as the requisite integrated electron
 contributions are likely only to arise 
 from a long extragalactic path through an ionized medium \citep{lorimer:2007}. % https://ui.adsabs.harvard.edu/abs/2007Sci...318..777L/abstract 
With the reasonable approximation that all intergalactic and circumgalactic gas is ionized,
 the extragalactic DM is thus a probe of \textit{all} the cosmic baryons, regardless of
 the exact phase of the gas \citep{mcquinn:2014}.

Even though $> 600$ FRBs have now been detected\footnote{\url{http://frbcat.org}} at the time of writing \citep[e.g.,][]{petroff:2016,the-chime/frb-collaboration:2021}, there are 
several uncertainties that need to be overcome in order to use FRBs to probe the cosmic baryon distribution.
While there should exist, on average, a monotonic relationship between the extragalactic DM contribution and the redshift of the FRB
\citep{ioka:2003, inoue:2004, mcquinn:2014, pol:2019}, 
 the redshift of the FRB needs to be measured in order to anchor constraints that relate models of the cosmic baryon distribution
 to the resulting DM. In practice, this requires first measuring the position of the FRB to sufficient accuracy ($\sim$ arcsecond) in order
 to associate it with a host galaxy, and then measuring the host galaxy's spectroscopic redshift. 
 With the first generation of FRB surveys conducted primarily using large
 single-dish radio telescopes such as Parkes (also known as \emph{Murriyang}), Arecibo, or the Green Bank Telescope, FRB positions could only be measured to
 within several arcminutes --- error circles within which hundreds of galaxies could be the FRB host.
 
 The discovery of repeating FRBs \citep{spitler:2016} enabled 
 the first localizations of FRBs to specific host galaxies thanks to follow-up observations with interferometric arrays. 
 However, repeating FRBs appear to comprise only a small fraction of all observed FRBs \citep{chime/frb-collaboration:2019,james:2020}, 
 possibly representing a different source population \citep[e.g.,][]{hashimoto:2020, heintz:2020}, and by nature are time-consuming to confirm and follow-up.
 The numbers of repeating FRBs that can be accurately localized to host galaxies therefore are low, and will take time to build up significant samples that can be used
 for studies of the cosmic baryons.
 
 % Some references to the repeating FRB population is the CHIME paper, which FRBs repeat (James+2020).
 
 % CHIME/FRB (2019)
 % https://ui.adsabs.harvard.edu/abs/2019ApJ...885L..24C/abstract
 
 % James+(2020)
 % https://ui.adsabs.harvard.edu/abs/2020MNRAS.495.2416J/abstract
 
 Instead, wide-field interferometric arrays such as the Australian Square Kilometre Array Pathfinder \citep[ASKAP;][]{mcconnell:2016,shannon:2018} and 
 the Deep Synoptic Array \citep[DSA;][]{kocz:2019} are starting to detect FRBs with sufficient resolution ($\sim 0.1 -1 \arcsec$) to match
  a plausible host galaxy within the error circle of the localization \citep[e.g.,][]{bannister:2019,bhandari:2020,ravi:2019a}. Using a sample
  of six FRBs that had been localized with ASKAP and subsequent redshift measurements of the host galaxies, \citet{macquart:2020} 
  demonstrated that the extragalactic DMs of these FRBs follow a DM-$z$ relationship that is consistent with the cosmic baryon fraction, $\Omega_\mathrm{b}$,
  expected from the standard $\Lambda$CDM cosmology --- the so-called ``Macquart Relation".
 
While the cosmic baryons are no longer ``missing'' thanks to the \citet{macquart:2020} result 
\citep[complemented by techniques
probing denser gas, e.g.][]{werk:2014,nicastro:2018}
%\textbf{as well as earlier results using different techniques
%such as \citealt{nicastro:2018}}, 
many questions remain regarding their distribution in the Universe. For example, the partition between
the diffuse cosmic web baryons and circumgalactic halo gas is currently unknown, 
as well as the spatial extent of CGM gas beyond their host galaxies. 
This question is reflected by the scatter in the Macquart Relation, 
in which the extragalactic DM at fixed FRB redshift exhibits
a variance from the diversity of possible paths through over- and under-densities in the intergalactic medium, as well as the possibility of
intersecting foreground galaxy or cluster halos. This scatter causes significant uncertainty in efforts to determine, for example, the fraction of cosmic 
baryons in the diffuse cosmic web as opposed to the CGM.

\citet{walters:2019} forecasted constraints on \figm, the fraction of cosmic baryons residing in the diffuse cosmic web, from observational samples of localized FRBs, 
and concluded that samples of $N \sim [100,1000]$ would be required to place $\sigma\sim [5\%, 2\%]$ constraints in combination with existing cosmological probes.
More observational information would allow tighter constraints on the cosmic baryon distribution: \citet{ravi:2019} argued that spectroscopic observations of foreground galaxies near FRB sightlines would allow $\sim 2\%$ constraints on \figm\ with samples of $\sim 50$ localized FRBs.
However, we note that both \citet{walters:2019} and \citet{ravi:2019} appear to have underestimated the scatter arising from the 
diffuse cosmic web by adopting $\sigma_\mathrm{igm} \approx 10\,\pccmcube$. This is a factor of $\sim 3-10\times$ smaller than
the corresponding quantity seen in hydrodynamical simulations \citep[e.g.,][]{jaroszynski:2019, takahashi:2020}, which make the 
forecasts in \citet{walters:2019} and \citet{ravi:2019} too optimistic;
this despite the fact that they already call 
for large numbers of localized FRBs, a number
unlikely to be achieved in the next 5 years.
In other words, even with large samples of localized FRBs, large-scale variance caused by the density fluctuations in the FRB foreground 
causes significant uncertainty in the cosmic baryon census that can be achieved using FRB lines-of-sight.

In this paper, we will argue that wide-field galaxy redshift surveys (covering several square degrees) targeted at the foreground of localized FRBs will allow us to address this sightline variance and make precision constraints on the
cosmic baryon distribution.  \citet{simha:2020} has carried out a preliminary implementation of this
idea using spectroscopic data from the Sloan Digital Sky Survey \citep[SDSS;][]{blanton:2005a,abazajian:2009} in front of the 
$z = 0.1167$ FRB190608 \citep{chittidi:2020}, 
although their analysis was %somewhat qualitative due to the 
very limited by the 
statistical power offered by just one FRB.
Assuming that expanded data sets consisting of dozens of localized FRBs and their foreground data will become available over the next few years, 
we will demonstrate an approach that can
be used to make precise constraints on the cosmic baryon distribution.
We will aim, in particular, to outline the numbers of galaxies and apparent magnitudes that would be needed to map out each FRB foreground field at various redshifts, in order
to demonstrate that the technique is viable with existing and upcoming multiplexed spectroscopic facilities.
 
 This paper is organized as follows: Section~\ref{sec:sims} will describe the simulations that we will use as a basis for our forecasts; 
 Section~\ref{sec:obs} will describe how we set up the mock observations that we will subsequently analyze; 
 Section~\ref{sec:recon} will describe the reconstruction algorithm used to estimate the density field from the foreground data,
 which will then feed into the parameter estimation and forecasts in Section~\ref{sec:fisher}. 
 Finally, in Section~\ref{sec:flimflam} we will give a preliminary description of FLIMFLAM, a newly-initiated spectroscopic survey that 
 aims to implement the techniques introduced in this paper.

% %%%%%%%%%%%%%%%%%%%%%%%%%%%%%%%%%%%%%%%%%
% %%%%%%%%%%%%%%%%%%%%%%%%%%%%%%%%%%%%%%%%%
\section{Simulations} \label{sec:sims}

In this paper, we perform mock analyses on redshift catalogs and density fields derived from numerical simulations
 in order to forecast the constraints on the cosmic baryon distribution in front of localized FRBs. Specifically, we use the Millennium N-body simulations\footnote{\url{http://gavo.mpa-garching.mpg.de/Millennium/}} \citep{springel:2005} and the associated semi-analytic lightcone catalogs from \citet{henriques:2015}, referred to hereafter H15.
 
Briefly, the H15 lightcone catalog is based the Millennium N-body simulation \citep{springel:2005}
which was run with $2160^3$ particles within a box size of $L=500\,\hmpc$ per side, using the smoothed particle hydrodynamics code GADGET-2 \citep{springel:2005a}.
To rescale the original simulation to be more consistent with the Planck13 cosmology \citep{planck-collaboration:2014}, H15 applied the technique of \citet{angulo:2015} which 
modified the effective box size from $L=500\,\hmpc$ to $L=480.279\,\hmpc$ and relabeled several of the snapshot redshifts in order to better match
the evolution of structure as expected from the Planck13 cosmology.

On the rescaled Millennium N-body data, H15 implemented an updated version of the ``Munich" semi-analytic galaxy formation model (\citealt{guo:2011}).
This involved changes to the modelling of various phenomena such as the star-formation thresholds, wind re-accretion, radio-mode feedback, stellar populations
and models. The 17 free parameters in the semi-analytic model were then tuned to match a set of observational data out to $z\sim 3$, in particular the abundances and passive fractions
of galaxies in the stellar mass range $8.0 \leq \log M_\star / M_\odot \leq 12.0$. They then calculated the galaxy rest-frame spectral energy distributions
assuming two different stellar population synthesis models; in this paper, we use the H15 catalogs generated with the \citet{maraston:2005} models. 

Lightcone catalogs were then generated using a method first introduced by \citet{blaizot:2005}, namely by calculating the coordinates and multi-wavelength apparent photometry of the galaxies as `observed' from several cartesian origin points
and random viewing angles in the simulation grid. They also used a method introduced by \citet{kitzbichler:2007} to chose the line-of-sight direction in order to minimize repetitions
of the same object within the lightcone as it periodically wraps through different simulation snapshots.
There were 24 separate lightcones each with a $\pi\,\mathrm{deg}^2 \simeq 3.14\,\mathrm{deg}^2$ field-of-view reaching to $z \sim 4$, as well as two ``all-sky" lightcones
that cover $4\pi$ steradian around the chosen origins but are limited to galaxies with apparent magnitudes of $i<21$. In this work, we will primarily use six of the $\pi$-square degree
lightcones that share the same simulation coordinate origins as the ``all-sky" catalogs in order to simultaneously simulate wide-surveys like 6dF or SDSS in conjunction with deeper dedicated
spectroscopic observations. For each of the narrow $\pi$-square degree lightcones, we 
downloaded the galaxy catalogs for $z<1.2$ objects down to a limiting halo mass of $M_\mathrm{vir} \geq 3\times 10^{11}\,M_\odot$, while for the all-sky lightcones we downloaded
the full $i<21$ catalog.

Following H15, we will use the Planck13 cosmology throughout this paper:
$\sigma_8 = 0.829$, $H_0 = 67.3\,\mathrm{km}\, \mathrm{s}^{-1} \mathrm{Mpc}^{-1}$,
$h=H_0/(100\,\mathrm{km}\, \mathrm{s}^{-1} \mathrm{Mpc}^{-1})=0.673$, $ \Omega_\Lambda = 0.685$, $\Omega_m = 0.315$,  
$\Omega_b = 0.0487$, and $n = 0.96$ \citep{planck-collaboration:2014}. These cosmological parameters are no longer up-to-date, but for the purposes of our mock
analyses this is not an issue so long as we remain internally consistent.

\section{Mock Observables} \label{sec:obs}

There are two overall components to the mock observations that we generate from the H15 lightcone and 
density field. The first is the cosmic dispersion measure traced along a given sightline of the lightcone, which would
be directly measured as part the initial FRB observation. The second component we  consider is the spectroscopic galaxy samples 
that need to be obtained as part of a dedicated observational program on optical telescopes (although some partial data might
already exist from prior wide-field surveys). 

First, we select an `FRB' at the desired redshift \zfrb\ by randomly choosing a halo with the correct redshift from one of the H15 lightcone catalogs,
that has a central halo virial mass of $M_\mathrm{vir} > 3 \times 10^{10}\,M_\odot$ or apparent magnitude of $i<23$. 
This is a somewhat arbitrary cut, but goes to sufficiently low masses and luminosities to reflect the fact that FRB host galaxies seem to span a wide range in properties \citep[e.g.][]{heintz:2020}.
We also ensure that each 
FRB position is at least 3 arcmin from any other FRB sightline selected within the same lightcone, 
in order to minimize correlations between adjacent lines-of-sight.

Secondly, we will define the mock spectroscopic samples of the foreground galaxies observed in front of each FRB.
These need to be at different depths and areal densities in order to probe different redshifts, and we will also describe the
observational requirements needed to achieve these samples.

\subsection{FRB Dispersion Measures}
\label{sec:dm_frb}

\begin{figure}
\includegraphics[width=0.48\textwidth, clip=true, trim=0 40 0 45]{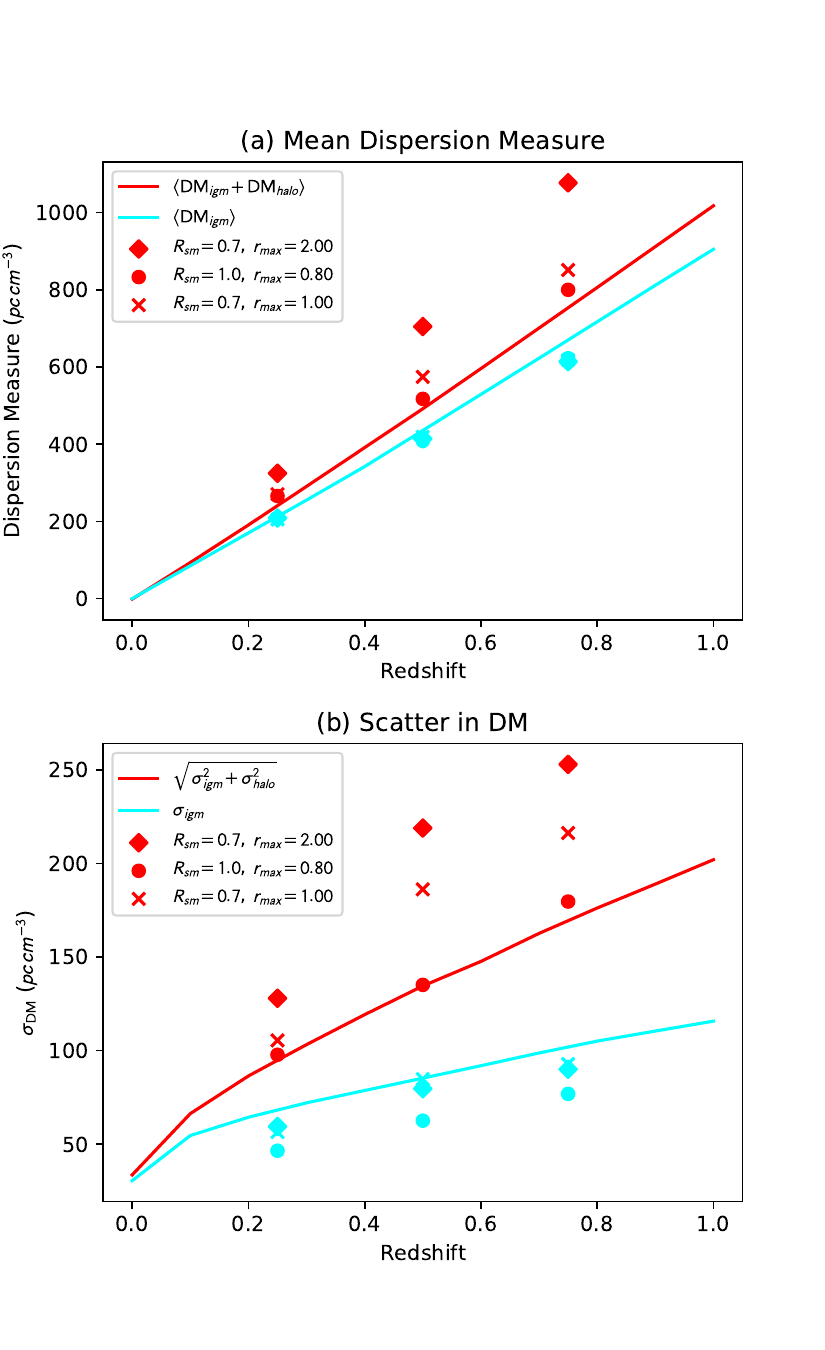}
\caption{\label{fig:sigigm}
Cosmic dispersion measure statistics compared between our models (symbols) against those from the Illustris hydrodynamical simulations analyzed by J19 (solid curves). Red curves and symbols indicate the combined IGM and halo quantities, while cyan is for the IGM only.
In the top panel (a), changes in our model parameters do not significantly modify the mean DM values, which are generally consistent with J19. 
At bottom (panel (b)), the scatter in the IGM contribution is affected by the Gaussian smoothing length, $R_{sm}$, applied to the matter field in our model. 
We will adopt the $R_{sm}$ value which best matches the \sigigm\ in J19.
In this figure, $\figm =0.85$ is assumed in our models, while $R_{sm}$ and \rmax\ in the legends are quoted in units of \hmpc{} and $r_{200}$, respectively.
}
\end{figure}

The first mock observable we need to define is the dispersion measure that is measured as part of the initial FRB detection in the radio.

%Note that in some of the literature, 
 %$\dmigm + \dmhalo$ are collectively termed `\dmigm' or `$\mathrm{DM}_\mathrm{cosmic}$', but in this paper it is our goal to
%demonstrate the separation of the smooth IGM and discrete halo components so we always refer to \dmigm\ and \dmhalo\ 
%as separate contributions.

We will separately calculate \dmigm, \dmhalo, and \dmhost\ contributions that would be observed for a simulated FRB sightline within the simulation.
First, we will define as the primary observable for our analysis, the extragalactic DM component:
\begin{equation}
\mathrm{DM}_\mathrm{eg}  =  \dm_\mathrm{FRB} - \dmmw,
\end{equation}
which represents all the DM contributions from beyond the Milky Way and its halo.
$\mathrm{DM}_\mathrm{eg}$ will therefore constitute one of the key observables in our subsequent forecast:
\begin{equation}\label{eq:dm_eg}
\mathrm{DM}_\mathrm{eg} = \dmigm + \dmhalo + \frac{\dmhost}{1+\zfrb} + \delta_N,
\end{equation}
where a random deviate arising from noise and Milky Way subtraction, $\delta_N$, has also been added.

To compute the \dmigm\ component arising from large-scale IGM, we use the gridded $256^3$ Millennium density field snapshots with a side length of
$L=480.279\,\hmpc$, that we smoothed with a Gaussian kernel with size $R_{sm}$. First, we computed the path between the FRB coordinate and lightcone origin\footnote{The 24 lightcones made
available by H15 all originate at different simulation coordinates even though they have all been rotated to center every field at [0,0] deg in [RA, Dec]. 
For each lightcone, we triangulated the simulation coordinates of several galaxies to determine the cartesian lightcone origin.}, keeping track
of the exact path length $l_s$ traced by the sightline through each grid cell $s$ as well as the redshift $z_i$ at to the simulation snapshot corresponding
to the line-of-sight comoving distance from the origin for the traversed cell.
We calculate a discretized version of $\dmigm= \int n_e^{\mathrm{igm}}/(1+z)\,\mathrm{d}s$ as follows:
\begin{equation} \label{eq:dmigm_sim}
\dmigm   = \bar{n}_e^{\mathrm{igm}} (z_i) \sum\limits_{s}^{} (1+\delta^{sm}_{m,s} )\, l_s (1+z_s)^{-1},
\end{equation}
where is $\delta^{sm}_{m,s}$ and $l_s$ are the smoothed matter overdensity and path length through the grid cell $s$, respectively; $z_s$ is the redshift of the grid cell calculated
from the distance-redshift relation\footnote{Note the difference between $z_s$ and $z_i$: $z_s$ is the redshift calculated for each intersected
grid cell along the line-of-sight, while $z_i$ is the redshift of the simulation density snapshot that corresponds closest to $z_s$.}, and \nbarigm\ is the mean
cosmic electron density at the redshift of the snapshot $z_i$, defined as:
\begin{equation}\label{eq:ne_figm}
\nbarigm(z_i) = \figm \Omega_\mathrm{b} \bar{\rho_c}(z_i)\left[ \frac{\mhe(1-Y) + 2Y\mh}{\mh\mhe}\right],
\end{equation}
where \mh\ and \mhe\ are the atomic masses of 
hydrogen and helium, respectively; $Y=0.243$ is the cosmic mass fraction in helium
(assumed to be double-ionized);  $\Omega_\mathrm{b}=0.044$ is the fraction of the 
cosmic critical density in baryons, $\bar{\rho}_c(z_i)$ is the critical density of the universe at the redshift $z_i$, and \figm\ is the
fraction of cosmic baryons residing in the diffuse IGM. 
The summation carried out in Equation~\ref{eq:dmigm_sim} traverses different density snapshots of the Millennium simulation
 according to the snapshot redshift $z_s$ which is
closest to the cell redshift $z_i$: for example, a sightline originating at $\zfrb=0.5$ would pass through 14 different redshift snapshots.

Next, we calculate the \dmhalo\ contribution from intervening galaxy halos close to the FRB line-of-sight, 
assuming that the halo gas is described by a modified Navarro-Frank-White (mNFW) profile as elucidated in \citet{prochaska:2019}.
The baryon density as a function of radius is given by:
\begin{equation} \label{eq:rhob_halo}
\rho_\mathrm{b} = \frac{\rho^0_\mathrm{b}}{y^{1-\alpha}(y_0 + y)^{2+\alpha}},
\end{equation}
where $\rho^0_\mathrm{b}$ is the central density, $y$ is a rescaled radius (see \citealt{mathews:2017} for the full definitions), and we fix the profile parameters $\alpha=2$ and $y_0=2$. 
This choice of parameters is designed to produce halo gas profiles similar to those derived by \cite{maller:2004} for multi-phase circumgalactic media.
For a halo with mass \mhalo, the total mass in baryons is:
\begin{equation} \label{eq:mhalo_bar}
 \mhalo^b \equiv \fhot (\Omega_b/\Omega_m) \mhalo, 
 \end{equation}
 where
\fhot\ represents the fraction of the halo baryons that are in a hot gas phase tracing the mNFW profile out to 1 virial radius. 
We truncate the mNFW profile at a given radius \rmax, which is usually of comparable magnitude to the virial radius \rvir, and so we will usually quote \rmax\ in units of \rvir. 
\citet{simha:2020} showed that varying the assumed \rmax\ for the foreground galaxies of FRB190608 from \rvir\ to $2\times \rvir$ more than
doubled the \dmhalo\ contribution, so this will be one of the free parameters that we will consider in our mock analysis.

To speed up the computation we only calculate the \dmhalo\ contribution for foreground galaxies within $10'$ of each simulated FRB sightline. 
This corresponds approximately to the transverse separation at $r_{200}$ of a $10^{13}\,M_\odot$ halo at $z=0.05$ --- for lower-mass
or higher redshift halos, the transverse extent $r_{200}$ will be correspondingly smaller. 
Since we are defining our mock observables at this point, there is no possibility of halos outside of this separation biasing our forecast so long as we remain internally consistent. 
In a real observation, we would have to carefully consider the possible contributions of more massive, cluster-sized, halos that might contribute at larger transverse
separations if they are low redshift. In this regime, \dmhalo{} and \dmigm{} will begin to overlap as group- and cluster-sized halos will also represent clear overdensities on $\sim$Mpc-scales. When analyzing real observations, we will need to carefully keep track
of these massive halos to avoid double-counting their DM contributions, but in our mock analysis
the separation is clear. This is because both mock `observations' and fitting functions are derived from
exactly the same model as described in this section.

In addition to computing the \dmhalo{} for each sightline, we also output a catalog listing the redshift, separation, halo mass, and DM contribution of
all galaxies at $<10\arcmin$ within each sightline. This will be used in the subsequent Fisher matrix parameter forecast.

Now that we have defined \dmigm\ and \dmhalo, we will remark on the effect of the Gaussian smoothing length, $R_{sm}$, applied to the 
N-body density field before calculating \dmigm\ from Equation~\ref{eq:dmigm_sim}. In principle, at fixed cosmology and for a given
cosmic baryon distribution, there is no effect from varying $R_{sm}$ on $\langle \dmigm + \dmhalo \rangle (z)$ 
(i.e. the IGM and halo dispersions averaged over the Universe
at fixed FRB redshift). 
However, the scatter in \dmigm, which we refer to as \sigigm, will vary as a function of $R_{sm}$ --- with large $R_{sm}$ the variance
in \dmigm\ between different sightlines will get washed out.
As a comparison, we use the results of \citet{jaroszynski:2019}, hereafter J19, who studied the statistics of
the \dmigm\ distribution in the Illustris cosmological hydrodynamical simulations (see also \citealt{takahashi:2020}; \citealt{zhu:2020}; \citealt{zhang:2021}, \citealt{batten:2021}).
We calculated $\langle \dmigm \rangle (z)$ at several redshifts using our model (Equation~\ref{eq:dmigm_sim}), for a few different values of $R_{sm}$.
In each case, we averaged over 750 sightlines drawn from the 24 separate H15 lightcones, 
and set $\figm=0.85$ to match the average $z<1$ value found by J19. 
The $\langle \dmigm \rangle (z)$ from our models are shown as cyan symbols in Figure~\ref{fig:sigigm}a, in comparison with the J19 \dmigm\ curve shown as a cyan line.

Our model --- derived from the density field of the non-hydrodynamical Millennium
simulation --- has a reasonable agreement with J19, and as expected varying $R_{sm}$ does not significantly modify $\langle \dmigm \rangle (z)$.

We also computed $\langle \dmigm + \dmhalo \rangle (z)$ 
 (red symbols in Figure~\ref{fig:sigigm}a) for a range of $[\rmax,R_{sm}]$,
 which does exhibit 
significant changes with respect to parameter choice. Since $\langle \dmigm (z)\rangle$ alone shows little variation respect to $R_{sm}$,
this variation in $\langle \dmigm + \dmhalo \rangle (z)$  must come from the response of \dmhalo{} with respect to \rmax{}.
For some of the larger values of \rmax{}, the overall $\langle \dmigm + \dmhalo \rangle (z)$ 
becomes large enough that it exceeds the total baryon budget of the Universe. 
However, in this paper we will treat \rmax{} as a free parameter for our forecast, 
therefore we will not seek to match \rmax{} with J19 or any other
hydrodynamical simulation. 
However, we note in passing that this comparison with J19 implies that the halo CGM in the Illustris
simulation has characteristic scales of comparable to the halo virial radius.

\begin{figure}
\includegraphics[width=0.48\textwidth]{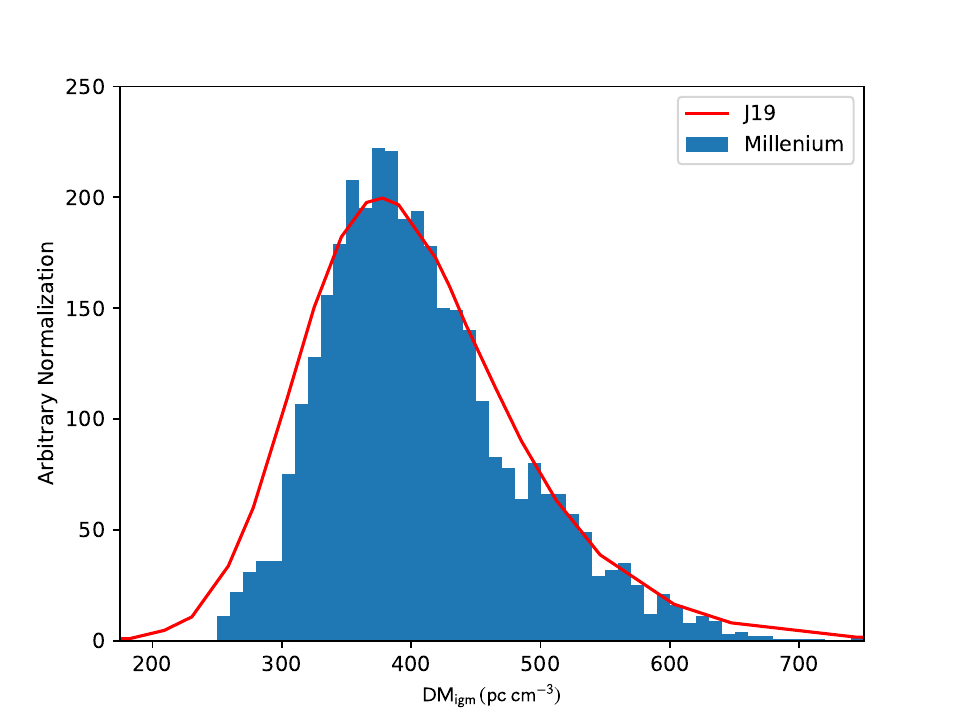}
\caption{\label{fig:pdf_dmigm}
Distribution of \dmigm{} at $z=0.5$, shown for both our model with $R_{sm}= 0.7\,\hmpc$ (histogram) as well as that from J19 (red curve). Our model provides a reasonable
match with that from the hydrodynamical simulation. The mean and scatter for our model distribution shown here is $\langle \dmigm = 412\,\pccmcube$ and $\sigigm = 79\,\pccmcube$. 
}
\end{figure}

The scatter of the \dmigm{} distribution (Figure~\ref{fig:sigigm}b), however, shows considerably more variation with respect to the choice of parameters
in our model, especially with respect to 
$R_{sm}$. Given the cell size of $1.875\,\hmpc$ in the Millennium binned density field, we find that Gaussian smoothing by $R_{sm}=0.7\,\hmpc$ provides an adequate match for \sigigm\ in our model 
in comparison with J19.
In Figure~\ref{fig:pdf_dmigm}, we show the probability distribution of \dmigm{} only computed at $z=0.5$ both for our model, and 
that from the Illustris hydrodynamical simulation by J19. We have set $R_{sm} = 0.7\,\hmpc$ and $\figm= 0.85$ to get the best match with J19. 
The two distributions are similar, with a long non-Gaussian tail toward high \dmigm{}, although the J19 distribution extends toward lower \dmigm{}
compared with ours.

Note that these \sigigm\ values, in both our model and J19, are higher at all redshifts than the $\sigigm = 10\,\pccmcube$ adopted by \citet{walters:2019} and \citet{ravi:2019}.
Considering the variance in both \dmigm\ and \dmhalo, our model slightly overestimates the
overall scatter of $(\dmigm + \dmhalo)$ relative to the hydrodynamical model. Since the parameters governing
\dmhalo{} will be adopted as free parameters in our framework, we do not consider it necessary for our model to perfectly reproduce 
the overall $(\dmigm + \dmhalo)$ scatter of J19, 
but deem the statistical properties in our \dmigm{} model accurate enough for realistic parameter forecasts.

The final free parameter in our forecast framework is the FRB host contribution to the DM signal, \dmhost, which in principle is
the sum of dispersions caused by electrons surrounding the progenitor object as well
as the host galaxy. This is likely the most uncertain component of our model given our lack of knowledge
on FRB progenitors, 
In this paper, we assume that the \dmhost{} of the FRBs is drawn from a Gaussian distribution
with a mean of $\bardmhost = 200\,\pccmcube$ and r.m.s.\ scatter of $\sigma_\mathrm{host} = 50\,\pccmcube$ in the FRB restframe. 
This is a distribution that is consistent with the emerging consensus for the general population of FRBs \citep{james:2021, cordes:2021}, 
although for the purpose of our
forecast it is not crucial that we come up with an accurate model for the \dmhost{} distribution.
When this paper was in an advanced stage of preparation, new results emerged of
FRBs with extremely high DMs given their host redshift (\citealt{niu:2021} and other as-yet
unpublished results), implying $\dmhost \sim 500-1000\,\pccmcube$ that is much larger than the `normal' population of FRBs with $\dmhost \sim 100-200\, \pccmcube$.
However, these high-\dmhost{} FRBs exhibit properties that would likely allow them to be excluded from the kind of 
cosmic baryon analysis that we develop in this paper --- we will discuss this aspect more in the Conclusions.

Finally, we assume that the Milky Way contribution, \dmmw, can be modeled with 
reasonable accuracy and subtracted from the overall signal, leaving only the IGM, halo, and host contributions.
We therefore do not explicitly model \dmmw\ but assume that its subtraction introduces an uncertainty of $\sigmw = 15\; \pccmcube$ in the measurement of 
the extragalactic FRB DM, which we add as a random Gaussian deviate, $\delta_N$, with standard deviation \sigmw\ to the $\mathrm{DM}_\mathrm{eg}$ (Equation~\ref{eq:dm_eg}). In reality, \dmmw{} varies across the sky and \sigmw{} for each sightline 
would vary as a proportional uncertainty on \dmmw. Our framework can easily incorporate individual sightline errors on \sigmw{}, but for convenience we simply pick the same value for
\sigmw{} for all of the mock sightlines.

\subsection{Wide-Field Mock Samples}
\label{sec:specmocks}

 \begin{deluxetable*}{lccccc}[t]
 \label{tbl:mocks}
 \tablecaption{Mock Large-Scale Catalogs used for FRB Foreground Reconstruction}
 \tablehead{
 \colhead{Catalog\tablenotemark{a}} & 
 \colhead{Redshift} &
 \colhead{Limiting $r$-mag} &
 \colhead{No.\ of Galaxies} &
 \colhead{Area (deg$^2$)} &
 \colhead{Area Density (deg$^{-2}$)}
 }
 \startdata
 2m-z01 & $z\leq 0.1$ & 16.4 & 8400 & 700 & 12 \\
 4m-z03 & $0.1<z<0.35$ & 19.8 & 2400 & 3.1 & 770 \\
 8m-z10  & $0.35<z<1.0$ & 22.75 & 7500 & 1.25 & 6000
 \enddata
 \tablenotetext{a}{The codenames denote the telescope size and upper redshift range,
e.g. `4m-z03' denotes a survey on a 4m-class telescope covering up to $z\sim 0.3$.}
 \end{deluxetable*}

 % GeneralSurveyDesign.ipynb
 \begin{figure}
\includegraphics[width=0.48\textwidth]{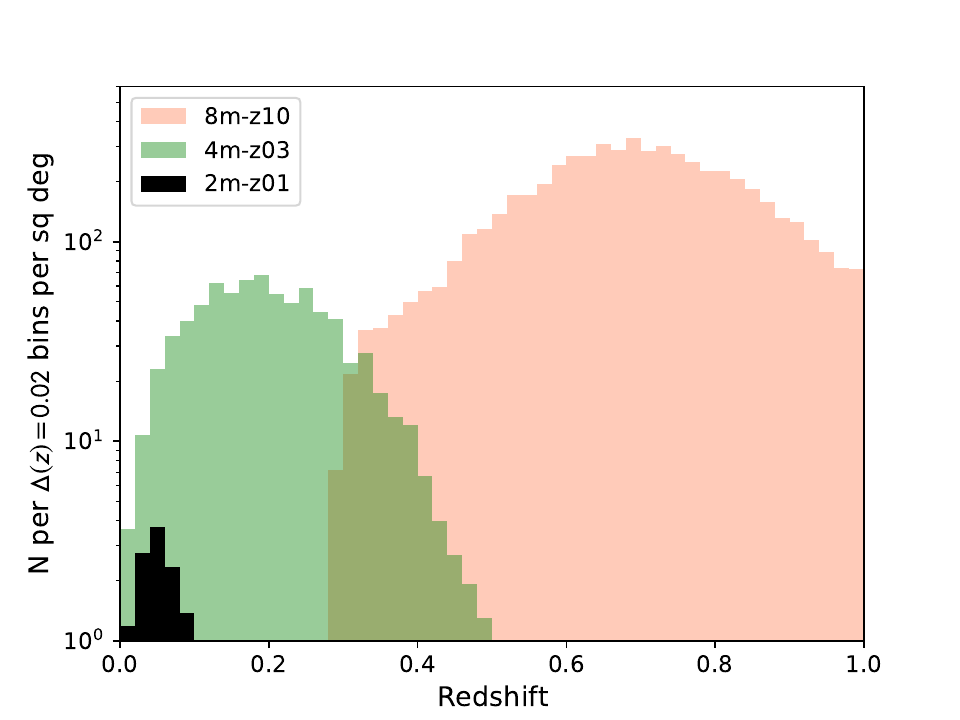}
\includegraphics[width=0.48\textwidth]{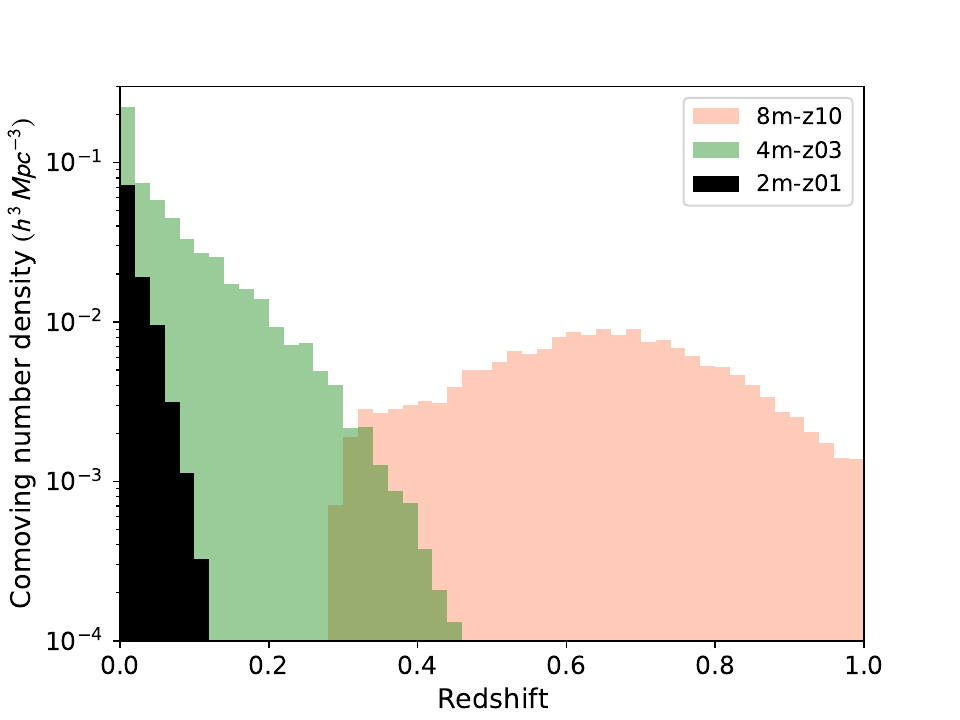}
\caption{\label{fig:surveys}
The area density (top) and comoving number density (bottom) of galaxies of the various mock galaxy redshift catalogs used as input for
FRB foreground density reconstructions analyzed in this paper. The catalogs will be combined to build up foreground galaxy samples to
cover a given FRB redshift. Note that even though the 2m-z01 catalog appears to provide a negligible contribution, its large footprint {on the sky} enables accurate reconstruction of very low-redshift structures that are otherwise ill-sampled by the narrow high-redshift catalogs.}
\end{figure}

As seen in the previous section, the IGM component is typically the dominant component of the extragalactic FRB dispersion mesure (see Equation~\ref{eq:dm_eg} and Figure~\ref{fig:sigigm}) in both the mean value and scatter. The scatter, \sigigm, is driven primarily by 
fluctuations in the underlying large-scale ($\gtrsim$Mpc) cosmic web traversed by the FRB sightlines. We therefore hypothesize that spectroscopic
galaxy surveys in the FRB foreground will allow us to significantly reduce \sigigm{} and improve our ability to constrain the parameters governing
the CGM and IGM.
% JXP -- polished the wording
 The H15 lightcone catalog includes synthetic source magnitudes calculated for a variety of photometric filters. 
 Therefore, it is
 straightforward to select samples of galaxies that simulate various 
 spectroscopic surveys 
 and then study the foreground DM 
 contribution in front 
 of hypothetical FRBs.

\begin{figure*}
\begin{center}
\includegraphics[width=0.85\textwidth]
{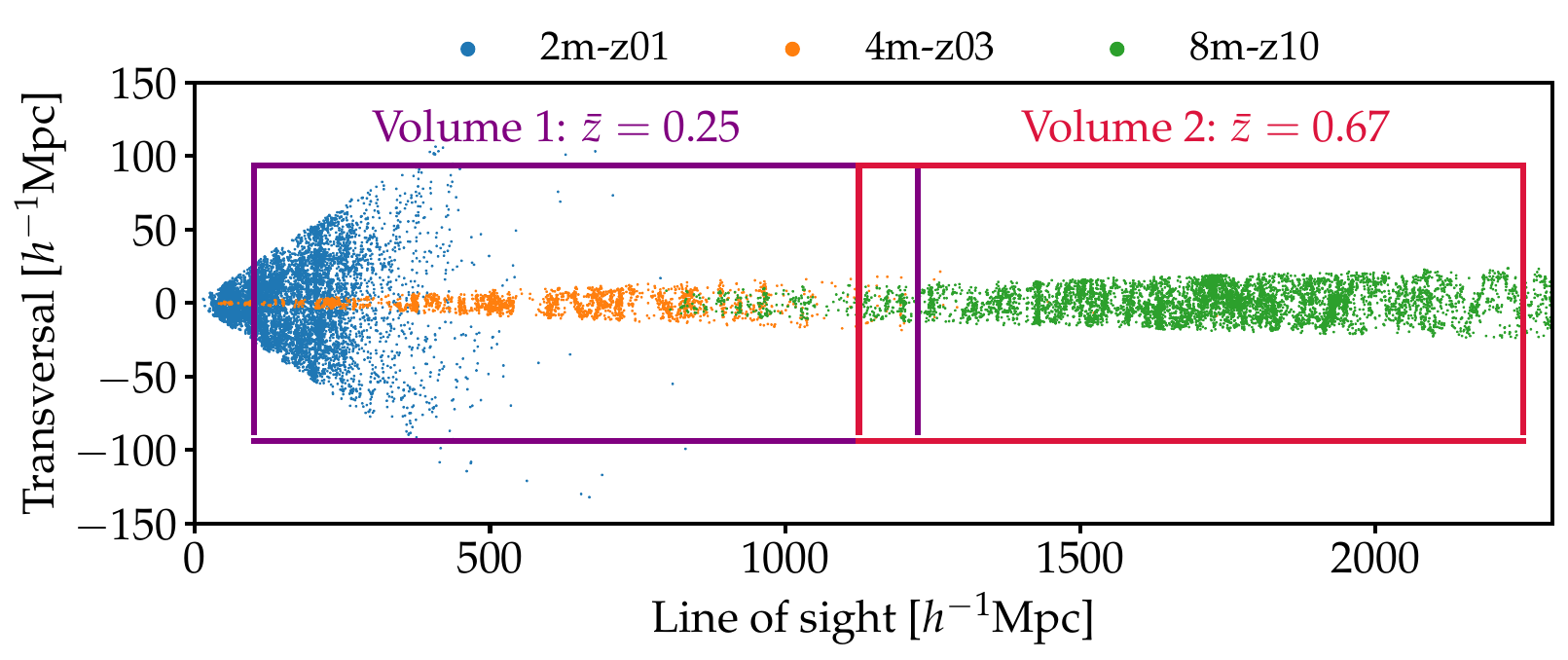}
\caption{\label{fig:lightcones}
Galaxy positions of the 3 different surveys 2m-z01; 4m-z03; and 8m-z10; used for reconstructions for one mock lightcone. We perform the reconstructions in 2 rectangular boxes consecutively placed along line of sight, dubbed as Volume 1 and Volume 2, leaving a $100\, h^{-1} \rm Mpc$ overlap region. The mean redshifts of the 2 volumes are $\bar{z}_1 = 0.25$ and $\bar{z}_2 = 0.67$, respectively. Note that the aspect ratio is different between the abscissa and ordinate axes in this figure.}
\end{center} \end{figure*}

We defined several mock spectroscopic catalogs up to $\zfrb \leq 0.8$, utilizing multiplexed wide-field spectrographs on a range of telescope sizes. These are described in detail in Appendix~\ref{app:specmocks} and summarized in Table~\ref{tbl:mocks}, {while Figure~\ref{fig:lightcones} 
provides a visual impression of the various surveys' coverage for a single lightcone}. Depending on the \zfrb, we will assume a combination of surveys to build up to the desired redshift: for example, for a simulated $\zfrb=0.3$ sightline we will assume the availability of the 2m-z01 and 4m-z03 data in the foreground.
These mock surveys are coarse-grained in terms of the redshift ranges they cover.
Thus, for the mock analysis on  
a $\zfrb=0.6$ FRB there might be many superfluous background galaxies at 
$\zgal > 0.6$ because 8m-z10 observations will cover up to $z=0.8$ (see below). 
In real observations, 
the target selection will be fine-tuned individually for the specific \zfrb\ fields that are being targeted in order to maximize the telescope efficiency. 
Nevertheless, these mock catalogs have been invaluable in helping to design the observational strategy for the FLIMFLAM Survey, which we briefly
describe in Section~\ref{sec:flimflam}.

The area densities and comoving number densities of the different surveys are shown in Figure~\ref{fig:surveys}, in which we have 
averaged all the 24 available H15 lightcones from the Millennium database in order to obtain a smooth redshift distribution --- the distributions
in the individual 3.1 deg$^2$ lightcones are much more uneven due to variations in large-scale structure. While it may appear from Figure~\ref{fig:surveys} as if the 2m-z01 catalog
contributes only negligibly to the foreground sampling, the wide-field coverage over hundreds of square degrees is crucial for accurately recovering transverse structures at very low redshifts. Similar data sets to 2m-z01 already exist for much of the extragalactic sky thanks to all-sky spectroscopic surveys such as SDSS \citep{abazajian:2009} and 6dF \citep{jones:2009}.

For the density reconstructions in the rest of this paper, we used mock catalogs from only 6 of the 24 lightcones released by H15, which were the subsets in which
the all-sky catalogs were available --- the other 18 lightcones had no data beyond the central 3.1$\,\mathrm{deg}^{2}$ footprints.

% %%%%%%%%%%%%%%%%%%%%%%%%%%%%%%%%%%%%%%%%%%%%%%%%%%%
\subsection{Intervening Galaxy Catalogs}

In addition to the wide-field survey redshift data defined above, for each mock FRB sightline
we also output an `observational' list of all galaxies within $<10\arcmin$ and $M_\mathrm{halo} \geq 3 \times 10^{11} \,\mathrm{M}_\odot$, listing their redshift, angular separation from 
the sightline, and estimated halo mass. To mimic a realistic observational scenario, we include all galaxies
with the aforementioned criteria whether or not they actually contribute to the sightline \dmhalo{} given our fiducial parameters. Also, since there are significant uncertainties
in estimating galaxy halo masses, we introduce random errors of $\sigma (\log_{10} M_\mathrm{halo}/M_\odot) = 0.3$ to the listed halo masses. Separately from the `observational'
intervening galaxy list, we also kept track of the true halo masses and \dmhalo{} contributions
for each sightline to calculate the `true' values.

\begin{figure}
    \centering
    \includegraphics[width=0.48\textwidth]{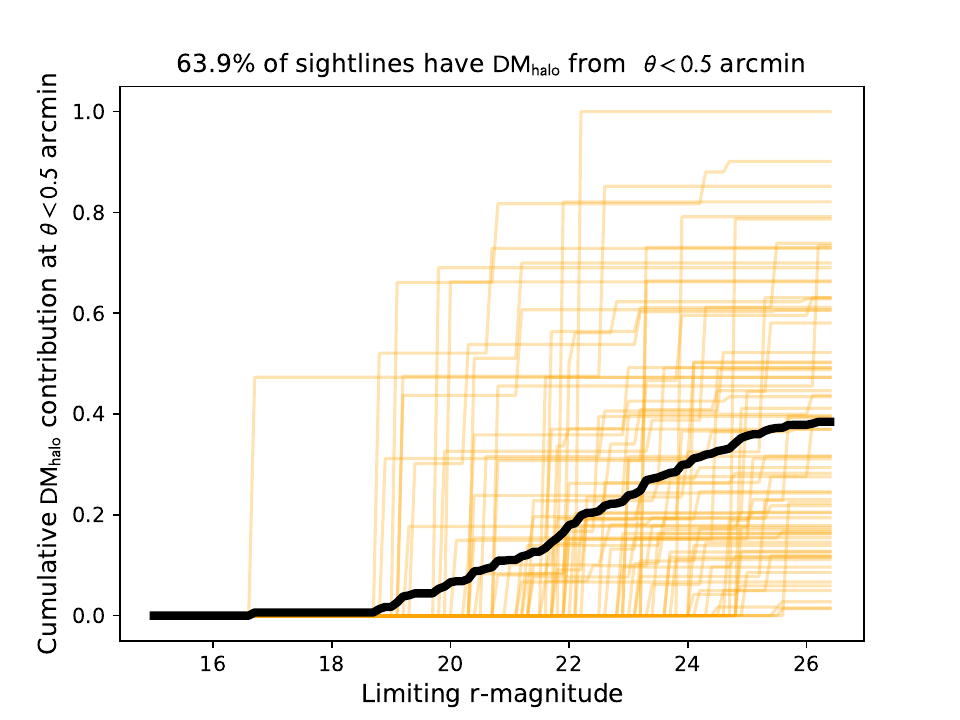}
    \includegraphics[width=0.48\textwidth]{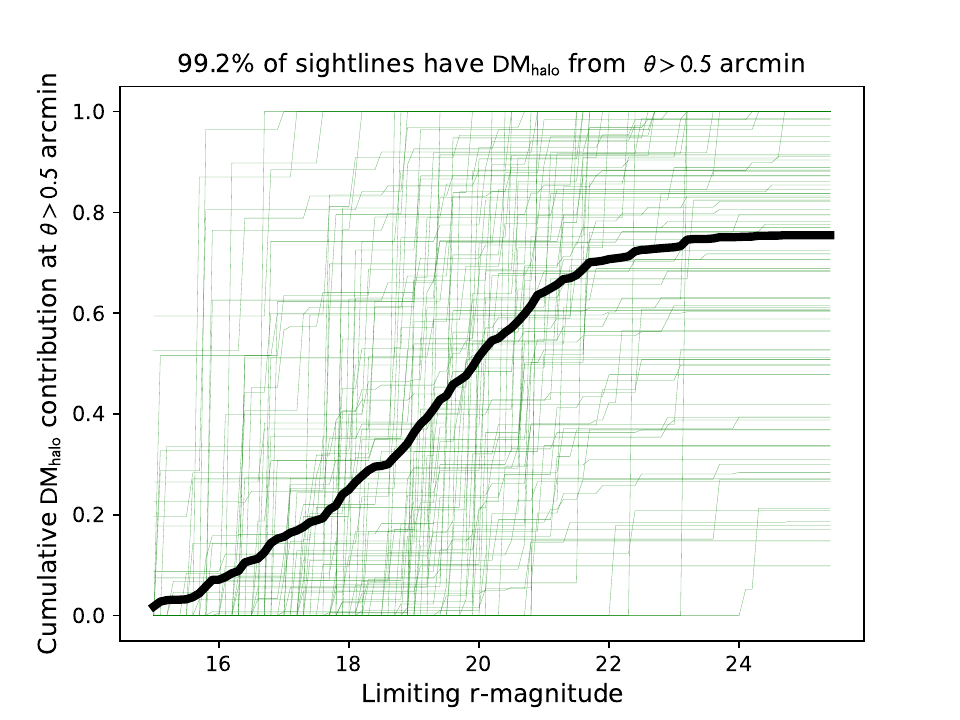}
    \caption{Colored lines indicate cumulative contributions to \dmhalo{} from mock sightlines extracted from our H15 catalogs, showing the contribution from (top) angular
    separations of $\theta < 0.5\arcmin$, and (bottom) separations of $0.5\arcmin < \theta < 5\arcmin$. In each panel, the black curve indicates the average cumulative contribution
    from all sightlines.}
    \label{fig:dmhalo_vs_mag}
\end{figure}

We now comment on the observations required to account for these intervening
 galaxies, which would be expected to be fainter than the large-scale tracer
galaxies described in the previous Section and thus need to be observed with larger telescopes. 
We analyze an ensemble of 120 random sightlines with $\zfrb \leq 0.5$ in Figure~\ref{fig:dmhalo_vs_mag}
to characterize the typical separations and galaxy magnitudes at which intervening halos
are likely to contribute to the sightline DM budget. 
In the top panel of Figure~\ref{fig:dmhalo_vs_mag}, we see that $64\%$ of
the mock sightlines have \dmhalo{} contributions from angular separations of $\theta <0.5\arcmin$,
of which on average $\sim 40\%$ of the total \dmhalo{} contribution is from these small separations.
But these closely-intervening galaxies can reach down to relatively faint magnitudes of $r>24$,
with about $\sim 10\%$ of the \dmhalo{} contribution missing from these sightlines if 
$r>24$ galaxies are unaccounted for. 

At slightly larger separations of $0.5\arcmin < \theta< 5\arcmin$, we see in the bottom panel of 
Figure~\ref{fig:dmhalo_vs_mag} that nearly all (99\%) sightlines have contributions
from larger angular distances from FRB sightlines. At these larger separations, however, 
the contributions come from galaxies with typical magnitudes of $r<23$. Indeed, an observational limit of $r<22.5$ would only miss out on a few percent of the overall \dmhalo{} contribution on
these larger separations. We find that on average, there are $\sim 4$ galaxies (up to a maximum of $\sim 12$) with $r<22.5$ that would contribute to each sightline. However, within a radius of $\theta<5\arcmin$ there
are likely to be $\sim 100-200$ galaxies at this magnitude limit. Photometric redshifts would
therefore be invaluable to pre-select $\sim 20-30$ candidate foreground galaxies that could
then be observed with a typical
imaging spectrograph on an 8-10m class telescope, to confirm their redshifts.

These results are consistent with the estimates of \citet{ravi:2019} who found that
observing galaxies down to $r<24$ would detect $>90\%$ of the \dmhalo{} contribution, but our findings motivate a tiered observing strategy to target intervening galaxies. 
First, one desires an integral field unit (IFU) spectrograph observations with 8-10m class
telescope (e.g., VLT/MUSE or Keck-II/KCWI) within 
$\theta \lesssim 0.5\arcmin$ with several-hour integration to reach depths of $r>24$.
For larger radii, 
($0.5\lesssim \theta \lesssim 5\arcmin$), 
multiple short pointings ($<1$ hr) with spectrographs on 8-10m class telescopes
such as Keck-I/LRIS, VLT/FORS2, or Gemini/GMOS 
would confirm any intervening galaxies. 
Note that it is possible for halos to 
contribute at larger separations of $\theta>5\arcmin$, but these tend to be galaxy 
group-sized halos or larger ($M \gtrsim 10^{13}\,M_\odot$), for which the central galaxies are typically massive and bright enough to be
identified through the wide-area surveys described in Section~\ref{sec:obs}.

\section{Density Reconstructions}
\label{sec:recon}
In this section we describe the method used to reconstruct the matter density field given the distribution of the mock galaxy redshifts extracted from H15 as described in the previous section. This reconstructed matter density will be used to place constraints on the possible
range of \dmigm{} that arises from large-scale diffuse gas in the foreground of individual sightlines.

We will use the most recent version of the \texttt{ARGO} Code \citep{ata:2015,ata:2017}, which is based on \citet{kitaura:2008,jasche:2010,kitaura:2010}.
First, we describe the reconstruction algorithm \texttt{ARGO}, presenting the necessary formulae and modifications to the algorithm. 
Then, we describe how the input mock galaxies have been organized and the reconstructions have been performed. 

As described in Section~\ref{sec:specmocks}, the different mock catalogs are biased tracers of the same underlying density field. In order to jointly reconstruct the density field from these distinct mock surveys, a multi-survey approach is required. We follow the multi-tracer/multi-survey approach introduced by \citet{ata:2021}, which was implemented in the \texttt{COSMIC BIRTH} algorithm \citep{kitaura:2021}.  
With this approach we can combine separate survey footprints, radial selection functions, galaxy biases and number densities.

\subsection{Bayesian Inference Model}

The goal of Bayesian inference using Monte Carlo Markov Chains (MCMC) is to draw samples of a posterior probability density $\mathcal{P}(\mbi{\Theta}|\mbi{d})$, representing a set of parameters $\mbi{\Theta}$ given a data vector $\mbi{d}$. The posterior consists of a prior $\pi$ and a likelihood function $\mathcal{L}$
\ba
\label{eq:bayes}
\cal P(\mbi \Theta|\mbi d)\propto\pi(\mbi \Theta)\times\cal L(\mbi d|\mbi \Theta)\, .
\ea
In our framework the variable of interest is the linearized matter density $\mbi \Theta = \mbi{\delta}_{\rm L}$, that we infer on a rectangular grid with $N_{\rm C}$ cells. The transformation between the physical density contrast $\mbi \delta = \mbi \varrho/\bar{\mbi \varrho}-1$ (matter density $\varrho$ ) and the linearized density $\mbi{\delta}_{\rm L}$ \citep{coles:1991} is 
\ba
\mbi\delta_{\rm L}=\log(1+\mbi\delta)-\mbi\mu\,,
\ea
where the mean field $\mbi \mu$ \citep{kitaura:2012a} ensures $\langle\mbi\delta_{\rm L} \rangle =0 $.
As prior distribution for $\mbi \delta_{\rm L}$ we utilize  a multivariate Gaussian 
\ba
\label{prior}
\pi(\mbi\delta_{\rm L})=\frac{1}{\sqrt{(2\pi)^{N_{c}}\det(\mat C_{\rm L})}}\exp\left(-\frac{1}{2}\mbi \delta^{\rm T}_{\rm L}\mat C_{\rm L}^{-1}\mbi\delta_{\rm L}\right)
\ea
where $\mat C_{\rm L}=\langle\mbi\delta^{\rm T}_{\rm L}\mbi\delta_{\rm L}\rangle$ is the co-variance matrix of the linearized density.

The information content of the data is encoded in the likelihood function that depends on the expected number of galaxies $\mbi \lambda$ given the underlying matter field $\mbi \delta$ and the observed number of galaxies $\mbi{N}_{\rm G}$.
We model the expectation number per cell $i$ as
\ba
\label{eq:expect}
\lambda_{i}=f_{N}w_{i}(1+G_i\delta_{i})^{b_i}\,,
\ea
with the normalization $f_{N}$ ensuring correct number density $\bar{N}$, $w_i$ the completeness at cell $i$, and  $b_i$ the galaxy bias parameter, accounting for the different clustering of galaxies as compared to the total matter. $G_i$ is the growth factor ratio between a reference redshift $z_{\rm ref}$ , and an arbitrary redshift $z_i$
\be
G(z_{\rm ref},z_i) = D(z_i)/D(z_{\rm ref})\, ,
\ee
where 
\be
D(z_i) = \frac{H(z_i)}{H_0} \int\limits_{z_i}^\infty{\rm d}z\frac{1+z}{H^3(z)} \bigg/ \int\limits_{0}^\infty{\rm d}z\frac{1+z}{H^3(z)} \, .
\ee
The factor $G_i$ is necessary to account for the redshift dependent clustering evolution if the density field is reconstructed from galaxy positions on a light cone \citep{ata:2017} when compared to a reference redshift. We will explain this important point in the next section.

We can now write the Poisson likelihood expression for a single survey as
\ba
\label{likelihood}
\mathcal{L} (\mbi{N}_{{\rm G}} \vert\mbi\lambda)=\prod_{i}\frac{\lambda_{i}^{{N}_{{\rm G}i}}{\rm e}^{-\lambda_{i}}}{{N}_{{\rm G}i}!}\,.
\ea

As stated above, in this paper we seek a reconstruction given multiple surveys. Thus we formulate a multi-survey likelihood with $k$ surveys as

\be
\label{multilikelihood}
\mathcal{L}^{\rm multi} = \prod_k \mathcal{L}^{(k)} (\mbi {N}^{(k)}_{\rm G}\vert\mbi {\lambda}^{(k)} )
\ee

This likelihood is then sampled using a Hybrid Monte-Carlo technique \citep[HMC,][]{duane:1987,neal:2011}, in which 
artificial momentum variables are introduced to avoid inefficient random walks in our high-dimensional space of matter density grid cells.
Our method is based on that first introduced in \citet{ata:2015}, but with a generalization for the multi-survey likelihood. 
Due to its technical nature, we will describe it in Appendix~\ref{app:hmc}.

\subsection{Velocity Reconstructions}
Another important aspect of the density reconstructions is the iterative correction of so-called redshift space distortions \citep{kaiser:1987,hamilton:1997}, where galaxy positions are displaced along line of sight due to their peculiar motions.

Let us denote the apparent line-of-sight position of a galaxy with $\mbi s$, which is the real space position $\mbi r$ biased by the 
radial component of the galaxies' peculiar motion\footnote{The factor $1/(H(a)\,a)$ sometimes is omitted in literature if the distance measure is conveniently expressed in velocity units.}
\be
\label{eq:rsd}
\mbi s = \mbi r + \frac{\mbi{v}_{\rm r}}{H(a)\,a} \, ,
\ee
with
\be
\mbi{v}_{\rm r} = (\mbi v \cdot \mbi r)\mbi{e}_r \, ,
\ee
where $\mbi{e}_r$ is the radial unit vector.

Solving Equation \ref{eq:rsd} for $\mbi r$ demands the knowledge of $\mbi v $, which we can compute from the density contrast using linear perturbation theory as
\be
\mbi v = -f H(a)\, a \mbi \nabla \nabla^{-2} \delta\, ,
\ee
with $f={\rm d} (\log D)/{\rm d } (\log a)$ being the logarithmic derivative of the growth factor, called growth rate.
We solve Equation \ref{eq:rsd} iteratively in a Gibbs sampling scheme for the real space positions given the inferred density contrast of the previous iteration (see \citealt{kitaura:2016}).

\subsection{Data Preparation \& Reconstruction Setup}\label{sec:recon_setup}

The mock galaxy positions are given on the sky with right ascension $\alpha$ and declination $\delta$, and observed redshift $z^{\rm obs}$.
We map them to comoving Cartesian coordinates as
\ba \label{eq:coord_to_xyz}
x &=& d_{\rm com}\cos{\alpha}\cos{\delta}\, , \nonumber \\
y &=& d_{\rm com}\sin{\alpha}\cos{\delta}\, , \nonumber \\
z &=& d_{\rm com}\sin{\delta}\, , 
\ea
where the comoving distance $d_{\rm com}$ is given by
\be \label{eq:dcom}
d_{\rm com} = \frac{c}{H_0}  \int\limits_0^{z^{\rm obs}} \frac{{\rm d}z}{\sqrt{\Omega_{\rm M}(1+z)^3+\Omega_\Lambda} } \, .
\ee

We organize the reconstructions with two consecutive rectangular volumes (dubbed as Volume 1 and Volume 2), each with $N_{\rm C} = 600\times 100 \times 100$ cells. 
With a cell resolution of $d_{\rm Cell} = 1.875 \,h^{-1} \rm Mpc$ (to match the H15 density field; see Section \ref{sec:sims}),
this leads to a line-of-sight comoving length of $d_{\rm LOS} = 1125 \,h^{-1} \rm Mpc$, while in transverse dimension it is $d_{\rm Trans} = 187.5 \,h^{-1} \rm Mpc$. 
In preliminary tests, we found an unsatisfactory performance for the reconstructions at nearby comoving distances of $d_\mathrm{com} \lesssim 100\,\hmpc$, which 
we suspect is caused by the 700$\,\mathrm{deg}^2$ footprints of the 2m-z01 samples being insufficiently wide. In a real reconstruction, 
we expect to be able to use all-sky reconstructions of the Local Universe to cover $d_\mathrm{com} \lesssim 100\,\hmpc$ \citep[e.g.,][]{erdogdu:2006},
so we begin our reconstructions at $d_\mathrm{com} = 100\,\hmpc$.
Volume 1 thus spans a line of sight range of $L_{\rm LOS} =  [100 - 1225] \,h^{-1} \rm Mpc$, while Volume 2 covers $L_{\rm LOS} = [1125 - 2250] \,h^{-1} \rm Mpc$. There is an overlap of $ 100 \,h^{-1} \rm Mpc$ between the two volumes for consistency. {The coverage of the two reconstruction
volumes on one of our mock lightcones is illustrated in Figure~\ref{fig:lightcones}}.
We reconstruct the individual volumes at their mean redshifts (see Equation \ref{eq:expect}), which are $\bar{z}_1 = 0.25$ for Volume 1 and $\bar{z}_2 = 0.67$ for Volume 2.

For Volume 1 we combine the data of the 3 surveys 2m-z01, 4m-z03, and 8m-z10, whereas for Volume 2 we only use 8m-z10 data, as shown in Figure \ref{fig:lightcones}.

\subsection{Survey Completeness}
\begin{figure}
\includegraphics[width=0.48\textwidth]
{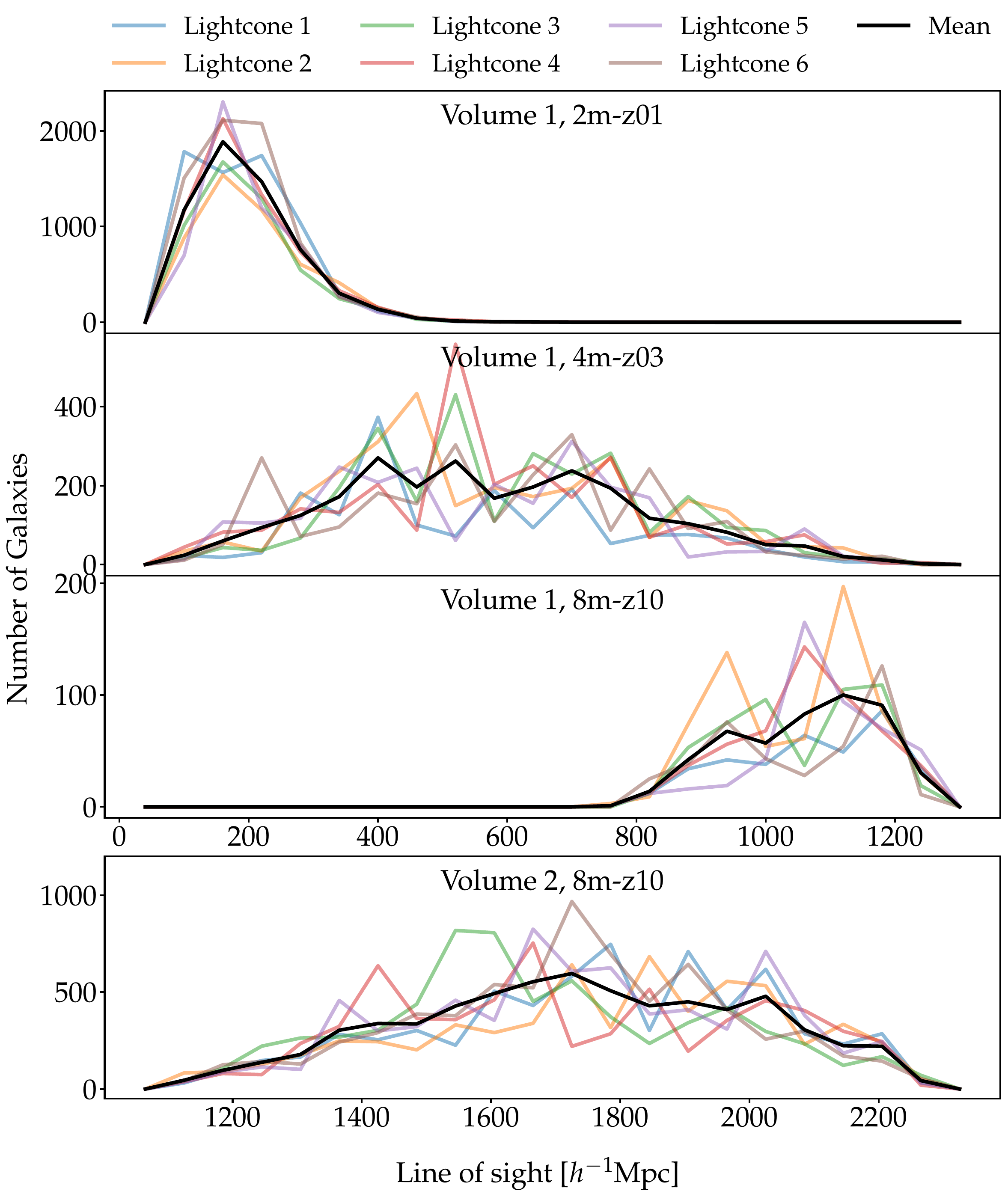}
\caption{\label{fig:radialdist}
Radial distribution of the different mock surveys. The colors indicate the 6 different lightcones extracted from H15, the black solid line is the mean over the lightcones. The 3 upper panels panels show the contributions of the individual surveys for Volume 1, the bottom panel for Volume 2.}
\end{figure}

We model the angular and radial selection function of each mock survey which we then combine to the survey completeness (see Equation \ref{eq:expect}).
In this paper we consider the angular footprint to not have any selection effect, i.e. we assume the angular footprints to be entirely uniform across their entire
area as summarized in Table \ref{tbl:mocks}.
The radial selection function, on the other hand, is more challenging to model. A natural way to model the expected galaxy numbers in radial direction can be done using the Schechter luminosity function (e.g. \citealt{erdogdu:2004}).
This is accurate if averaged over the whole sky. In our case, dealing with footprints of $\mathcal{O}(1{\,\rm deg}^2)$, we are prone to fluctuations in the large-scale structure. Therefore, we average the radial galaxy distribution from the 6 mock lightcones extracted from H15 and calculate the radial selection function from it, see Figure \ref{fig:radialdist}.

\begin{figure*}\begin{center}
\includegraphics[width=1.\textwidth, clip=true, trim=60 0 40 0]{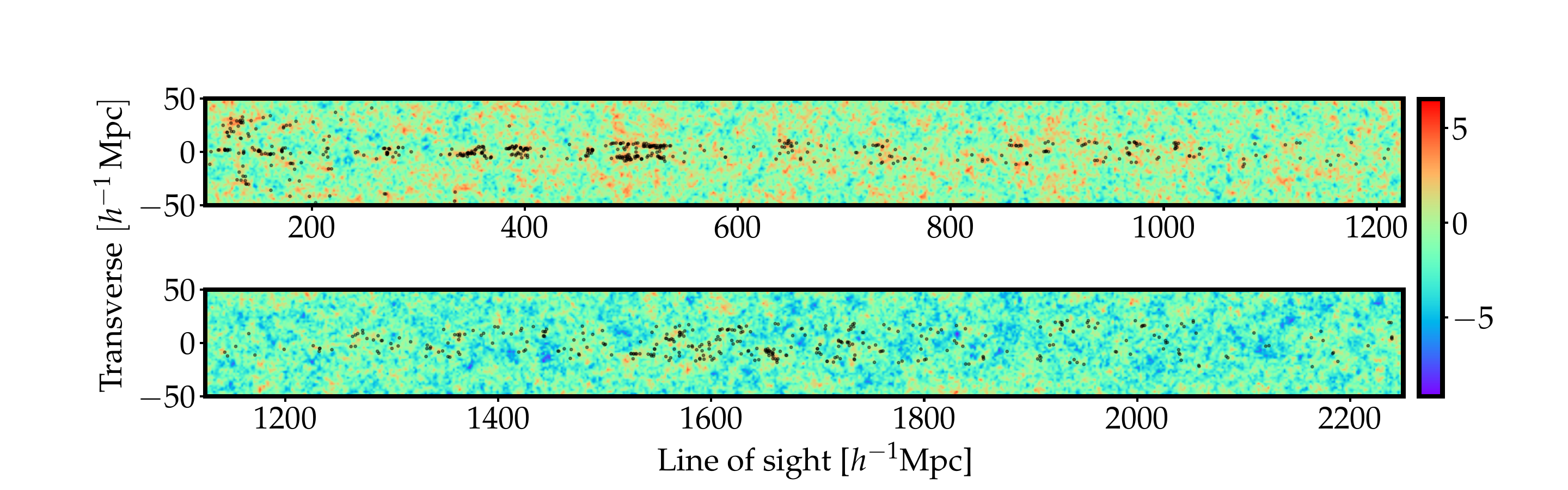}
\caption{\label{fig:slice}
Slice plots of the inferred real space density field $\log(1+\mbi \delta)$ from \argo{}, with ``observed" galaxy positions overplotted {as black dots}. Top panel shows Volume 1, bottom panel Volume 2.
As expected, the density field is consistent with the galaxy distribution.}
\end{center}\end{figure*}

\subsection{Galaxy Bias Fits}

We seek a single bias fitting function to obtain unbiased density fields for all lightcones extracted from H15. 
For an initial estimate we use linear bias measurements from completed galaxy surveys, in particular the bias values measured from SDSS \citep{tegmark:2004} for the low redshift contribution and VIPERS measurements \citep{di-porto:2016} for the high redshift contribution.

We find that a moderate increase of the galaxy bias with redshift $z$ fits the data sufficiently well, yielding nearly unbiased density fields with respect to the H15 simulations.
\be
b(z) = b(z=0.33) + 0.22\, z \,
\ee
where $b(z=0.33) = 1.24$. 

However, when we compute the integrated line-of-sight density, we still find a small bias in the resulting values in comparison with the `truth'. 
Due to the time-consuming computational requirements to carry out the reconstructions, it is 
prohibitive to iterate over different bias models to find a more accurate bias model. 
Instead, we will correct for the residual bias as a post-processing step which we describe in the following sub-section.

\subsection{Reconstruction Results}\label{sec:recon_results}
Examples of the resulting reconstructed real-space matter density fields are shown in Figure \ref{fig:slice}. The Volume 1 and Volume 2 maps are shown on the top and the bottom panel, respectively.
On top of the density slices we plot the inferred real space positions of the galaxies. For each reconstruction, we run the HMC sampler for up to 10,000 iterations. However 
the posterior samples are found to have correlation lengths of $\sim 100-150$, thus we extract 50 iterations of each density field for analysis with the
selected samples separated by at least 150 iterations on the chain. This allows us to not only estimate
the integrated line-of-sight dispersion measure, but also its uncertainty.

\argo{} reconstructs the real-space matter density field based on input catalogs of galaxy positions and redshifts, but for a given localized
FRB with known redshift, we do not have {\it a priori} information regarding its real-space position along the line-of-sight.
We therefore have to convert the FRB redshift into a line-of-sight comoving distance using the Hubble Law in order to determine its coordinate within the reconstruction box. 
In other words, we are forced to ignore the line-of-sight peculiar velocity of the FRB {host galaxy} when computing its integrated line-of-sight dispersion measure. This, however,
is typically of order several hundred kilometers per second, corresponding to an error of up to several comoving Mpc along the line-of-sight. 
For FRBs at cosmological distances, the error from ignoring the peculiar velocity contribution is negligible in comparison with the 
overall line-of-sight paths of hundreds or
thousands of megaparsecs.

 % DM_Recon.ipynb and DM_Recon_lowz.ipynb
 \begin{figure}
\includegraphics[width=0.48\textwidth]{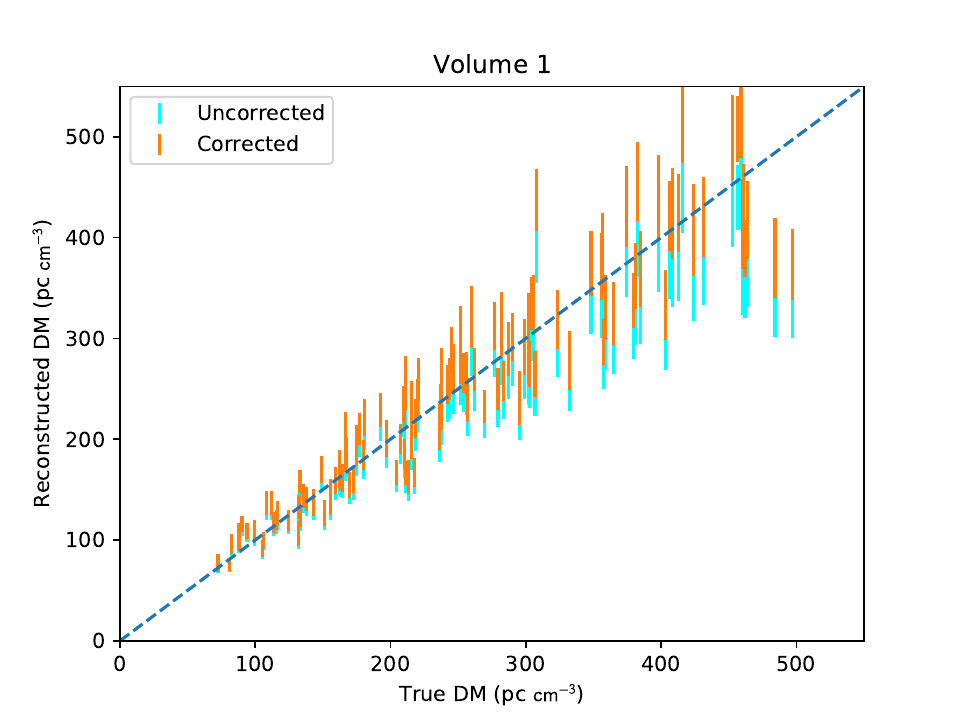}
\includegraphics[width=0.48\textwidth]{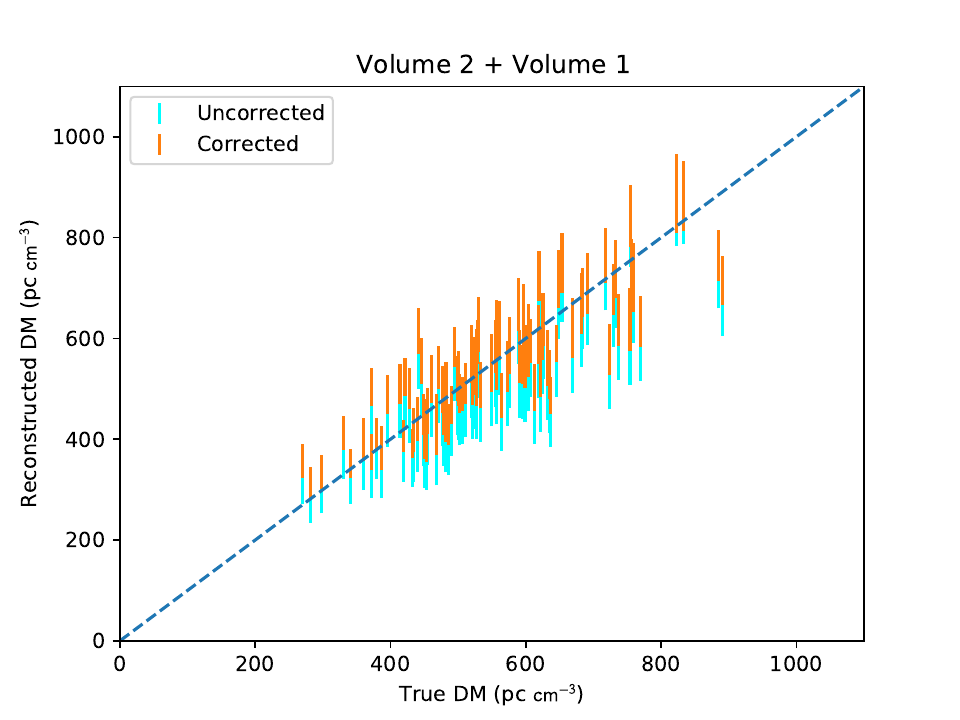}
\caption{\label{fig:dm_vs_true}
The reconstructed \dmigm{} values from the \argo{} reconstructions compared with the `true' values in the Millennium simulation, shown in cyan. The top panel shows the 
results from an ensemble of random sightlines at $0.1<z<0.4$ that are entirely contained within reconstruction Volume 1, while the bottom panel shows the same
for $0.45<z<0.8$ sightlines that span both Volumes 2 and 1. The error bars indicate the standard deviation of the reconstructed \dmigm{} evaluated over 50 \argo{}
realizations. The orange bars show the results after correcting the reconstructed \dmigm. }
\end{figure}

We first smooth the reconstructed density field by $R_{sm}=0.7\,\hmpc$ to match the
smoothing scale we adopt for the Millennium density field. Next, we calculate \dmargo{}, a version of the discretized line-of-sight integral in Equation~\ref{eq:dmigm_sim} in which the smoothed matter density $\delta^{sm}_{m}$ is evaluated on the reconstructed matter density grid from \argo{} rather than the `true' Millennium density field as in Equation~\ref{eq:dmigm_sim}. For the purposes of the analysis in this Section only, we will adopt $\figm=1$ for simplicity, although it will be considered a free model parameter in our forecasts of Section~\ref{sec:fisher}. As noted earlier, our reconstructions skip the first $100\,\hmpc$ of the line-of-sight from the observer towards the FRB,  so we simply add in a fixed contribution of $\dmigm( d_\mathrm{com}<100\,\hmpc\, | \, \figm = 1) = 33 \,\pccmcube$ to all the derived \dmigm{} values.

Since we know the `true' \dmigm{} values for each mock FRB sightline from the H15 Millennium simulations, we can directly assess the accuracy and precision
of the \dmargo{} relative to the true \dmigm{}. These are shown in Figure~\ref{fig:dm_vs_true}, where the dispersion measures are evaluated for
an ensemble of random `FRB' sightlines at $0.1<z<0.8$ with random angular positions within 12 arcmin of the lightcone field centers. We see that the reconstructed \dmigm{} deviate
slightly from an exact one-to-one relationship with the true \dmigm{}, while also having some scatter. We therefore fitted a quadratic offset as a function of
the median reconstructed \dmigm{} of each sightline to correct this small bias --- this was done separately for sightlines contained within Volume 1, 
and for the combined \dmigm{}  of Volume 1 and 2 in the case of sightlines that originate in Volume 2. The orange points in Figure~\ref{fig:dm_vs_true} 
show the reconstructed \dmigm{} after implementing this correction. 
In real observations, similar corrections can also be carried out if realistic simulations and mock catalogs
that reflect the observational data are available.

Since it is the goal of this paper to carry out Fisher parameter forecasts, it is necessary to check whether the distribution of the reconstructed \dmigm{} from
the \argo{} realizations reflect the uncertainty of the line-of-sight integrated density. To do this, we compute the {``pull''} distribution, $(\mathrm{DM_{igm,argo} }- \mathrm{DM_{igm,true}})/\sigma_\mathrm{argo}$.
If the scatter of the \argo{} realizations accurately reflects the true error in our estimate of \dmigm{}, 
then the pull should be distributed as a Gaussian with a standard deviation of 1. 
We find the pull distributions of $\mathrm{DM_{igm,argo} }$ to be approximately Gaussian, but with standard deviations of 1.63 and 1.55 for Volume 1 and Volume 1+2, respectively. 
{There is a small skew to the pull distributions, with slightly more outliers at negative values where $\mathrm{DM_{igm,argo} }$ is underestimated 
relative to $\mathrm{DM_{igm,true}}$. This suggests that the \argo{} reconstructions do not currently recover the full amplitude in some of the density peaks. In an actual likelihood analysis on real data, this would need to be corrected to avoid biasing the results. 
For the purpose of our subsequent Fisher forecast, however, we will treat the pull distribution as Gaussian} but broaden the $\mathrm{DM_{igm,argo} }$ distributions of each sightline around its median by the aforementioned
factors of 1.63 and 1.35. This rescales the scatter in each sightline as to be more accurate representations of the uncertainty in the \dmigm{} measurement from the 
foreground galaxy redshift measurement.

%Test_DM_Recon.ipynb
\begin{figure*}\centering
\includegraphics[width=0.8\textwidth]{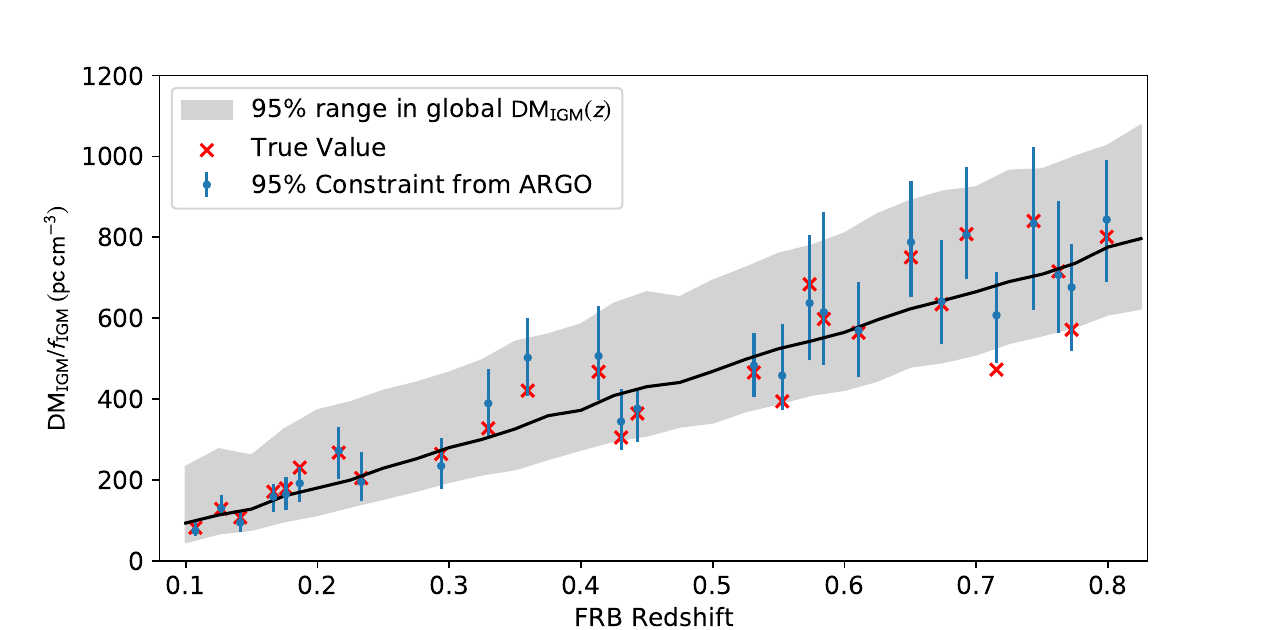}
\caption{\label{fig:dm_vs_z}  Solid line and gray regions show the mean and central 95\% range of the the global \dmigm{} distribution within the Millennium 
simulation as a function of redshift, assuming $\figm=1$. Red crosses show the \dmigm{} toward randomly-selected sightlines within the H15 lightcone catalogs, 
the blue error bars show the \dmigm{} constraint from the \argo{} matter density reconstructions of the same catalogs. 
}
\end{figure*}

We illustrate the efficacy of the \argo{} reconstructions in Figure~\ref{fig:dm_vs_z}, which shows the global $\langle \dmigm \rangle$ as well as its
central 95\% distribution (i.e.\ the range between the 2.5th and 97.5th percentiles of the distribution), as a function of FRB redshift. We also show as crosses the \dmigm{} corresponding to some randomly selected sightlines at $0.1<z<0.8$ selected from the lightcones. The error bars
show the corresponding constraints on those sightlines obtained by applying \argo{} on the foreground spectroscopic surveys described in Section~\ref{sec:specmocks}. 
The \argo{} reconstructions are effective in bracketing the true \dmigm{} to within an uncertainty that is $\sim 2-3\times$ smaller than that from cosmic variance.
The efficacy does appear to degrade somewhat toward the higher-redshift end ($z\gtrsim 0.7$), with broader errors as well as catastrophic failures where the
reconstructed range falls outside the true value. There is ongoing work to improve the performance of \argo{}, but in the meantime we will proceed with
assuming this performance for our Fisher forecasts.

In all sightlines, the \argo{} reconstructions allow us to constrain \dmigm{}, assuming a fixed $\figm = 1$, with better precision than if the global
\dmigm{} distribution were to be assumed. Conversely, if \figm{} was an unknown parameter, the density fluctuations estimated by \argo{} would then be used
to fit for the most likely \figm{} that leads to the observed \dmigm{}. This is the scenario that will be adopted in the Fisher forecasts
of the following section.

% %%%%%%%%%%%%%%%%%%%%%%%%%%%%%%%%%%%%%%%%%%%%%%%%%%%%%%%%%
% %%%%%%%%%%%%%%%%%%%%%%%%%%%%%%%%%%%%%%%%%%%%%%%%%%%%%%%%%
% %%%%%%%%%%%%%%%%%%%%%%%%%%%%%%%%%%%%%%%%%%%%%%%%%%%%%%%%%
\section{Fisher matrix forecast}
\label{sec:fisher}

To predict how accurately we can infer the model parameters from future datasets, we perform a Fisher matrix analysis. Given our four model parameters $p = \{ f_{\rm igm}, r_{\rm max}, f_{\rm hot}, \bardmhost \}$, and the observable $f_m\left( p \right) \equiv {\rm DM^{model}_{eg}}$,
where the subscript `eg' denotes `extragalactic', and the elements of the Fisher matrix are given by the sum over $m$ FRBs:

\begin{equation}\label{eq:fisher}
    \mathcal{F}_{i,j} = \sum_m \frac{1}{\sigma_m^2}\frac{\partial f_m }{\partial p_i}~\frac{\partial f_m }{\partial p_j}.
\end{equation}

The model DM, $f_m\equiv {\rm DM}^\mathrm{model}_\mathrm{eg,m}$, is calculated for each individual FRB sightline $m$ in our mock sample:
\begin{align}
{\rm DM}^\mathrm{model}_{\mathrm{eg},m} =& \langle {\rm DM}_{\mathrm{argo},m}(\figm) \rangle  \nonumber \\ &+ \mathrm{DM}_{\mathrm{halos},m}(\rmax, \fhot) \nonumber \\ &+  \mathrm{DM_{host}}(1+z_m)^{-1},
\end{align}
where $\langle {\rm DM}_{\mathrm{argo},m}(\figm) \rangle$ is the median \dmigm{} value
computed based on the \argo{} density reconstruction for sightline $m$, 
$\mathrm{DM}_{\mathrm{halos},m}(\rmax, \fhot)$ is the \dmhalo{} computed given the 
list of intervening halos in sightline $m$ (including 0.3 dex halo mass error), while
$z_m$ is the redshift of the FRB (Compare with Equation~\ref{eq:dm_eg}).
{In the model calculation we also make the approximation $\dmhost = \bardmhost$, i.e.\ the host contribution
is represented by its population mean.}

The uncertainty $\sigma_m$ on each observable $f_m$ is calculated by adding up in quadrature: \sigmw, the observational uncertainty on the extragalactic dispersion measure after subtraction of the Milky Way component; \sigigm, uncertainty on the intergalactic contribution from the \argo{} reconstructions; { \sighost{}, the intrinsic scatter in the distribution of rest-frame \dmhost}; and
\sighalo, the error induced by the uncertain halo mass of the intervening ($\sim$ arcmin-separation) galaxies in the line-of-sight of each mock FRB:
\begin{equation}\label{eq:fishersig}
    \sigma_m = \sqrt{ \sigmw^2 + \sigigm^2 + \sighalo^2 + \sighost^2}.
\end{equation}

For all FRBs in each mock sample, we assume $\sigmw = 15\,\pccmcube$ ensues from subtraction of the Milky Way component {as discussed in Section~\ref{sec:dm_frb}}. The \sigigm{} component is given by the standard deviation of DM$_\mathrm{igm, argo}$ along individual sightlines based on the Monte Carlo realizations of the \argo{} density reconstructions. This is 
subsequently corrected to be unbiased and have accurate error properties (Section~\ref{sec:recon_results}). 
The \sighalo{} component, on the other hand, arises primarily from the halo mass uncertainty which goes into our modified NFW halo gas model 
(Equations~\ref{eq:rhob_halo} and \ref{eq:mhalo_bar}) for intervening galaxies within 10\arcmin{} of each sightline. 
We assume that the halo masses can be estimated to within {$\pm$0.3 dex, which incorporates uncertainties in 
estimating galaxy stellar masses and subsequent conversion to halo mass using the stellar mass-halo mass relationship \citep[see, e.g.,][]{wechsler:2018}. }
Therefore, for the $<10\arcmin{}$ intervening galaxy catalog of each sightline, we
generated 100 Monte Carlo realizations of the \dmhalo{} where we randomly drew combinations of different halo masses for the same set of intervening galaxies with
$\sigma(\log_{10} \mhalo/M_\odot) = 0.3$. The standard deviation of the resulting Monte Carlo realizations of \dmhalo{} is then adopted as \sighalo{} for that sightline. {Finally, we adopt a fixed scatter of $\sighost = 50\,\pccmcube$ for the host contribution, 
which is a guess based on the currently-known sample of localized FRBs \citep{cordes:2021}. In future analyses of real data, we might aim to treat this
scatter as a free or unknown parameter.}

While the true distributions of the errors above may be non-Gaussian (especially $\sigma_{\rm halo}$), for simplicity we  take the standard deviations. To improve the flow of this text, we divert the description on the partial derivatives $\partial f_m/\partial p_i$ in Equation~\ref{eq:fisher} to the Appendix~\ref{app:fish_der}. We assume that all cosmological parameters, including $\Omega_\mathrm{b}\,h^2$, are fixed.
The covariance matrix of parameter constraints from a given set of FRBs and their associated foreground data is then given by the inverse of the Fisher matrix, $\mathcal{F}_{i,j}^{-1}$. 

\begin{table}[]
\centering
\caption{\label{tab:fiducial_par} 68.3\% Limits From Fisher Forecast}
\begin{tabular}{lccc}
Parameter & Fiducial Value  & $N=30$  & $N=96$  \\
\tableline 
\figm{}      & 0.8      & $\pm 0.11$ & $\pm 0.06$      \\
$r_\mathrm{max}$/$r_{200}$      & 1.4    & $\pm 0.35$  & $\pm 0.15$        \\
\fhot{}      & 0.75    &  $\pm 0.15$ & $\pm 0.09$       \\
\dmhost{} (\pccmcube)   & 100   & $\pm 29.21$ & $\pm 18.36$     \\
\tableline   
\end{tabular}
\end{table}

{A Fisher forecast yields only the estimated errors from an experiment, i.e.\ it is not a mock MCMC likelihood analysis which would
actually estimate the parameters from a simulated data set. However, many of the partial derivatives in the Fisher matrix (Equation~\ref{eq:fisher}), and hence the
forecasted errors, do have a dependence on the assumed parameters in the calculation (Appendix~\ref{app:fish_der})
}
We therefore adopt the fiducial set of parameters in Table~\ref{tab:fiducial_par}, and also impose flat priors such that $0\leq\figm \leq 1$, $0 \leq \rmax/r_{200} \leq 5$, $0\leq \fhot \leq 1$, and $0\,\pccmcube \leq \bardmhost \leq 1000\,\pccmcube$.

\subsection{Foreground Mapping Constraints}

We will make specific forecasts for two hypothetical FRB samples: (i) $\nfrb=30$ spanning $0.1<z<0.5$, and (ii) $\nfrb=96$ over $0.1<z<0.8$. 
 The $\nfrb=30$ sample approximately reflects the goals for the ongoing FLIMFLAM survey as described in Section~\ref{sec:flimflam}, 
 while foreground data sets covering $\nfrb=96$ should be achievable with the advent of 8m-class massively-multiplexed spectrographs. 
 
{Early samples of localized FRBs show a roughly uniform redshift distribution \citep[e.g.,][]{heintz:2020}.
However, the number of known localized FRBs is increasing rapidly and will soon outstrip the number that can be targeted for foreground mapping. 
At that point, 
the redshift distribution of FRBs to be targeted for foreground mapping will become a deliberate choice. 
We therefore choose to adopt a uniform redshift distribution for our mock samples, which is a tradeoff between more expensive observational
requirements of higher-redshift FRBs and better constraints on \bardmhost{} offered by a wider redshift range.} {
In the longer term ($\gtrsim 2-3$ years) as samples of localized FRBs increase into the hundreds or thousands, the true underlying
FRB redshift distribution might turn out not to be uniform with redshift.}

\begin{figure*}
\center
\par\bigskip
\begin{overpic}[width=0.42\textwidth]{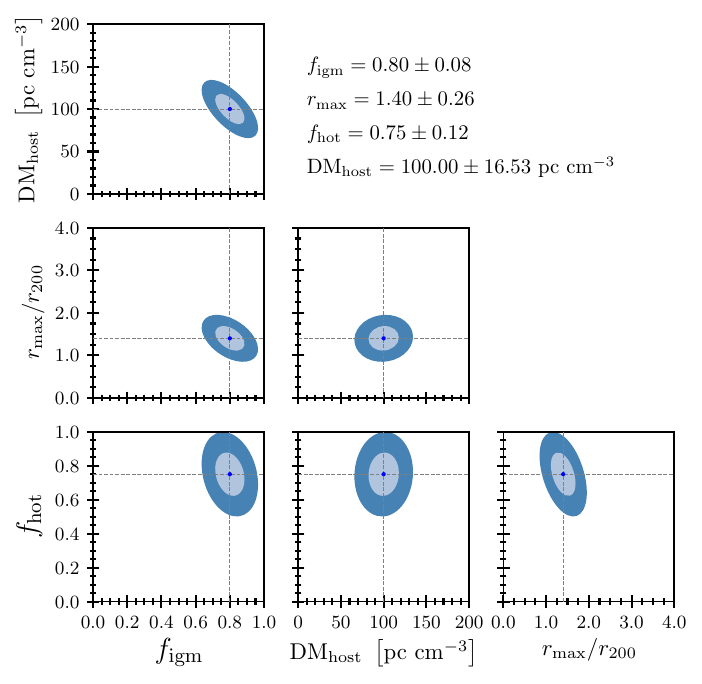}
\put(16, 98){\textsf{(a) $\nfrb=30$ with Foreground Mapping}}
\end{overpic}
\begin{overpic}[width=0.42\textwidth]{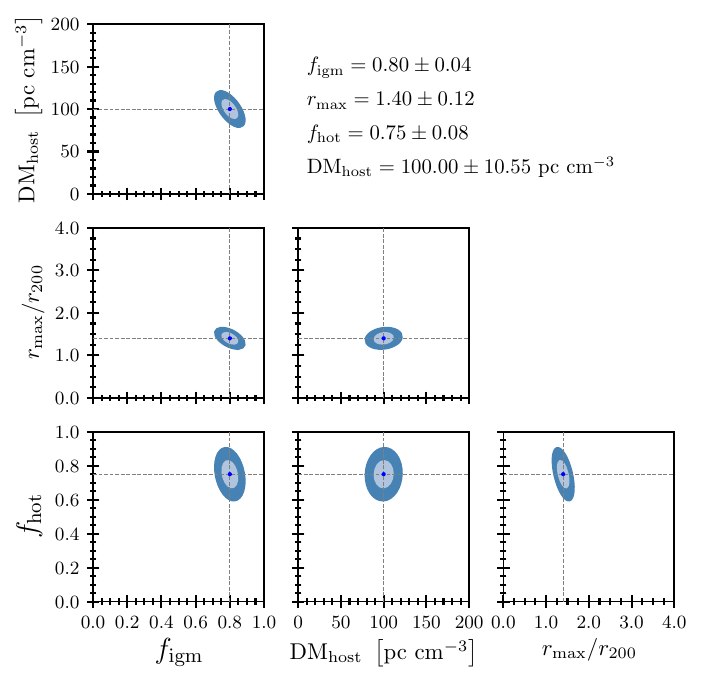}
\put(16, 98){\textsf{(b) $\nfrb=96$ with Foreground Mapping}}
\end{overpic}
\caption{\label{fig:forecasts}
Fisher matrix forecast on the parameter constraints given by (a) $\nfrb=30$ FRB sightlines
at $0.1<z<0.5$; and (b) $\nfrb=96$ FRB sightlines at $0.1<z<0.8$. The light-blue contours show the
$68\%$ confidence intervals, while the dark blue indicates the $95\%$ intervals.}
%\kg{Ilya please update these figures to the versions with the prior.}}

\end{figure*}

{From the resulting Fisher matrices, we compute the 68\% and 95\% error ellipses using
standard techniques (usefully summarized in \citealt{coe:2009}).}
Figure~\ref{fig:forecasts}a shows the forecasted confidence intervals for the $\nfrb=30$ sample. With foreground spectroscopy from both wide-field surveys (described in Section~\ref{sec:specmocks}) as well
as deeper but narrow-field observations of intervening galaxies, we find that the 68.3\% confidence constraints are $\figm = 0.80 \pm 0.11$, $\rmax/r_{200} = 1.40 \pm 0.35$, $\fhot = 0.75 \pm 0.15$, and $\bardmhost = 100.0 \pm 29.21\; \pccmcube$. 

We also experimented with drawing several different random mock samples with the same number of sightlines spanning 
the same redshift range, and find that there is some heterogeneity in the forecast between the different mock samples with $\nfrb = 30$.
This is caused by the variance in the number of intervening halos in the mock sightlines: `clean' sightlines with few intervening halos 
tend to be more useful for constraining $\figm$ and $\dmhost$, whereas sightlines with more intervening galaxies are better
for constraining \rmax{} and \fhot.
This is reflected in the Fisher terms contributed by the individual sightlines (c.f.\ Equation~\ref{eq:fisher} and Appendix~\ref{app:fish_der}), where we
find that sightlines with more intervening halos have larger values of 
$\partial \dmhalo/\partial \rmax$ and $\partial \dmhalo/\partial \fhot$ than those without.
We also note, from experimenting with the redshift range of the sightline distribution, 
that a broader redshift distribution helps break the degeneracy between \figm{} and \dmhost{} thanks to the $(1+z_\mathrm{host})$ 
denominator associated with \dmhost{} (Equation~\ref{eq:dm_eg}). 
As expected, lower redshift FRBs have greater sensitivity toward \dmhost{}. These trends should help guide the choice of FRB sightlines to 
follow-up with foreground spectroscopy once upcoming surveys start delivering large numbers of localized FRBs.

With $\nfrb=96$ sightlines over $0.1<z<0.8$ (Figure~\ref{fig:forecasts}b), we forecast 68\% constraints of $\figm = 0.80 \pm 0.06$, $\rmax/r_{200} = 1.40 \pm 0.15$, $\fhot = 0.75 \pm 0.09$, and $\bardmhost = 100.0 \pm 18.36\; \pccmcube$, i.e.\ all the IGM and CGM parameters are constrained to within $\sim 10\%$. This shows the high precision that should, in principle, be achievable
once the next generation of massive-multiplexed spectroscopic facilities become available. 

\subsection{Constraints without Foreground Maps}

\begin{figure}
    \centering
    \vspace{2em}
    \begin{overpic}[width=0.42\textwidth]{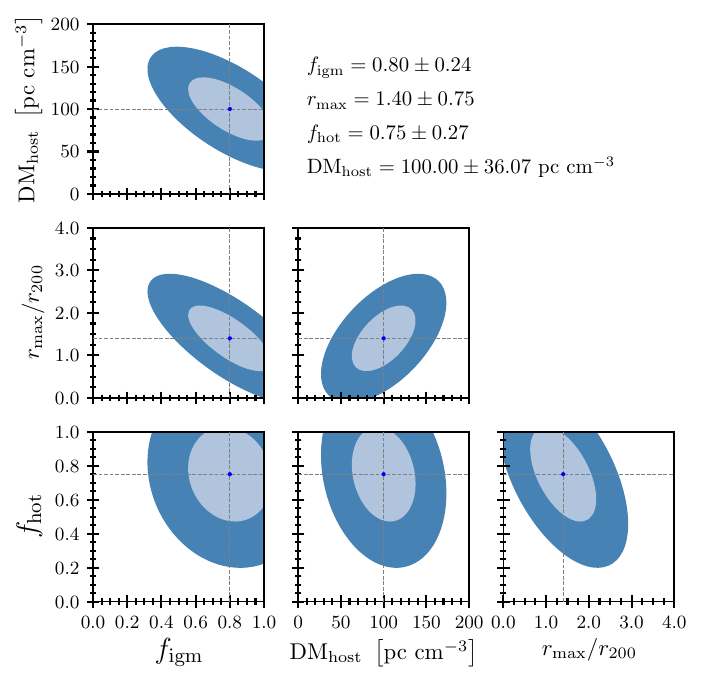}
        \put(18, 98){\textsf{$\nfrb=96$ without Foreground Data}}
    \end{overpic}
    \caption{Parameter forecast in the case of $\nfrb=96$ localized FRB sightlines,
    for which no foreground data is available. The constraints are clearly considerably weakened from the lack of foreground data (c.f.\ Figure~\ref{fig:forecasts}).}
    \label{fig:n96_no_fg}
\end{figure}

We can also make Fisher forecasts of constraints on the same set of parameters assuming \textit{only} the redshift and $\mathrm{DM_{eg}}$ from the FRBs, i.e.\ without
any knowledge of the foreground structures or galaxies. In practical terms, this is equivalent
to {trying to fit a theoretical Macquart relation to the extragalactic dispersion measures and host galaxy redshifts from an observed sample of  
localized FRBs. }
The parametric model we need to differentiate for the Fisher matrix then becomes:
\begin{align}
\label{eq:obs_no_fg}
f_m\left(p\right) \equiv {\rm DM}^\mathrm{model}_{\mathrm{eg},m} =& \langle {\rm DM}_\mathrm{igm}(\figm) \rangle (z_m) \nonumber \\ &+ \langle \mathrm{DM}_\mathrm{halos}(\rmax, \fhot) \rangle (z_m) \nonumber \\ &+ \bardmhost(1+z_m)^{-1}.
\end{align}
Instead of the per-sightline model calculations that are feasible with foreground spectroscopic data, we are now forced to use the global $\langle \dmigm \rangle(z)$ and 
$\langle \dmhalo \rangle (z)$ (i.e. Figure~\ref{fig:sigigm}a), evaluated at the redshift of each FRB (see Appendix~\ref{app:fish_der} for the partial derivatives). Similarly, for the $\sigma_m$ calculation (Equation~\ref{eq:fishersig}) we insert the
global scatter for $\sigigm$ and $\sighalo$ (see Figure~\ref{fig:sigigm}b) instead of 
bespoke calculations based on the foreground data from each sightline. 
The forecasted parameters for the same mock sample of $\nfrb = 96$ --- but without any foreground
information --- is shown in Figure~\ref{fig:n96_no_fg}. The constraining power clearly is much weakened in the 
absence of foreground data, with confidence intervals $\sim 2-5\times$ broader
in all the parameters. In addition, the degeneracy between $\rmax$ and \dmhost{} appears to have increased when foreground data is unavailable.

How much information does the foreground data bring to cosmic
baryon census, in addition to that offered by the localized FRBs and
their measured $\mathrm{DM_{eg}}$ values alone? We can attempt to quantify this by rescaling the elements of the no-foreground Fisher matrix (which yields the forecast in Figure~\ref{fig:n96_no_fg})
until the resulting constraints are roughly the same as if foreground data is available (Figure~\ref{fig:forecasts}b) for the
same sightlines. This is equivalent to increasing the number of sightlines available for analysis by that factor. 
We find that to approximately match the constraints of $\nfrb=96$ in Figure~\ref{fig:forecasts}b, we had to rescale the no-foreground Fisher matrix
elements by $\sim 25\times$. {In other words, the foreground data improves the Fisher information content such
that every localized sightline that has a foreground map is equivalent to $\sim 25$ localized FRBs where only their redshift and total DM is known}. Without
foreground data, samples
of $\nfrb \gtrsim 2000$ localized FRBs with known redshifts would be required to 
achieve similar constraints as $\nfrb \sim 100$ localized FRBs with foreground data.
{This is qualitatively consistent with the recent analysis of \citet{batten:2021a}, who found that $\sim 10^3$ localized FRBs are needed
to distinguish between their no-feedback simulations and those with feedback (i.e.\ in effect, detect the existence of the CGM);
whereas $\sim 10^4$ localized FRBs would be 
needed to distinguish between galaxy feedback models which lead to different CGM properties.}

\subsection{Comparison with Previous Forecasts}
Our forecasts are, on the surface, more pessimistic than previous efforts in the literature. \citet{walters:2019}, for example, carried out a MCMC likelihood
forecast and argued that  $\nfrb=100$ localized FRB sightlines without any foreground data should be able to constrain total diffuse
baryon fraction (i.e.\ combined IGM and CGM components) $f_d$ to within a few percent 
at 95\% confidence level. Meanwhile, \citet{ravi:2019} assumed the spectroscopic observations would be available for intervening galaxies within
several arcminutes of each FRB sightline, and argued that \figm{} could be constrained to within $\sim 2-5\%$ at 95\% confidence with $\nfrb=100$.
Both of these previous forecasts were built on analytic models. In particular, they both assumed that the uncertainty in the DM contribution of the cosmic web
or diffuse IGM is given by a standard deviation of $\sigigm=10\,\pccmcube$. 
{This value, which was taken from analytic estimates by \cite{shull:2018}},
is a significant underestimate of the \sigigm{} compared with simulation results by a factor of $>5$ (see
 Figure~\ref{fig:sigigm}, and see also \citealt{batten:2021}).
 Such a small $\sigigm$ is indeed even tighter than our \argo{} constraints
on individual line-of-sight dispersion measures as seen in Figure~\ref{fig:dm_vs_z}. 
Apart from this discrepancy, it is hard to make direct comparison with the earlier works since we have also adopted simultaneous constraints on galaxy halo parameters
as well as the host DM contribution. Both \citet{walters:2019} and \citet{ravi:2019} assumed a fixed \dmhost{} with some uncertainty, while \citet{ravi:2019}
assumed fixed halo parameters for the intervening foreground galaxies. We therefore argue that our forecasts are a more realistic determination of the precision
that can be achieved with upcoming FRB samples.

\section{FRB Foreground Mapping on the AAT}\label{sec:flimflam}

\begin{figure*}
\centering   % Plot_FRB180924_prelim.ipynb
\includegraphics[width=1\textwidth,clip=true,trim=55 0 55 0]{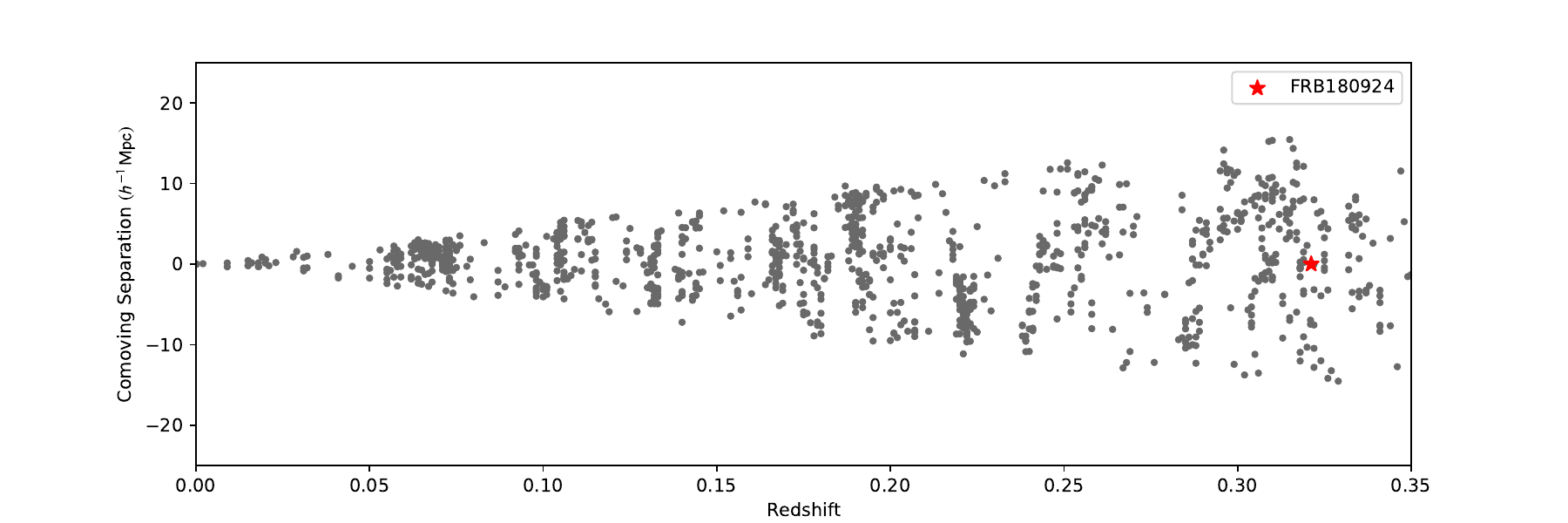}
\caption{\label{fig:frb180924}
Positions of galaxies in the foreground of FRB180924, based on a preliminary reduction of our 2dF-AAOmega data.
The grey dots show comoving coordinates of galaxies within a $10\,\hmpc$ slice along the plane of real ascension, 
with the position of the FRB host galaxy indicated by the red star.
}
\end{figure*}

As part of an effort to actualize the technique of FRB foreground mapping introduced in this paper, 
we have recently commenced the FRB Line-of-sight Ionization Measurement From Lightcone AAOmega Mapping (FLIMFLAM) Survey.
This is a nominally $\sim$40-night observational survey on the 2dF-AAOmega fiber spectrograph on the 3.9m Anglo-Australian Telescope (AAT; \citealt{lewis:2002}, \citealt{sharp:2006}) to map out the large-scale cosmic web in front of FRBs. 
Due to the preliminary stage of our data collection and analysis,
we simply give a brief introduction to the project in this section and defer a more detailed description to future papers.

The nominal goal of FLIMFLAM is to spectroscopically map the foreground of $\nfrb \sim 30$ localized FRBs over the redshift range
 $0.1<z<0.5$ in order to achieve
$\sim 15-20\%$ constraints on \figm{} as forecasted in the previous section. The lower redshift limit is because $z\lesssim 0.1$ foreground galaxies are typically covered by publicly-available redshift surveys like 2MASS, 6dF, and SDSS that cover large fractions of the extragalactic sky, 
while the upper limit of $z<0.5$ was deemed the 
highest practical redshift for which we can obtain reasonable samples of foreground galaxies within $\sim$ 2-3 nights of observing time
per field. Given the location of the AAT in Siding Spring, Australia, it is well-situated to follow up the increasing numbers of 
localized Southern Hemisphere FRBs 
being discovered predominately by the Commensal Real-Time
ASKAP Fast Transient \citep[CRAFT,][]{shannon:2018} survey located also in Australia, 
with subsequent host galaxy follow-up being conducted by the F$^4$ collaboration\footnote{\url{https://sites.google.com/ucolick.org/f-4}}.
{This collaboration is, at the time of writing, the largest single source of localized FRBs.}
We are in close collaboration with both CRAFT and F$^4$ teams which allows rapid access to FRB coordinates and redshifts 
as they are identified and localized,
although we will also target FRBs from other sources that have publicly available localizations and redshifts.

For each FRB field, we will typically observe multiple plate configurations over a single pointing of 2dF-AAOmega's $3.1\;\mathrm{deg}^2$ 
field-of-view. Our target selection for a fiducial $z=0.3$ FRB field is similar to the ``4m-z03" mock sample described in Section~\ref{sec:specmocks},
namely a nominally complete targeting of $r_\mathrm{AB} < 19.8$ galaxies. Given the $\sim 350$ science fibers\footnote{The actual number of science fibers that are available for use fluctuates on any given observing night.}, it therefore takes seven repeated visits
on 2dF-AAOmega to target all $\sim 2400$ galaxies with this selection. Assuming 40 minutes of on-sky integration time and 20 minutes
of overhead time (detector readout, calibration exposures etc), it takes 7 hours of AAT time to observe the foreground field for a $z=0.3$ FRB. For FRBs at different redshifts, we adjust the $r$-magnitude limit of the foreground galaxies: for FRBs at $z<0.2$, we will adopt 
$r_\mathrm{AB}<19.2$, while conversely at $z>0.4$ we will target $r_\mathrm{AB}<20.6$ galaxies. This leads to roughly 1500
and 4800 foreground galaxies for the $z<0.2$ and $z>0.4$ fields, respectively, while for the fainter galaxies we have to 
increase on-sky exposures up to 90 minutes per galaxy. We use the 580V grating on the blue camera and 
385R grating on the red in conjunction with the $5700\,\mathrm{\AA}$
 dichroic in order to achieve a spectral resolution of $R \equiv \lambda/\Delta \lambda \sim 1100-1500$
over $3750\,\mathrm{\AA} <\lambda <8850\,\mathrm{\AA}$ --- a well-established setup for measuring galaxy redshifts on 2dF-AAOmega.

Assuming a roughly uniform redshift distribution of FRBs within $0.1<z<0.5$, this leads to an average requirement of $\sim 1$ night 
of on-sky observations per FRB. For $\nfrb = 30$, this leads to a total requirement of $\sim$40 nights if we add on a 33\% weather overhead. We commenced pilot observations in October 2020 and so far have had a total of 17.5 nights allocated through semester 2021B. 
We show some preliminary data on the foreground field of FRB180924 in Figure~\ref{fig:frb180924}, where we had first
transformed the celestial coordinates and redshifts of the foreground galaxies into comoving Cartesian coordinates using the transformations
described by Equations~\ref{eq:coord_to_xyz} and \ref{eq:dcom}. The foreground galaxies clearly trace out the cosmic web 
of large-scale structure even by eye, and should allow us to confirm whether the low DM values of FRB180924 is indeed
due to an underdense foreground \citep{simha:2021}. % https://ui.adsabs.harvard.edu/abs/2021arXiv210809881S/abstract

FLIMFLAM observations are not the only piece of the puzzle for successfully carrying out FRB foreground mapping. 
For FRB fields that do not have pre-existing imaging coverage from publicly-available surveys such as the DES or Pan-STARRS, 
we are using the Dark Energy Camera to obtain pre-imaging on our targeted FRB fields. 
In addition, we are also pursuing a coordinated
 campaign to observe the FLIMFLAM fields with 8-10m class telescopes in order to obtain spectra of intervening galaxies at $\sim 5\arcmin$ separations of the sightline
 down to limiting magnitudes of $\sim 24$.

\section{Conclusion}
In this paper, we analyzed the ability of spectroscopic foreground data, including wide-field galaxy surveys, to improve on the ability of localized 
fast radio bursts to constrain the cosmic baryon distribution. Using semi-analytic models for the cosmic dispersion measure that are consistent with
those from hydrodynamical simulations, we created mock data samples of both FRB dispersion measures and foreground galaxy catalogs,.
We then applied \argo{}, a state-of-the-art Bayesian density reconstruction algorithm, on the foreground galaxy catalogs to reconstruct the matter density
field. This allows us to constrain the line-of-sight matter density integrand of \dmigm{} to $\sim 2-3\times$ better precision than allowed by cosmic variance. In conjunction with spectroscopic data on the intervening galaxy halos within $<10\arcmin$, we make Fisher
forecast of the simultaneous constraints on parameters governing the IGM and CGM baryons that could be attainable with upcoming
samples of localized FRBs. We find that the foreground data enhances the constraining power
on our parameters of interest by a factor of $\sim 25\times$ over samples of localized FRBs
without foreground data.

Based on our analysis, we are now conducting various observations to gather foreground
observations on $\sim 30$ FRBs sightlines. This observational campaign includes FLIMFLAM, a spectroscopic galaxy redshift survey on the AAT 
designed to map out the foreground cosmic web in these fields. In addition to the observations, we are also actively building on the methods presented in
this paper to create a parameter 
estimation framework based on Bayesian Markov Chain Monte Carlo (MCMC).

{
In this paper, we have assumed a \dmhost{} distribution in the range of $100\,\pccmcube \lesssim \dmhost \lesssim 300\,\pccmcube$ in the FRB
restframe.
However, recent results \citep{niu:2021} hint at the existence of FRBs with much larger host dispersions ($\dmhost \sim 1000\,\pccmcube$).
FRBs with such large \dmhost{} could potentially contaminate a FRB sample intended for cosmic baryon analysis, leading to biased parameter
constraints. 
There are several possible ways to ensure that only `normal' FRBs with $\dmhost \sim 200\,\pccmcube$ are included in a cosmic
baryon analysis sample. Methods are emerging to model the FRB host contribution, including estimating the host galaxy DM using resolved
observations of FRB locations within their host galaxy \citep[e.g.,][]{chittidi:2020,mannings:2020,marcote:2020}, or by modelling the Balmer
line in the host galaxy spectra \citep[e.g.,][]{tendulkar:2017,bassa:2017,niu:2021}.
Evidence is also beginning to emerge that FRBs with unusually high \dmhost{} also exhibit various properties
that differ from `normal' FRBs, such as repeatability, scattering, Faraday rotation measure, and others (R.\ Shannon, private communication).
We therefore expect our framework to study cosmic baryons to be largely
unaffected by the existence of 
high-\dmhost{} sources.
}

Recently, \citet{cordes:2021} published an analysis of 14 localized FRBs in which they combined the observed DM together with scattering times to assess 
the reported host redshifts. They found that adopting a value of $\figm = 0.85 \pm 0.05 $ lead to the most optimal redshift constraints, 
leading them to question the host galaxy associations of two FRBs in the sample. This suggests that, for a sample of FRBs with high-confidence
host galaxy associations and redshifts (e.g.\ using the PATH method; \citealt{aggarwal:2021}), the scattering time of the FRB is a valuable piece 
of information that can in principle be combined with foreground spectroscopic data to further tighten constraints beyond those forecasted in this paper.

By the middle of the 2020s, the number of localized FRBs will almost certainly increase into the hundreds, {especially once 
Northern Hemisphere radio arrays such as the Deep Synoptic Array \citep[DSA;][]{kocz:2019} and CHIME Fast Radio Burst Project
\citep{chime/frb-collaboration:2018} acquire the capability to localize FRBs.
By then, next-generation massively-multiplexed 4m-class spectroscopic facilities such as DESI \citep{levi:2013} and
WEAVE \citep{dalton:2012} will map large numbers of FRB foregrounds more efficiently than FLIMFLAM at $z<0.5$, while
8m-class spectroscopic facilities such as Subaru PFS \citep{sugai:2015} and MOONS \citep{cirasuolo:2014} will also be available to push 
FRB foreground mapping to higher redshifts. }
It will then become feasible to comprehensively
constrain the cosmic partition of baryons in diffuse gas, and indeed potentially push towards the epoch of Helium-II reionization ($z\sim 3$) with sufficient allocations
of telescope time.

\appendix

\section{Detailed Description of Wide-Field Spectroscopic Surveys}
\label{app:specmocks}

In this Appendix, we explain how we designed the mock wide-field foreground spectroscopic surveys defined in Section~\ref{sec:specmocks}. Note that the galaxy photometry and number counts selected from the \citet{henriques:2015} catalog are somewhat 
 dependent on the assumptions of the simulation and semi-analytic model, and might depart from reality especially
 at higher redshifts. Thus, while we make \textit{ad hoc} photometric cuts on these mock catalogs to select the desired number density of galaxies 
 at specific redshift ranges, we expect real observations to use more realistic and refined selection criteria to achieve comparable galaxy
 number densities. The observing requirements we outline in here should, however, be approximately correct in terms of exposure times as a function of magnitude, 
 since they are based on the known performance of existing facilities.

\begin{description}
\item[\hspace{20pt} \MakeLowercase{2m-z01}]
 For much of the extra-galactic sky, there already exists good coverage of galaxy redshifts out to $z\lesssim 0.1$ 
that were obtained by 1-2m diameter class survey telescopes. In the 
Northern Hemisphere, the Sloan Digital Sky Survey (SDSS) Main Galaxy Survey \citep{abazajian:2009, blanton:2005a} on the 2.5m SDSS Telescope \citep{gunn:2006} has provided good spectroscopic coverage out to $z\lesssim 0.2$. 
Meanwhile, the 6dF Galaxy Redshift Survey \citep[6dFGRS, ][]{jones:2009} has covered the Southern Hemisphere with the 1m UK Schmidt Telescope, albeit with slightly lower redshfit ($z \lesssim 0.1$). The wide-field coverage of these surveys helps mitigate the boundary effects caused by limited field-of-view in more targeted surveys with larger telescopes. For example, at $z=0.03$, two degrees on the sky corresponds to only 3 Mpc in the transverse dimension, which makes it difficult to accurately reconstruct structures or indeed sample any galaxies from said structures. We note that while \citet{simha:2020}
used SDSS data to reconstruct the foreground structures in front of the $\zfrb = 0.1178$ FRB190608, this was in the relatively narrow Stripe 82 equatorial region 
and thus suffers from narrow-angle effects at the lowest redshifts ($z \lesssim 0.05$) --- they did not use 6dFGRS data which has wide-area coverage in the 
same region.
This narrow-angle aliasing introduces additional uncertainty to the density reconstructions unless wide-field data is incorporated at low redshifts. 
We therefore use the `all-sky' H15 lightcones from the Millennium database to generate this catalog, with an \textit{ad hoc} magnitude selection of $r<16.4$ to define a sample peaked at $\zgal \sim 0.05$, similar to 6dFGRS \citep{jones:2009}. This yields an area density of $12\,\persqdeg$, with which we generated a catalog spanning 700 sq deg in order to ensure wide-area coverage in the Local Universe.

\item[\hspace{20pt}\MakeLowercase{4m-z03}] Out to $z\sim 0.35$, it is straightforward to obtain large numbers of redshifts using both existing and upcoming 4m-class
spectroscopic facilities over at least several square degrees. For example, the GAMA survey \citep{driver:2011, liske:2015} has covered 286 sq
deg at this redshift range using the AAOmega spectrograph on the 4m Anglo-Australian Telescope --- conveniently, the individual H15 lightcones have the same field-of-view (3.1 sq deg) as AAOmega. We adopt the same magnitude cut
as GAMA  ($r < 19.8$) to generate this mock sample, which yields $770\,\persqdeg$ of galaxies --- with the $\sim 320$ useable fibers on AAOmega \citep{liske:2015}, this would require about
4-5hrs on-sky per FRB field assuming 30-45min exposure times per object.
In the near future, the Bright Galaxy Survey carried out with the 4m Dark Energy Spectroscopic Instrument \citep[DESI, ][]{levi:2013,desi-collaboration:2016} will cover the entire Northern Hemisphere to similar depth, redshift range,
and number density as GAMA, making it a powerful complement for future Northern radio experiments capable of FRB localization.

\item[\hspace{20pt}\MakeLowercase{8m-z10}]
At higher redshifts ($z\gtrsim 0.4$), 8m-class telescopes are required to achieve the necessary depths to obtain redshifts of typical galaxies,
while simultaneously having sufficient multiplexing and field-of-view to observe large numbers of galaxies,
thus the Prime Focus Spectrograph (PFS, \citealt{sugai:2015}) on the 8.2m Subaru Telescope is the ideal instrument for this regime. 
Since the Subaru Telescope is in the Northern Hemisphere, we assume that both SDSS
and DESI Bright Galaxy Survey data will be available to cover lower redshifts, and define a target sample to cover $0.3 \lesssim z \lesssim 1$. 
We adopt a photometric color-color selection inspired by that of the VIPERS survey \citep{guzzo:2014}:

\begin{eqnarray}
 (r-i) > 2\:  &\mathrm{or}& \: (r-i)  > 0.5\, (u-g),\nonumber \\
i &<& 22.75, \nonumber \\
r &>& 22.8, \nonumber  
\end{eqnarray}
where we have made \textit{ad hoc} modifications to get the desired redshift range within the H15 lightcones.
Also, 
the transverse comoving distance increases with redshift fixed angular separation (e.g.\ at $z=0.7$, one angular degree on the sky subtends
44 cMpc in the transverse dimension), it is possible to decrease the assumed field-of-view at reduced loss to our ability
to recover transverse structures. We therefore reduce the footprint of the 8m-z10 mock catalog from 3.1 sq deg in the lower-redshift
samples, in order to match the 1.25 sq deg of Subaru PFS.
These selections lead to a target density of $5700\;\persqdeg$, implying that Subaru PFS should be able to observe
each $\zfrb \sim 1$ field within approximately 3hrs, assuming 45min exposure times per galaxy like VIPERS which was also a 8.2m-class instrument. 
For FRBs at intermediate redshifts such as $z\sim 0.5$, it is possible to define target selections to cover those redshifts at a depth that is
accessible to 4m-class facilities like AAT/2dF-AAOmega, albeit with deeper integrations than defined in 4m-z03 above. However, due to the computational
expense of running the density reconstructions, in the mock analysis for this paper we will use the reconstruction volumes using 8m-z10 catalog even for sightlines at $z\sim0.5$. 

\end{description}

\section{Hamiltonian Monte-Carlo for Multi-Surveys}\label{app:hmc}

In this Appendix, we will generalize the Hamiltonian Monte-Carlo formalism from \citet{ata:2015} to sample the multi-survey
likelihood defined in Equation~\ref{likelihood}.

Firstly, we define the Hamiltonian $\mathcal{H}(\mbi{q},\mbi p)$, a function of the canonical coordinates of position $\mbi q$ and  momentum $\mbi p$. Introducing the potential energy $\mathcal{U}(\mbi{q})$ and the kinetic energy $\mathcal{K}(\mbi{p})$, we write
\ba
\label{eq:hamil}
\mathcal{H}(\mbi{q},\mbi{p})=\mathcal{U}(\mbi{q})+\mathcal{K}(\mbi{p})\,,
\ea
with the  kinetic term
\ba
\label{eq:mom}
\mathcal{K}(\mbi{p})=\frac{1}{2}\mbi{p}^{\rm T}\mat{M}^{-1}\mbi{p}\,,
\ea
where $\mat{M}$ is the positive definite mass matrix, representing the co-variance of the momenta $\mat M=\langle \mbi{p}^{\rm T} \mbi{p} \rangle$.

The posterior probability density then written as
\ba
\label{eq:canonical}
\mathcal{P}(\mbi{q},\mbi{p})=\frac{1}{Z}\,{\rm e}^{-\mathcal{H}(\mbi{q},\mbi{p})}\,,
\ea
where $Z$ acts as normalization of the posterior. 
Plugging in Equation \ref{eq:hamil} into Equation \ref{eq:canonical}, we get
\ba
\label{eq:factorial}
\mathcal{P}(\mbi{q},\mbi{p})=\frac{1}{Z}\,\mathcal{P}(\mbi{q})\mathcal{P}(\mbi{p})=\frac{1}{Z}\,{\rm e}^{-\mathcal{U}(\mbi{q})}\,{\rm e}^{-\mathcal{K}(\mbi{p})}\,.
\ea
We can see that the posterior now factorized into two separated probabilities depending solely on the potential energy: $\mathcal{P}(\mbi{q})$, and the kinetic energy: $\mathcal{P}(\mbi{p})$. 
Considering only the potential term of Equation \ref{eq:factorial} , we get:
\ba
\label{eq:pos}
\mathcal{U}(\mbi{q})=-\ln \mathcal{P}(\mbi{q})\,,
\ea
and finally for the kinetic term $K(\mbi{p})$:
\ba
\mathcal{P}(\mbi{p})\propto {\rm e}^{-\mathcal{K}}={\rm e}^{-\frac{1}{2}\mbi{p}^{\rm T} \mat M^{-1}\mbi{p}}\,,
\ea
which resembles a multivariate Gaussian distribution.

Now, in HMC the spatial positions are identified as parameters that we want to sample  $\mbi q = \mbi \delta_{\rm L}$, in our case the linearized density per cell.
In contrast,  the momentum variables, $\mbi  p$, are artificially introduced in each sampling step so that the HMC algorithm avoids ineffective random walks to move through the phase space. 
We use the Hamiltonian equations of motion to evolve the system in (pseudo-)time $t$. 
The partial derivatives of the Hamiltonian determine how $\mbi{q}$ and $\mbi{p}$ change with time, $t$, according to Hamilton's equations of motion 
\ba
\label{eq:ham1}
\frac{\rm d \mbi q}{\rm d t} & = & \frac{\partial {\mathcal H}(\mbi{q},\mbi{p})}{\partial \mbi p}=   \mat{M}^{-1}\mbi p  \,,\\
\frac{\rm d \mbi p}{\rm d t} & = & -\frac{\partial {\mathcal H}(\mbi{q},\mbi{p})}{\partial \mbi q}= -\frac{\partial \mathcal{U}(\mbi{q})}{\partial \mbi q} \,.
\ea
We use the leap-frog discretization scheme, which preserves phase space volume and is time reversible. For a single iteration we calculate the positions and momenta from $t$ to  $t+\epsilon$ as follows
\ba
\mbi p\left(t+\frac{\epsilon}{2}\right)&=&\mbi p(t)-\frac{\epsilon}{2}\frac{\partial\mathcal{U}}{\partial\mbi q}(\mbi q(t))\\
\mbi q(t +\epsilon)&=&\mbi q(t)+\epsilon \,\mat M^{-1} \,\mbi p\left(t +\frac{\epsilon}{2}\right)\\
\mbi p(t +\epsilon)&=&\mbi p\left(t +\frac{\epsilon}{2}\right)-\frac{\epsilon}{2}\frac{\partial \mathcal{U}}{\partial \mbi q}(\mbi q(t +\epsilon))\,.
\ea

The leap-frog scheme introduces numerical errors of order $\mathcal{O}(\epsilon^3)$ so that one has to introduce a Metropolis-Hastings rejection step. The new state obtained after  $N_\epsilon$ steps forward with step size $\epsilon$ will be accepted with a  probability of
\ba
{\cal P}_{\rm acceptance}=\min\left[1,{\rm e}^{-\Delta \mathcal{H}(\mbi q,\mbi p)}\right]\,,
\ea
where  $\Delta \mathcal{H}(\mbi q,\mbi p)=\mathcal{H}(\mbi q',\mbi p')-\mathcal{H}(\mbi q,\mbi p)$ is the  difference in the Hamiltonian between the old $(\mbi q, \mbi p)$ and new $(\mbi q', \mbi p')$ state.

According to Equations \ref{eq:pos} and \ref{eq:ham1} we need the negative natural logarithm of the posterior of Equation \ref{eq:bayes} and its gradient with respect to the linearized density $\mbi \delta_{\rm L}$:

\be
\label{eq:posterior}
-\ln \mathcal{P} =-\ln \pi-\ln \mathcal{L}^{\rm multi} \, ,
\ee
where
\be
-\ln \mathcal{L}^{\rm multi} = \sum_k -\ln \mathcal{L}^{(k)}\, .
\ee

Thus, we write for the prior of Equation \ref{prior} 
\be
\label{lnprior}
-\ln \pi(\delta_{\rm L})=\frac{1}{2}\mbi \delta^{\rm T}_{\rm L}\mat C_{\rm L}^{-1}\mbi \delta_{\rm L}+\rm const\,,
\ee
where $\rm const$ does not depend on $\mbi \delta_{\rm L}$.

For the likelihood defined in Equation \ref{likelihood}, we get
\be
\label{lnlikelihood}
-\ln \mathcal{L}^{\rm multi} (\mbi{N}_{{\rm G}}\vert\mbi\lambda)=\sum_{k}\sum_{i}\lambda^{(k)}_{i}-{N}^{(k)}_{{\rm G}i}\ln\lambda^{(k)}_{i}+\rm const\,,
\ee
where $\rm const$ is a constant that does not depend on $\mbi \delta_{\rm L}$.
After defining the logarithms, we write the gradients as
\ba
\label{eq:grad1}
-\frac{\partial \ln \pi}{\partial \mbi\delta_{\rm L}}=\mat C_{\rm L}^{-1} \mbi\delta_{\rm L}\,,
\ea
and for the individual $k$ survey likelihoods at cell $i$
\ba
\label{eq:grad2}
-\frac{\partial \ln \mathcal{L}^{(k)}}{\partial \delta_{{\rm L} i}}=b\left(\lambda^{(k)}_{i}-{N}^{(k)}_{{\rm G}i}\right)\,.
\ea

\section{Fisher Matrix Partial Derivatives and Priors}
\label{app:fish_der}

In this Appendix, we derive the partial derivatives of the our model relative to the free parameters $ \{p_1, p_2, p_3, p_4\} \equiv \{f_{\rm igm}, {\rm DM_{host}}, r_{\rm max}, f_{\rm hot}\}$, $\partial f_m/\partial p_i$, which we use in the Fisher matrix calculations (see Equation~\ref{eq:fisher}) to forecast the precision on the  parameter estimate.

\subsection{Case of available foreground data}

First, we calculate the partial derivatives in our default case, where we have information on the foreground matter distribution and intervening galaxies. In this case, the observable $f_{m}\left( p \right)$ for each individual FRB in the sample is given by the model of the total extragalactic DM component
\begin{equation}
f_m\left( p \right) = {\rm DM}_{\rm igm}\left( f_{\rm igm} \right) + {\rm DM}_{\rm halo}\left( r_{\rm max}, f_{\rm hot} \right) + {\rm DM_{host}}\times\left( 1 + z \right)^{-1}, 
\end{equation}

For the first parameter $f_{\rm igm}$, we find 
\begin{equation}
    \frac{\partial f_m }{\partial p_1} = \frac{\partial ~{\rm DM}_{\rm igm}\left( f_{\rm igm}\right)  }{\partial f_{\rm igm}} = \frac{ \langle {\rm DM_{argo}} \rangle}{\figm}
\end{equation}
where $\langle {\rm DM_{argo}} \rangle $ is the mean of ${\rm DM_{igm}}$ values calculated from the $N=50$ \argo{} reconstructed density realizations. Since $\langle \dmargo \rangle$ is linear in $\figm$, taking its derivative is equivalent to setting $f_{\rm igm} = 1$.

Next, the partial derivative for ${\rm DM_{host}}$ is given by

\begin{equation}
\label{eq:b21}
    \frac{\partial f_m }{\partial p_2} = \frac{ \partial \left( {\rm DM_{host}} \times \left( 1 + z_m \right)^{-1} \right) }{ \partial ~{\rm DM_{host}} } = \left(1 + z_m\right)^{-1},
\end{equation}
where $z_m$ is the redshift of the FRB.

In order to find partial derivative for $r_{\rm max}$, 
\begin{equation}
    \frac{\partial f_{m}}{\partial p_3} = \frac{\partial {\rm DM_{\rm halos}}\left( r_{\rm max}, f_{\rm hot} \right)}{\partial r_{\rm max}},
\end{equation}
where we numerically differentiate the modified NFW profiles (see Section~\ref{sec:dm_frb}) of each mock foreground galaxy in the line-of-sight of the FRB in question with respect to $r_{\rm max}$. 

Lastly, for $f_{\rm hot}$ the final partial derivative is given by
\begin{equation}
        \frac{\partial f_{m}}{\partial p_4} = \frac{\partial {\rm DM_{\rm halos}}\left( r_{\rm max}, f_{\rm hot} \right)}{\partial f_{\rm hot}} = \frac{ {\rm DM_{halos}}(r_{\rm max}) }{f_{\rm hot}},
\end{equation}
computed at the fiducial value of $r_{\rm max} = 1.40$ for all the halos within the sightline. Similar to the relation between $f_{\rm igm}$ and ${\rm DM_{igm}}$, ${\rm DM_{halos}}$ is linear with respect to $f_{\rm hot}$ and, therefore, this is equivalent to setting $f_{\rm hot} = 1$.

%%%%%%%%%%%%%%%%%%%%%%%%%%%%%%%%%%%%%%%%%%%%%%%%%%%%%%
\subsection{Case of no foreground data}

When no information on the foreground IGM structure and distribution of galaxies along the sightline is available, we have to rely on global estimates $\langle \dmigm \rangle(z)$ and $\langle \dmhalo \rangle (z)$, evaluated at the redshift of each FRB (see Section~\ref{sec:fisher} for details). In this case, the model $f_m\left(p\right)$ is given by 

\begin{equation}
    f_m\left(p\right) = \langle {\rm DM}_\mathrm{igm}(\figm) \rangle (z_m) \nonumber  + \langle \mathrm{DM}_\mathrm{halos}(\rmax, \fhot) \rangle (z_m) \nonumber +  \mathrm{DM_{host}}(1+z_m)^{-1}.
\end{equation}

Consequently, we find that the partial derivatives of $f_m$ with respect to each model parameter are as follows.

For $f_{\rm igm}$, the partial derivative becomes 
\begin{equation}
    \frac{\partial f_m}{\partial p_1} = \frac{\partial \langle {\rm DM}_{\rm igm} \left( f_{\rm igm} \right) \rangle (z_m)}{ \partial f_{\rm igm}} =  \frac{\langle {\rm DM}_{\rm igm} \rangle (z_m)}{\figm}
\end{equation}
which again is equivalent to setting 
the global IGM contribution to a value of $f_{\rm igm} = 1$ (see Figure~\ref{fig:sigigm})
since $\langle \dmigm \rangle$ linear with respect to \figm.

The partial derivative over ${\rm DM_{host}}$ does not change with respect to our default case (Equation~\ref{eq:b21}), i.e., $\partial f_m \slash \partial p_2 = \left( 1 + z_m \right)^{-1}$.

Finally, in order to calculate the partial derivatives over parameters $r_{\rm max}$ and $f_{\rm hot}$, we estimate the global contribution of the foreground galaxies along the FRB sightline as a function of FRB redshift, $\langle {\rm DM_{halos}}\left( r_{\rm max}, f_{\rm hot} \right) \rangle (z_m)$ (see discussion of global DM contributions in Section~\ref{sec:dm_frb}). 

Firstly, we estimate $\langle {\rm DM_{halos}}\left( r_{\rm max}, f_{\rm hot} \right) \rangle (z_m)$ at 3 values of $r_{\rm max} = \{ 1.38, 1.40, 1.42 \} $, and compute the partial derivative with respect to $r_{\rm max}$, $\partial f_m \slash \partial r_{\rm max}$, using the finite differences method. We evaluated this numerically at a grid of redshifts separated by $\Delta (z) = 0.1$ then fit the result as a with a linear function to find 
\begin{equation}
    \frac{\partial f_m}{ \partial p_3} = \frac{\partial \langle {\rm DM_{halos}}\left(r_{\rm max}, f_{\rm hot}\right)\rangle (z_m) }{\partial r_{\rm max}} = 371.25 \cdot z_m - 30.88
\end{equation}

Secondly, since parameter $f_{\rm hot}$ enters Equation~\ref{eq:mhalo_bar} as a multiplicative constant, the partial derivative over the parameter $f_{\rm hot}$ is given by

\begin{equation}
    \frac{\partial f_m}{ \partial p_4} = \frac{ \partial \langle {\rm DM_{halos}}\left(r_{\rm max}, f_{\rm hot}\right)\rangle (z_m) }{\partial f_{\rm hot}} = \frac{ \langle {\rm DM_{halos}}\left(r_{\rm max}, \fhot \right)\rangle (z_m)}{\fhot}
\end{equation}
where $\langle {\rm DM_{halos}}\left(r_{\rm max} \right)\rangle (z_m)$ is the global estimate of the ${\rm DM}_{\rm halos}$ calculated as a function of FRB redshift. Given our fiducial value $r_{\rm max} = 1.40$, we average over large numbers of mock sightlines at a redshift grid of $\Delta(z) = 0.1$ to a find a reasonable fitting function of:

\begin{equation}
     \frac{\partial f_m}{ \partial p_4} \equiv \frac{\langle {\rm DM_{halos}}\left(r_{\rm max} \right)\rangle (z_m)}{f_{\rm hot}} = 150.21 \cdot z_m^2 + 520.07 \cdot z_m - 1.21  
\end{equation}

\subsection{Prior information on the model parameters}

We also include priors on each of the parameters into the Fisher matrix calculations described in Section~\ref{sec:fisher}. In order to do that, we simply add the corresponding variances $\sigma^2_{{\rm prior}, p}$ of the parameter $p$ estimated from its prior distribution, to the corresponding diagonal elements of the $4\times 4$ Fisher matrix given by Equation~\ref{eq:fisher}. Because the free parameters that we use in our model have not been previously well constrained, we adopt flat uniform priors on each of them within the following limits

\begin{gather*}
    0 \leq f_{\rm igm} \leq 1\\
    0 \leq {\rm DM_{host}} \leq 500\ \left( {\rm pc\ cm^{-3}}\right)\\
    0 \leq r_{\rm max}\slash r_{200} \leq 5\\
    0 \leq f_{\rm hot} \leq 1.
\end{gather*}
The corresponding variance $\sigma^2_{{\rm prior},p}$ for each parameter $p$ is then given by

\begin{equation}
    \sigma^2_{{\rm prior}, p} = \frac{1}{12}\left( p_{\rm max} - p_{\rm min}\right)^2
\end{equation}
where $p_{\rm min}$ and $p_{\rm max}$ are the lower and upper bounds of the uniformly distributed priors for a given parameter $p$.

\begin{acknowledgements}

We thank Jingjing Shi, Ryuichi Takahashi, Ryan Shannon, Nicolas Tejos, and Sunil Simha for useful discussions.
KGL acknowledges support from JSPS Kakenhi Grants JP18H05868 and JP19K14755.
MA was supported by JSPS Kakenhi Grant JP21K13911. Parts of this research were supported by the Australian Research Council Centre of Excellence for All Sky Astrophysics in 3 Dimensions (ASTRO 3D), through project number CE170100013.
Based in part on data acquired at the Anglo-Australian Telescope, under program A/2020B/04. We acknowledge the traditional custodians of the land on which the AAT stands, the Gamilaraay people, and pay our respects to elders past and present.
\end{acknowledgements}

\section*{Software}
In this work we used  NumPy \citep{harris:2020}, Scipy \citep{virtanen:2020}, Astropy \citep{astropy-collaboration:2013,astropy-collaboration:2018},
 Matplotlib \citep{hunter:2007}, emcee \citep{foreman-mackey:2019}, and corner \citep{foreman-mackey:2016}.

\bibliography{references_kg}

\begin{thebibliography}{}
\expandafter\ifx\csname natexlab\endcsname\relax\def\natexlab#1{#1}\fi
\providecommand{\url}[1]{\href{#1}{#1}}
\providecommand{\dodoi}[1]{doi:~\href{http://doi.org/#1}{\nolinkurl{#1}}}
\providecommand{\doeprint}[1]{\href{http://ascl.net/#1}{\nolinkurl{http://ascl.net/#1}}}
\providecommand{\doarXiv}[1]{\href{https://arxiv.org/abs/#1}{\nolinkurl{https://arxiv.org/abs/#1}}}

\bibitem[{{Abazajian} {et~al.}(2009){Abazajian}, {Adelman-McCarthy},
  {Ag{\"u}eros}, {Allam}, {Allende Prieto}, {An}, {Anderson}, {Anderson},
  {Annis}, {Bahcall}, \& et~al.}]{abazajian:2009}
{Abazajian}, K.~N., {Adelman-McCarthy}, J.~K., {Ag{\"u}eros}, M.~A., {et~al.}
  2009, \apjs, 182, 543, \dodoi{10.1088/0067-0049/182/2/543}

\bibitem[{{Aggarwal} {et~al.}(2021){Aggarwal}, {Budav{\'a}ri}, {Deller},
  {Eftekhari}, {James}, {Prochaska}, \& {Tendulkar}}]{aggarwal:2021}
{Aggarwal}, K., {Budav{\'a}ri}, T., {Deller}, A.~T., {et~al.} 2021, \apj, 911,
  95, \dodoi{10.3847/1538-4357/abe8d2}

\bibitem[{{Angulo} \& {Hilbert}(2015)}]{angulo:2015}
{Angulo}, R.~E., \& {Hilbert}, S. 2015, \mnras, 448, 364,
  \dodoi{10.1093/mnras/stv050}

\bibitem[{{Astropy Collaboration} {et~al.}(2013){Astropy Collaboration},
  {Robitaille}, {Tollerud}, {Greenfield}, {Droettboom}, {Bray}, {Aldcroft},
  {Davis}, {Ginsburg}, {Price-Whelan}, {Kerzendorf}, {Conley}, {Crighton},
  {Barbary}, {Muna}, {Ferguson}, {Grollier}, {Parikh}, {Nair}, {Unther},
  {Deil}, {Woillez}, {Conseil}, {Kramer}, {Turner}, {Singer}, {Fox}, {Weaver},
  {Zabalza}, {Edwards}, {Azalee Bostroem}, {Burke}, {Casey}, {Crawford},
  {Dencheva}, {Ely}, {Jenness}, {Labrie}, {Lim}, {Pierfederici}, {Pontzen},
  {Ptak}, {Refsdal}, {Servillat}, \& {Streicher}}]{astropy-collaboration:2013}
{Astropy Collaboration}, {Robitaille}, T.~P., {Tollerud}, E.~J., {et~al.} 2013,
  \aap, 558, A33, \dodoi{10.1051/0004-6361/201322068}

\bibitem[{{Astropy Collaboration} {et~al.}(2018){Astropy Collaboration},
  {Price-Whelan}, {Sip{\H{o}}cz}, {G{\"u}nther}, {Lim}, {Crawford}, {Conseil},
  {Shupe}, {Craig}, {Dencheva}, {Ginsburg}, {VanderPlas}, {Bradley},
  {P{\'e}rez-Su{\'a}rez}, {de Val-Borro}, {Aldcroft}, {Cruz}, {Robitaille},
  {Tollerud}, {Ardelean}, {Babej}, {Bach}, {Bachetti}, {Bakanov}, {Bamford},
  {Barentsen}, {Barmby}, {Baumbach}, {Berry}, {Biscani}, {Boquien}, {Bostroem},
  {Bouma}, {Brammer}, {Bray}, {Breytenbach}, {Buddelmeijer}, {Burke},
  {Calderone}, {Cano Rodr{\'\i}guez}, {Cara}, {Cardoso}, {Cheedella}, {Copin},
  {Corrales}, {Crichton}, {D'Avella}, {Deil}, {Depagne}, {Dietrich}, {Donath},
  {Droettboom}, {Earl}, {Erben}, {Fabbro}, {Ferreira}, {Finethy}, {Fox},
  {Garrison}, {Gibbons}, {Goldstein}, {Gommers}, {Greco}, {Greenfield},
  {Groener}, {Grollier}, {Hagen}, {Hirst}, {Homeier}, {Horton}, {Hosseinzadeh},
  {Hu}, {Hunkeler}, {Ivezi{\'c}}, {Jain}, {Jenness}, {Kanarek}, {Kendrew},
  {Kern}, {Kerzendorf}, {Khvalko}, {King}, {Kirkby}, {Kulkarni}, {Kumar},
  {Lee}, {Lenz}, {Littlefair}, {Ma}, {Macleod}, {Mastropietro}, {McCully},
  {Montagnac}, {Morris}, {Mueller}, {Mumford}, {Muna}, {Murphy}, {Nelson},
  {Nguyen}, {Ninan}, {N{\"o}the}, {Ogaz}, {Oh}, {Parejko}, {Parley}, {Pascual},
  {Patil}, {Patil}, {Plunkett}, {Prochaska}, {Rastogi}, {Reddy Janga},
  {Sabater}, {Sakurikar}, {Seifert}, {Sherbert}, {Sherwood-Taylor}, {Shih},
  {Sick}, {Silbiger}, {Singanamalla}, {Singer}, {Sladen}, {Sooley},
  {Sornarajah}, {Streicher}, {Teuben}, {Thomas}, {Tremblay}, {Turner},
  {Terr{\'o}n}, {van Kerkwijk}, {de la Vega}, {Watkins}, {Weaver}, {Whitmore},
  {Woillez}, {Zabalza}, \& {Astropy Contributors}}]{astropy-collaboration:2018}
{Astropy Collaboration}, {Price-Whelan}, A.~M., {Sip{\H{o}}cz}, B.~M., {et~al.}
  2018, \aj, 156, 123, \dodoi{10.3847/1538-3881/aabc4f}

\bibitem[{{Ata} {et~al.}(2021){Ata}, {Kitaura}, {Lee}, {Lemaux}, {Kashino},
  {Cucciati}, {Hern{\'a}ndez-S{\'a}nchez}, \& {Le F{\`e}vre}}]{ata:2021}
{Ata}, M., {Kitaura}, F.-S., {Lee}, K.-G., {et~al.} 2021, \mnras, 500, 3194,
  \dodoi{10.1093/mnras/staa3318}

\bibitem[{{Ata} {et~al.}(2015){Ata}, {Kitaura}, \& {M{\"u}ller}}]{ata:2015}
{Ata}, M., {Kitaura}, F.-S., \& {M{\"u}ller}, V. 2015, \mnras, 446, 4250,
  \dodoi{10.1093/mnras/stu2347}

\bibitem[{{Ata} {et~al.}(2017){Ata}, {Kitaura}, {Chuang},
  {Rodr{\'\i}guez-Torres}, {Angulo}, {Ferraro}, {Gil-Mar{\'\i}n}, {McDonald},
  {Hern{\'a}ndez Monteagudo}, {M{\"u}ller}, {Yepes}, {Autefage}, {Baumgarten},
  {Beutler}, {Brownstein}, {Burden}, {Eisenstein}, {Guo}, {Ho}, {McBride},
  {Neyrinck}, {Olmstead}, {Padmanabhan}, {Percival}, {Prada}, {Rossi},
  {S{\'a}nchez}, {Schlegel}, {Schneider}, {Seo}, {Streblyanska}, {Tinker},
  {Tojeiro}, \& {Vargas-Magana}}]{ata:2017}
{Ata}, M., {Kitaura}, F.-S., {Chuang}, C.-H., {et~al.} 2017, \mnras, 467, 3993,
  \dodoi{10.1093/mnras/stx178}

\bibitem[{{Bannister} {et~al.}(2019){Bannister}, {Deller}, {Phillips},
  {Macquart}, {Prochaska}, {Tejos}, {Ryder}, {Sadler}, {Shannon}, {Simha},
  {Day}, {McQuinn}, {North-Hickey}, {Bhandari}, {Arcus}, {Bennert}, {Burchett},
  {Bouwhuis}, {Dodson}, {Ekers}, {Farah}, {Flynn}, {James}, {Kerr}, {Lenc},
  {Mahony}, {O'Meara}, {Os{\l}owski}, {Qiu}, {Treu}, {U}, {Bateman}, {Bock},
  {Bolton}, {Brown}, {Bunton}, {Chippendale}, {Cooray}, {Cornwell}, {Gupta},
  {Hayman}, {Kesteven}, {Koribalski}, {MacLeod}, {McClure-Griffiths},
  {Neuhold}, {Norris}, {Pilawa}, {Qiao}, {Reynolds}, {Roxby}, {Shimwell},
  {Voronkov}, \& {Wilson}}]{bannister:2019}
{Bannister}, K.~W., {Deller}, A.~T., {Phillips}, C., {et~al.} 2019, Science,
  365, 565, \dodoi{10.1126/science.aaw5903}

\bibitem[{{Bassa} {et~al.}(2017){Bassa}, {Tendulkar}, {Adams}, {Maddox},
  {Bogdanov}, {Bower}, {Burke-Spolaor}, {Butler}, {Chatterjee}, {Cordes},
  {Hessels}, {Kaspi}, {Law}, {Marcote}, {Paragi}, {Ransom}, {Scholz},
  {Spitler}, \& {van Langevelde}}]{bassa:2017}
{Bassa}, C.~G., {Tendulkar}, S.~P., {Adams}, E.~A.~K., {et~al.} 2017, \apjl,
  843, L8, \dodoi{10.3847/2041-8213/aa7a0c}

\bibitem[{{Batten} {et~al.}(2021{\natexlab{a}}){Batten}, {Duffy}, {Flynn},
  {Gupta}, {Ryan-Weber}, \& {Wijers}}]{batten:2021a}
{Batten}, A.~J., {Duffy}, A.~R., {Flynn}, C., {et~al.} 2021{\natexlab{a}},
  arXiv e-prints, arXiv:2109.13472.
\newblock \doarXiv{2109.13472}

\bibitem[{{Batten} {et~al.}(2021{\natexlab{b}}){Batten}, {Duffy}, {Wijers},
  {Gupta}, {Flynn}, {Schaye}, \& {Ryan-Weber}}]{batten:2021}
{Batten}, A.~J., {Duffy}, A.~R., {Wijers}, N.~A., {et~al.} 2021{\natexlab{b}},
  \mnras, 505, 5356, \dodoi{10.1093/mnras/stab1528}

\bibitem[{{Bhandari} {et~al.}(2020){Bhandari}, {Sadler}, {Prochaska}, {Simha},
  {Ryder}, {Marnoch}, {Bannister}, {Macquart}, {Flynn}, {Shannon}, {Tejos},
  {Corro-Guerra}, {Day}, {Deller}, {Ekers}, {Lopez}, {Mahony}, {Nu{\~n}ez}, \&
  {Phillips}}]{bhandari:2020}
{Bhandari}, S., {Sadler}, E.~M., {Prochaska}, J.~X., {et~al.} 2020, \apjl, 895,
  L37, \dodoi{10.3847/2041-8213/ab672e}

\bibitem[{{Blaizot} {et~al.}(2005){Blaizot}, {Wadadekar}, {Guiderdoni},
  {Colombi}, {Bertin}, {Bouchet}, {Devriendt}, \& {Hatton}}]{blaizot:2005}
{Blaizot}, J., {Wadadekar}, Y., {Guiderdoni}, B., {et~al.} 2005, \mnras, 360,
  159, \dodoi{10.1111/j.1365-2966.2005.09019.x}

\bibitem[{{Blanton} {et~al.}(2005){Blanton}, {Schlegel}, {Strauss},
  {Brinkmann}, {Finkbeiner}, {Fukugita}, {Gunn}, {Hogg}, {Ivezi{\'c}}, {Knapp},
  {Lupton}, {Munn}, {Schneider}, {Tegmark}, \& {Zehavi}}]{blanton:2005a}
{Blanton}, M.~R., {Schlegel}, D.~J., {Strauss}, M.~A., {et~al.} 2005, \aj, 129,
  2562, \dodoi{10.1086/429803}

\bibitem[{{Bregman}(2007)}]{bregman:2007}
{Bregman}, J.~N. 2007, \araa, 45, 221,
  \dodoi{10.1146/annurev.astro.45.051806.110619}

\bibitem[{{Cen} \& {Ostriker}(2006)}]{cen:2006}
{Cen}, R., \& {Ostriker}, J.~P. 2006, \apj, 650, 560, \dodoi{10.1086/506505}

\bibitem[{{CHIME/FRB Collaboration} {et~al.}(2018){CHIME/FRB Collaboration},
  {Amiri}, {Bandura}, {Berger}, {Bhardwaj}, {Boyce}, {Boyle}, {Brar},
  {Burhanpurkar}, {Chawla}, {Chowdhury}, {Cliche}, {Cranmer}, {Cubranic},
  {Deng}, {Denman}, {Dobbs}, {Fandino}, {Fonseca}, {Gaensler}, {Giri},
  {Gilbert}, {Good}, {Guliani}, {Halpern}, {Hinshaw}, {H{\"o}fer}, {Josephy},
  {Kaspi}, {Landecker}, {Lang}, {Liao}, {Masui}, {Mena-Parra}, {Naidu},
  {Newburgh}, {Ng}, {Patel}, {Pen}, {Pinsonneault-Marotte}, {Pleunis}, {Rafiei
  Ravandi}, {Ransom}, {Renard}, {Scholz}, {Sigurdson}, {Siegel}, {Smith},
  {Stairs}, {Tendulkar}, {Vanderlinde}, \&
  {Wiebe}}]{chime/frb-collaboration:2018}
{CHIME/FRB Collaboration}, {Amiri}, M., {Bandura}, K., {et~al.} 2018, \apj,
  863, 48, \dodoi{10.3847/1538-4357/aad188}

\bibitem[{{CHIME/FRB Collaboration} {et~al.}(2019){CHIME/FRB Collaboration},
  {Andersen}, {Bandura}, {Bhardwaj}, {Boubel}, {Boyce}, {Boyle}, {Brar},
  {Cassanelli}, {Chawla}, {Cubranic}, {Deng}, {Dobbs}, {Fandino}, {Fonseca},
  {Gaensler}, {Gilbert}, {Giri}, {Good}, {Halpern}, {Hill}, {Hinshaw},
  {H{\"o}fer}, {Josephy}, {Kaspi}, {Kothes}, {Landecker}, {Lang}, {Li}, {Lin},
  {Masui}, {Mena-Parra}, {Merryfield}, {Mckinven}, {Michilli}, {Milutinovic},
  {Naidu}, {Newburgh}, {Ng}, {Patel}, {Pen}, {Pinsonneault-Marotte}, {Pleunis},
  {Rafiei-Ravandi}, {Rahman}, {Ransom}, {Renard}, {Scholz}, {Siegel}, {Singh},
  {Smith}, {Stairs}, {Tendulkar}, {Tretyakov}, {Vanderlinde}, {Yadav}, \&
  {Zwaniga}}]{chime/frb-collaboration:2019}
{CHIME/FRB Collaboration}, {Andersen}, B.~C., {Bandura}, K., {et~al.} 2019,
  \apjl, 885, L24, \dodoi{10.3847/2041-8213/ab4a80}

\bibitem[{{Chittidi} {et~al.}(2020){Chittidi}, {Simha}, {Mannings},
  {Prochaska}, {Rafelski}, {Neeleman}, {Macquart}, {Tejos}, {Jorgenson},
  {Ryder}, {Day}, {Marnoch}, {Bhandari}, {Deller}, {Qiu}, {Bannister},
  {Shannon}, \& {Heintz}}]{chittidi:2020}
{Chittidi}, J.~S., {Simha}, S., {Mannings}, A., {et~al.} 2020, arXiv e-prints,
  arXiv:2005.13158.
\newblock \doarXiv{2005.13158}

\bibitem[{{Cirasuolo} {et~al.}(2014){Cirasuolo}, {Afonso}, {Carollo}, {Flores},
  {Maiolino}, {Oliva}, {Paltani}, {Vanzi}, {Evans}, {Abreu}, {Atkinson},
  {Babusiaux}, {Beard}, {Bauer}, {Bellazzini}, {Bender}, {Best}, {Bezawada},
  {Bonifacio}, {Bragaglia}, {Bryson}, {Busher}, {Cabral}, {Caputi}, {Centrone},
  {Chemla}, {Cimatti}, {Cioni}, {Clementini}, {Coelho}, {Crnojevic}, {Daddi},
  {Dunlop}, {Eales}, {Feltzing}, {Ferguson}, {Fisher}, {Fontana}, {Fynbo},
  {Garilli}, {Gilmore}, {Glauser}, {Guinouard}, {Hammer}, {Hastings}, {Hess},
  {Ivison}, {Jagourel}, {Jarvis}, {Kaper}, {Kauffman}, {Kitching}, {Lawrence},
  {Lee}, {Lemasle}, {Licausi}, {Lilly}, {Lorenzetti}, {Lunney}, {Maiolino},
  {Mannucci}, {McLure}, {Minniti}, {Montgomery}, {Muschielok}, {Nandra},
  {Navarro}, {Norberg}, {Oliver}, {Origlia}, {Padilla}, {Peacock}, {Pedichini},
  {Peng}, {Pentericci}, {Pragt}, {Puech}, {Randich}, {Rees}, {Renzini}, {Ryde},
  {Rodrigues}, {Roseboom}, {Royer}, {Saglia}, {Sanchez}, {Schiavon},
  {Schnetler}, {Sobral}, {Speziali}, {Sun}, {Stuik}, {Taylor}, {Taylor},
  {Todd}, {Tolstoy}, {Torres}, {Tosi}, {Vanzella}, {Venema}, {Vitali},
  {Wegner}, {Wells}, {Wild}, {Wright}, {Zamorani}, \&
  {Zoccali}}]{cirasuolo:2014}
{Cirasuolo}, M., {Afonso}, J., {Carollo}, M., {et~al.} 2014, in Society of
  Photo-Optical Instrumentation Engineers (SPIE) Conference Series, Vol. 9147,
  Ground-based and Airborne Instrumentation for Astronomy V, ed. S.~K.
  {Ramsay}, I.~S. {McLean}, \& H.~{Takami}, 91470N, \dodoi{10.1117/12.2056012}

\bibitem[{{Coe}(2009)}]{coe:2009}
{Coe}, D. 2009, arXiv e-prints, arXiv:0906.4123.
\newblock \doarXiv{0906.4123}

\bibitem[{{Coles} \& {Jones}(1991)}]{coles:1991}
{Coles}, P., \& {Jones}, B. 1991, \mnras, 248, 1, \dodoi{10.1093/mnras/248.1.1}

\bibitem[{{Cordes} \& {Chatterjee}(2019)}]{cordes:2019}
{Cordes}, J.~M., \& {Chatterjee}, S. 2019, \araa, 57, 417,
  \dodoi{10.1146/annurev-astro-091918-104501}

\bibitem[{{Cordes} {et~al.}(2021){Cordes}, {Ocker}, \&
  {Chatterjee}}]{cordes:2021}
{Cordes}, J.~M., {Ocker}, S.~K., \& {Chatterjee}, S. 2021, arXiv e-prints,
  arXiv:2108.01172.
\newblock \doarXiv{2108.01172}

\bibitem[{{Dalton} {et~al.}(2012){Dalton}, {Trager}, {Abrams}, {Carter},
  {Bonifacio}, {Aguerri}, {MacIntosh}, {Evans}, {Lewis}, {Navarro}, {Agocs},
  {Dee}, {Rousset}, {Tosh}, {Middleton}, {Pragt}, {Terrett}, {Brock}, {Benn},
  {Verheijen}, {Cano Infantes}, {Bevil}, {Steele}, {Mottram}, {Bates},
  {Gribbin}, {Rey}, {Rodriguez}, {Delgado}, {Guinouard}, {Walton}, {Irwin},
  {Jagourel}, {Stuik}, {Gerlofsma}, {Roelfsma}, {Skillen}, {Ridings},
  {Balcells}, {Daban}, {Gouvret}, {Venema}, \& {Girard}}]{dalton:2012}
{Dalton}, G., {Trager}, S.~C., {Abrams}, D.~C., {et~al.} 2012, in Society of
  Photo-Optical Instrumentation Engineers (SPIE) Conference Series, Vol. 8446,
  Ground-based and Airborne Instrumentation for Astronomy IV, ed. I.~S.
  {McLean}, S.~K. {Ramsay}, \& H.~{Takami}, 84460P, \dodoi{10.1117/12.925950}

\bibitem[{{Danforth} \& {Shull}(2008)}]{danforth:2008}
{Danforth}, C.~W., \& {Shull}, J.~M. 2008, \apj, 679, 194,
  \dodoi{10.1086/587127}

\bibitem[{{DESI Collaboration} {et~al.}(2016){DESI Collaboration}, {Aghamousa},
  {Aguilar}, {Ahlen}, {Alam}, {Allen}, {Allende Prieto}, {Annis}, {Bailey},
  {Balland}, \& et~al.}]{desi-collaboration:2016}
{DESI Collaboration}, {Aghamousa}, A., {Aguilar}, J., {et~al.} 2016, arXiv
  e-prints, arXiv:1611.00036.
\newblock \doarXiv{1611.00036}

\bibitem[{{Di Porto} {et~al.}(2016){Di Porto}, {Branchini}, {Bel}, {Marulli},
  {Bolzonella}, {Cucciati}, {de la Torre}, {Granett}, {Guzzo}, {Marinoni},
  {Moscardini}, {Abbas}, {Adami}, {Arnouts}, {Bottini}, {Cappi}, {Coupon},
  {Davidzon}, {De Lucia}, {Fritz}, {Franzetti}, {Fumana}, {Garilli}, {Ilbert},
  {Iovino}, {Krywult}, {Le Brun}, {Le F{\`e}vre}, {Maccagni}, {Ma{\l}ek},
  {McCracken}, {Paioro}, {Polletta}, {Pollo}, {Scodeggio}, {Tasca}, {Tojeiro},
  {Vergani}, {Zanichelli}, {Burden}, {Marchetti}, {Martizzi}, {Mellier},
  {Nichol}, {Peacock}, {Percival}, {Viel}, {Wolk}, \&
  {Zamorani}}]{di-porto:2016}
{Di Porto}, C., {Branchini}, E., {Bel}, J., {et~al.} 2016, \aap, 594, A62,
  \dodoi{10.1051/0004-6361/201424448}

\bibitem[{{Driver} {et~al.}(2011){Driver}, {Hill}, {Kelvin}, {Robotham},
  {Liske}, {Norberg}, {Baldry}, {Bamford}, {Hopkins}, {Loveday}, {Peacock},
  {Andrae}, {Bland-Hawthorn}, {Brough}, {Brown}, {Cameron}, {Ching}, {Colless},
  {Conselice}, {Croom}, {Cross}, {de Propris}, {Dye}, {Drinkwater}, {Ellis},
  {Graham}, {Grootes}, {Gunawardhana}, {Jones}, {van Kampen}, {Maraston},
  {Nichol}, {Parkinson}, {Phillipps}, {Pimbblet}, {Popescu}, {Prescott},
  {Roseboom}, {Sadler}, {Sansom}, {Sharp}, {Smith}, {Taylor}, {Thomas},
  {Tuffs}, {Wijesinghe}, {Dunne}, {Frenk}, {Jarvis}, {Madore}, {Meyer},
  {Seibert}, {Staveley-Smith}, {Sutherland}, \& {Warren}}]{driver:2011}
{Driver}, S.~P., {Hill}, D.~T., {Kelvin}, L.~S., {et~al.} 2011, \mnras, 413,
  971, \dodoi{10.1111/j.1365-2966.2010.18188.x}

\bibitem[{{Duane} {et~al.}(1987){Duane}, {Kennedy}, {Pendleton}, \&
  {Roweth}}]{duane:1987}
{Duane}, S., {Kennedy}, A.~D., {Pendleton}, B.~J., \& {Roweth}, D. 1987,
  Physics Letters B, 195, 216, \dodoi{10.1016/0370-2693(87)91197-X}

\bibitem[{{Erdo{\v{g}}du} {et~al.}(2004){Erdo{\v{g}}du}, {Lahav}, {Zaroubi},
  {Efstathiou}, {Moody}, {Peacock}, {Colless}, {Baldry}, {Baugh},
  {Bland-Hawthorn}, {Bridges}, {Cannon}, {Cole}, {Collins}, {Couch}, {Dalton},
  {De Propris}, {Driver}, {Ellis}, {Frenk}, {Glazebrook}, {Jackson}, {Lewis},
  {Lumsden}, {Maddox}, {Madgwick}, {Norberg}, {Peterson}, {Sutherland}, \&
  {Taylor}}]{erdogdu:2004}
{Erdo{\v{g}}du}, P., {Lahav}, O., {Zaroubi}, S., {et~al.} 2004, \mnras, 352,
  939, \dodoi{10.1111/j.1365-2966.2004.07984.x}

\bibitem[{{Erdo{\v{g}}du} {et~al.}(2006){Erdo{\v{g}}du}, {Lahav}, {Huchra},
  {Colless}, {Cutri}, {Falco}, {George}, {Jarrett}, {Jones}, {Macri}, {Mader},
  {Martimbeau}, {Pahre}, {Parker}, {Rassat}, \& {Saunders}}]{erdogdu:2006}
{Erdo{\v{g}}du}, P., {Lahav}, O., {Huchra}, J.~P., {et~al.} 2006, \mnras, 373,
  45, \dodoi{10.1111/j.1365-2966.2006.11049.x}

\bibitem[{{Foreman-Mackey}(2016)}]{foreman-mackey:2016}
{Foreman-Mackey}, D. 2016, The Journal of Open Source Software, 1, 24,
  \dodoi{10.21105/joss.00024}

\bibitem[{{Foreman-Mackey} {et~al.}(2019){Foreman-Mackey}, {Farr}, {Sinha},
  {Archibald}, {Hogg}, {Sanders}, {Zuntz}, {Williams}, {Nelson}, {de
  Val-Borro}, {Erhardt}, {Pashchenko}, \& {Pla}}]{foreman-mackey:2019}
{Foreman-Mackey}, D., {Farr}, W., {Sinha}, M., {et~al.} 2019, The Journal of
  Open Source Software, 4, 1864, \dodoi{10.21105/joss.01864}

\bibitem[{{Fukugita} {et~al.}(1998){Fukugita}, {Hogan}, \&
  {Peebles}}]{fukugita:1998}
{Fukugita}, M., {Hogan}, C.~J., \& {Peebles}, P.~J.~E. 1998, \apj, 503, 518,
  \dodoi{10.1086/306025}

\bibitem[{{Fukugita} \& {Peebles}(2004)}]{fukugita:2004}
{Fukugita}, M., \& {Peebles}, P.~J.~E. 2004, \apj, 616, 643,
  \dodoi{10.1086/425155}

\bibitem[{{Gunn} {et~al.}(2006){Gunn}, {Siegmund}, {Mannery}, {Owen}, {Hull},
  {Leger}, {Carey}, {Knapp}, {York}, {Boroski}, {Kent}, {Lupton}, {Rockosi},
  {Evans}, {Waddell}, {Anderson}, {Annis}, {Barentine}, {Bartoszek}, {Bastian},
  {Bracker}, {Brewington}, {Briegel}, {Brinkmann}, {Brown}, {Carr},
  {Czarapata}, {Drennan}, {Dombeck}, {Federwitz}, {Gillespie}, {Gonzales},
  {Hansen}, {Harvanek}, {Hayes}, {Jordan}, {Kinney}, {Klaene}, {Kleinman},
  {Kron}, {Kresinski}, {Lee}, {Limmongkol}, {Lindenmeyer}, {Long}, {Loomis},
  {McGehee}, {Mantsch}, {Neilsen}, {Neswold}, {Newman}, {Nitta}, {Peoples},
  {Pier}, {Prieto}, {Prosapio}, {Rivetta}, {Schneider}, {Snedden}, \&
  {Wang}}]{gunn:2006}
{Gunn}, J.~E., {Siegmund}, W.~A., {Mannery}, E.~J., {et~al.} 2006, \aj, 131,
  2332, \dodoi{10.1086/500975}

\bibitem[{{Guo} {et~al.}(2011){Guo}, {White}, {Boylan-Kolchin}, {De Lucia},
  {Kauffmann}, {Lemson}, {Li}, {Springel}, \& {Weinmann}}]{guo:2011}
{Guo}, Q., {White}, S., {Boylan-Kolchin}, M., {et~al.} 2011, \mnras, 413, 101,
  \dodoi{10.1111/j.1365-2966.2010.18114.x}

\bibitem[{{Guzzo} {et~al.}(2014){Guzzo}, {Scodeggio}, {Garilli}, {Granett},
  {Fritz}, {Abbas}, {Adami}, {Arnouts}, {Bel}, {Bolzonella}, {Bottini},
  {Branchini}, {Cappi}, {Coupon}, {Cucciati}, {Davidzon}, {De Lucia}, {de la
  Torre}, {Franzetti}, {Fumana}, {Hudelot}, {Ilbert}, {Iovino}, {Krywult}, {Le
  Brun}, {Le F{\`e}vre}, {Maccagni}, {Ma{\l}ek}, {Marulli}, {McCracken},
  {Paioro}, {Peacock}, {Polletta}, {Pollo}, {Schlagenhaufer}, {Tasca},
  {Tojeiro}, {Vergani}, {Zamorani}, {Zanichelli}, {Burden}, {Di Porto},
  {Marchetti}, {Marinoni}, {Mellier}, {Moscardini}, {Nichol}, {Percival},
  {Phleps}, \& {Wolk}}]{guzzo:2014}
{Guzzo}, L., {Scodeggio}, M., {Garilli}, B., {et~al.} 2014, \aap, 566, A108,
  \dodoi{10.1051/0004-6361/201321489}

\bibitem[{{Hamilton}(1997)}]{hamilton:1997}
{Hamilton}, A.~J.~S. 1997, \mnras, 289, 285, \dodoi{10.1093/mnras/289.2.285}

\bibitem[{Harris {et~al.}(2020)Harris, Millman, van~der Walt, Gommers,
  Virtanen, Cournapeau, Wieser, Taylor, Berg, Smith, Kern, Picus, Hoyer, van
  Kerkwijk, Brett, Haldane, del R{\'{i}}o, Wiebe, Peterson,
  G{\'{e}}rard-Marchant, Sheppard, Reddy, Weckesser, Abbasi, Gohlke, \&
  Oliphant}]{harris:2020}
Harris, C.~R., Millman, K.~J., van~der Walt, S.~J., {et~al.} 2020, Nature, 585,
  357, \dodoi{10.1038/s41586-020-2649-2}

\bibitem[{{Hashimoto} {et~al.}(2020){Hashimoto}, {Goto}, {Wang}, {Kim}, {Ho},
  {On}, {Lu}, \& {Santos}}]{hashimoto:2020}
{Hashimoto}, T., {Goto}, T., {Wang}, T.-W., {et~al.} 2020, \mnras, 494, 2886,
  \dodoi{10.1093/mnras/staa895}

\bibitem[{{Heintz} {et~al.}(2020){Heintz}, {Prochaska}, {Simha}, {Platts},
  {Fong}, {Tejos}, {Ryder}, {Aggerwal}, {Bhandari}, {Day}, {Deller},
  {Kilpatrick}, {Law}, {Macquart}, {Mannings}, {Marnoch}, {Sadler}, \&
  {Shannon}}]{heintz:2020}
{Heintz}, K.~E., {Prochaska}, J.~X., {Simha}, S., {et~al.} 2020, \apj, 903,
  152, \dodoi{10.3847/1538-4357/abb6fb}

\bibitem[{{Henriques} {et~al.}(2015){Henriques}, {White}, {Thomas}, {Angulo},
  {Guo}, {Lemson}, {Springel}, \& {Overzier}}]{henriques:2015}
{Henriques}, B. M.~B., {White}, S. D.~M., {Thomas}, P.~A., {et~al.} 2015,
  \mnras, 451, 2663, \dodoi{10.1093/mnras/stv705}

\bibitem[{Hunter(2007)}]{hunter:2007}
Hunter, J.~D. 2007, Computing in Science \& Engineering, 9, 90,
  \dodoi{10.1109/MCSE.2007.55}

\bibitem[{{Inoue}(2004)}]{inoue:2004}
{Inoue}, S. 2004, \mnras, 348, 999, \dodoi{10.1111/j.1365-2966.2004.07359.x}

\bibitem[{{Ioka}(2003)}]{ioka:2003}
{Ioka}, K. 2003, \apjl, 598, L79, \dodoi{10.1086/380598}

\bibitem[{{James} {et~al.}(2021){James}, {Prochaska}, {Macquart},
  {North-Hickey}, {Bannister}, \& {Dunning}}]{james:2021}
{James}, C.~W., {Prochaska}, J.~X., {Macquart}, J.~P., {et~al.} 2021, \mnras,
  \dodoi{10.1093/mnrasl/slab117}

\bibitem[{{James} {et~al.}(2020){James}, {Os{\l}owski}, {Flynn}, {Kumar},
  {Bannister}, {Bhandari}, {Farah}, {Kerr}, {Lorimer}, {Macquart}, {Ng},
  {Phillips}, {Price}, {Qiu}, {Shannon}, \& {Spiewak}}]{james:2020}
{James}, C.~W., {Os{\l}owski}, S., {Flynn}, C., {et~al.} 2020, \mnras, 495,
  2416, \dodoi{10.1093/mnras/staa1361}

\bibitem[{{Jaroszynski}(2019)}]{jaroszynski:2019}
{Jaroszynski}, M. 2019, \mnras, 484, 1637, \dodoi{10.1093/mnras/sty3529}

\bibitem[{{Jasche} \& {Kitaura}(2010)}]{jasche:2010}
{Jasche}, J., \& {Kitaura}, F.~S. 2010, \mnras, 407, 29,
  \dodoi{10.1111/j.1365-2966.2010.16897.x}

\bibitem[{{Jones} {et~al.}(2009){Jones}, {Read}, {Saunders}, {Colless},
  {Jarrett}, {Parker}, {Fairall}, {Mauch}, {Sadler}, {Watson}, {Burton},
  {Campbell}, {Cass}, {Croom}, {Dawe}, {Fiegert}, {Frankcombe}, {Hartley},
  {Huchra}, {James}, {Kirby}, {Lahav}, {Lucey}, {Mamon}, {Moore}, {Peterson},
  {Prior}, {Proust}, {Russell}, {Safouris}, {Wakamatsu}, {Westra}, \&
  {Williams}}]{jones:2009}
{Jones}, D.~H., {Read}, M.~A., {Saunders}, W., {et~al.} 2009, \mnras, 399, 683,
  \dodoi{10.1111/j.1365-2966.2009.15338.x}

\bibitem[{{Kaiser}(1987)}]{kaiser:1987}
{Kaiser}, N. 1987, \mnras, 227, 1, \dodoi{10.1093/mnras/227.1.1}

\bibitem[{{Kitaura} \& {Angulo}(2012)}]{kitaura:2012a}
{Kitaura}, F.-S., \& {Angulo}, R.~E. 2012, \mnras, 425, 2443,
  \dodoi{10.1111/j.1365-2966.2012.21614.x}

\bibitem[{{Kitaura} {et~al.}(2016){Kitaura}, {Ata}, {Angulo}, {Chuang},
  {Rodr{\'\i}guez-Torres}, {Monteagudo}, {Prada}, \& {Yepes}}]{kitaura:2016}
{Kitaura}, F.-S., {Ata}, M., {Angulo}, R.~E., {et~al.} 2016, \mnras, 457, L113,
  \dodoi{10.1093/mnrasl/slw009}

\bibitem[{{Kitaura} {et~al.}(2021){Kitaura}, {Ata}, {Rodr{\'\i}guez-Torres},
  {Hern{\'a}ndez-S{\'a}nchez}, {Balaguera-Antol{\'\i}nez}, \&
  {Yepes}}]{kitaura:2021}
{Kitaura}, F.-S., {Ata}, M., {Rodr{\'\i}guez-Torres}, S.~A., {et~al.} 2021,
  \mnras, 502, 3456, \dodoi{10.1093/mnras/staa3774}

\bibitem[{{Kitaura} \& {En{\ss}lin}(2008)}]{kitaura:2008}
{Kitaura}, F.~S., \& {En{\ss}lin}, T.~A. 2008, \mnras, 389, 497,
  \dodoi{10.1111/j.1365-2966.2008.13341.x}

\bibitem[{{Kitaura} {et~al.}(2010){Kitaura}, {Jasche}, \&
  {Metcalf}}]{kitaura:2010}
{Kitaura}, F.-S., {Jasche}, J., \& {Metcalf}, R.~B. 2010, \mnras, 403, 589,
  \dodoi{10.1111/j.1365-2966.2009.16163.x}

\bibitem[{{Kitzbichler} \& {White}(2007)}]{kitzbichler:2007}
{Kitzbichler}, M.~G., \& {White}, S.~D.~M. 2007, \mnras, 376, 2,
  \dodoi{10.1111/j.1365-2966.2007.11458.x}

\bibitem[{{Kocz} {et~al.}(2019){Kocz}, {Ravi}, {Catha}, {D'Addario},
  {Hallinan}, {Hobbs}, {Kulkarni}, {Shi}, {Vedantham}, {Weinreb}, \&
  {Woody}}]{kocz:2019}
{Kocz}, J., {Ravi}, V., {Catha}, M., {et~al.} 2019, \mnras, 489, 919,
  \dodoi{10.1093/mnras/stz2219}

\bibitem[{{Levi} {et~al.}(2013){Levi}, {Bebek}, {Beers}, {Blum}, {Cahn},
  {Eisenstein}, {Flaugher}, {Honscheid}, {Kron}, {Lahav}, {McDonald}, {Roe},
  {Schlegel}, \& {representing the DESI collaboration}}]{levi:2013}
{Levi}, M., {Bebek}, C., {Beers}, T., {et~al.} 2013, ArXiv e-prints.
\newblock \doarXiv{1308.0847}

\bibitem[{{Lewis} {et~al.}(2002){Lewis}, {Balogh}, {De Propris}, {Couch},
  {Bower}, {Offer}, {Bland-Hawthorn}, {Baldry}, {Baugh}, {Bridges}, {Cannon},
  {Cole}, {Colless}, {Collins}, {Cross}, {Dalton}, {Driver}, {Efstathiou},
  {Ellis}, {Frenk}, {Glazebrook}, {Hawkins}, {Jackson}, {Lahav}, {Lumsden},
  {Maddox}, {Madgwick}, {Norberg}, {Peacock}, {Percival}, {Peterson},
  {Sutherland}, \& {Taylor}}]{lewis:2002}
{Lewis}, I., {Balogh}, M., {De Propris}, R., {et~al.} 2002, \mnras, 334, 673,
  \dodoi{10.1046/j.1365-8711.2002.05558.x}

\bibitem[{{Lim} {et~al.}(2020){Lim}, {Mo}, {Wang}, \& {Yang}}]{lim:2020}
{Lim}, S.~H., {Mo}, H.~J., {Wang}, H., \& {Yang}, X. 2020, \apj, 889, 48,
  \dodoi{10.3847/1538-4357/ab63df}

\bibitem[{{Liske} {et~al.}(2015){Liske}, {Baldry}, {Driver}, {Tuffs},
  {Alpaslan}, {Andrae}, {Brough}, {Cluver}, {Grootes}, {Gunawardhana},
  {Kelvin}, {Loveday}, {Robotham}, {Taylor}, {Bamford}, {Bland-Hawthorn},
  {Brown}, {Drinkwater}, {Hopkins}, {Meyer}, {Norberg}, {Peacock}, {Agius},
  {Andrews}, {Bauer}, {Ching}, {Colless}, {Conselice}, {Croom}, {Davies}, {De
  Propris}, {Dunne}, {Eardley}, {Ellis}, {Foster}, {Frenk}, {H{\"a}u{\ss}ler},
  {Holwerda}, {Howlett}, {Ibarra}, {Jarvis}, {Jones}, {Kafle}, {Lacey},
  {Lange}, {Lara-L{\'o}pez}, {L{\'o}pez-S{\'a}nchez}, {Maddox}, {Madore},
  {McNaught-Roberts}, {Moffett}, {Nichol}, {Owers}, {Palamara}, {Penny},
  {Phillipps}, {Pimbblet}, {Popescu}, {Prescott}, {Proctor}, {Sadler},
  {Sansom}, {Seibert}, {Sharp}, {Sutherland}, {V{\'a}zquez-Mata}, {van Kampen},
  {Wilkins}, {Williams}, \& {Wright}}]{liske:2015}
{Liske}, J., {Baldry}, I.~K., {Driver}, S.~P., {et~al.} 2015, \mnras, 452,
  2087, \dodoi{10.1093/mnras/stv1436}

\bibitem[{{Lorimer} {et~al.}(2007){Lorimer}, {Bailes}, {McLaughlin},
  {Narkevic}, \& {Crawford}}]{lorimer:2007}
{Lorimer}, D.~R., {Bailes}, M., {McLaughlin}, M.~A., {Narkevic}, D.~J., \&
  {Crawford}, F. 2007, Science, 318, 777, \dodoi{10.1126/science.1147532}

\bibitem[{{Macquart} {et~al.}(2020){Macquart}, {Prochaska}, {McQuinn},
  {Bannister}, {Bhandari}, {Day}, {Deller}, {Ekers}, {James}, {Marnoch},
  {Os{\l}owski}, {Phillips}, {Ryder}, {Scott}, {Shannon}, \&
  {Tejos}}]{macquart:2020}
{Macquart}, J.~P., {Prochaska}, J.~X., {McQuinn}, M., {et~al.} 2020, \nat, 581,
  391, \dodoi{10.1038/s41586-020-2300-2}

\bibitem[{{Maller} \& {Bullock}(2004)}]{maller:2004}
{Maller}, A.~H., \& {Bullock}, J.~S. 2004, \mnras, 355, 694,
  \dodoi{10.1111/j.1365-2966.2004.08349.x}

\bibitem[{{Manchester} {et~al.}(2005){Manchester}, {Hobbs}, {Teoh}, \&
  {Hobbs}}]{manchester:2005}
{Manchester}, R.~N., {Hobbs}, G.~B., {Teoh}, A., \& {Hobbs}, M. 2005, \aj, 129,
  1993, \dodoi{10.1086/428488}

\bibitem[{{Mannings} {et~al.}(2020){Mannings}, {Fong}, {Simha}, {Prochaska},
  {Rafelski}, {Kilpatrick}, {Tejos}, {Heintz}, {Bhandari}, {Day}, {Deller},
  {Ryder}, {Shannon}, \& {Tendulkar}}]{mannings:2020}
{Mannings}, A.~G., {Fong}, W.-f., {Simha}, S., {et~al.} 2020, arXiv e-prints,
  arXiv:2012.11617.
\newblock \doarXiv{2012.11617}

\bibitem[{{Maraston}(2005)}]{maraston:2005}
{Maraston}, C. 2005, \mnras, 362, 799, \dodoi{10.1111/j.1365-2966.2005.09270.x}

\bibitem[{{Marcote} {et~al.}(2020){Marcote}, {Nimmo}, {Hessels}, {Tendulkar},
  {Bassa}, {Paragi}, {Keimpema}, {Bhardwaj}, {Karuppusamy}, {Kaspi}, {Law},
  {Michilli}, {Aggarwal}, {Andersen}, {Archibald}, {Bandura}, {Bower}, {Boyle},
  {Brar}, {Burke-Spolaor}, {Butler}, {Cassanelli}, {Chawla}, {Demorest},
  {Dobbs}, {Fonseca}, {Giri}, {Good}, {Gourdji}, {Josephy}, {Kirichenko},
  {Kirsten}, {Landecker}, {Lang}, {Lazio}, {Li}, {Lin}, {Linford}, {Masui},
  {Mena-Parra}, {Naidu}, {Ng}, {Patel}, {Pen}, {Pleunis}, {Rafiei-Ravandi},
  {Rahman}, {Renard}, {Scholz}, {Siegel}, {Smith}, {Stairs}, {Vanderlinde}, \&
  {Zwaniga}}]{marcote:2020}
{Marcote}, B., {Nimmo}, K., {Hessels}, J.~W.~T., {et~al.} 2020, \nat, 577, 190,
  \dodoi{10.1038/s41586-019-1866-z}

\bibitem[{{Mathews} \& {Prochaska}(2017)}]{mathews:2017}
{Mathews}, W.~G., \& {Prochaska}, J.~X. 2017, \apjl, 846, L24,
  \dodoi{10.3847/2041-8213/aa8861}

\bibitem[{{McConnell} {et~al.}(2016){McConnell}, {Allison}, {Bannister},
  {Bell}, {Bignall}, {Chippendale}, {Edwards}, {Harvey-Smith}, {Hegarty},
  {Heywood}, {Hotan}, {Indermuehle}, {Lenc}, {Marvil}, {Popping}, {Raja},
  {Reynolds}, {Sault}, {Serra}, {Voronkov}, {Whiting}, {Amy}, {Axtens}, {Ball},
  {Bateman}, {Bock}, {Bolton}, {Brodrick}, {Brothers}, {Brown}, {Bunton},
  {Cheng}, {Cornwell}, {DeBoer}, {Feain}, {Gough}, {Gupta}, {Guzman},
  {Hampson}, {Hay}, {Hayman}, {Hoyle}, {Humphreys}, {Jacka}, {Jackson},
  {Jackson}, {Jeganathan}, {Joseph}, {Koribalski}, {Leach}, {Lensson},
  {MacLeod}, {Mackay}, {Marquarding}, {McClure-Griffiths}, {Mirtschin},
  {Mitchell}, {Neuhold}, {Ng}, {Norris}, {Pearce}, {Qiao}, {Schinckel},
  {Shields}, {Shimwell}, {Storey}, {Troup}, {Turner}, {Tuthill}, {Tzioumis},
  {Wark}, {Westmeier}, {Wilson}, \& {Wilson}}]{mcconnell:2016}
{McConnell}, D., {Allison}, J.~R., {Bannister}, K., {et~al.} 2016, \pasa, 33,
  e042, \dodoi{10.1017/pasa.2016.37}

\bibitem[{{McQuinn}(2014)}]{mcquinn:2014}
{McQuinn}, M. 2014, \apjl, 780, L33, \dodoi{10.1088/2041-8205/780/2/L33}

\bibitem[{{Neal}(2011)}]{neal:2011}
{Neal}, R. 2011, in Handbook of Markov Chain Monte Carlo, 113--162,
  \dodoi{10.1201/b10905}

\bibitem[{{Nicastro} {et~al.}(2018){Nicastro}, {Kaastra}, {Krongold},
  {Borgani}, {Branchini}, {Cen}, {Dadina}, {Danforth}, {Elvis}, {Fiore},
  {Gupta}, {Mathur}, {Mayya}, {Paerels}, {Piro}, {Rosa-Gonzalez}, {Schaye},
  {Shull}, {Torres-Zafra}, {Wijers}, \& {Zappacosta}}]{nicastro:2018}
{Nicastro}, F., {Kaastra}, J., {Krongold}, Y., {et~al.} 2018, \nat, 558, 406,
  \dodoi{10.1038/s41586-018-0204-1}

\bibitem[{{Niu} {et~al.}(2021){Niu}, {Aggarwal}, {Li}, {Zhang}, {Chatterjee},
  {Tsai}, {Yu}, {Law}, {Burke-Spolaor}, {Cordes}, {Zhang}, {Ocker}, {Yao},
  {Wang}, {Feng}, {Niino}, {Bochenek}, {Cruces}, {Connor}, {Jiang}, {Dai},
  {Luo}, {Li}, {Miao}, {Niu}, {Anna-Thomas}, {Sydnor}, {Stern}, {Wang}, {Yuan},
  {Yue}, {Zhou}, {Yan}, {Zhu}, \& {Zhang}}]{niu:2021}
{Niu}, C.~H., {Aggarwal}, K., {Li}, D., {et~al.} 2021, arXiv e-prints,
  arXiv:2110.07418.
\newblock \doarXiv{2110.07418}

\bibitem[{{Persic} \& {Salucci}(1992)}]{persic:1992}
{Persic}, M., \& {Salucci}, P. 1992, \mnras, 258, 14P,
  \dodoi{10.1093/mnras/258.1.14P}

\bibitem[{{Petroff} {et~al.}(2016){Petroff}, {Barr}, {Jameson}, {Keane},
  {Bailes}, {Kramer}, {Morello}, {Tabbara}, \& {van Straten}}]{petroff:2016}
{Petroff}, E., {Barr}, E.~D., {Jameson}, A., {et~al.} 2016, \pasa, 33, e045,
  \dodoi{10.1017/pasa.2016.35}

\bibitem[{{Planck Collaboration} {et~al.}(2014){Planck Collaboration}, {Ade},
  {Aghanim}, {Armitage-Caplan}, {Arnaud}, {Ashdown}, {Atrio-Barandela},
  {Aumont}, {Baccigalupi}, {Banday}, \& et~al.}]{planck-collaboration:2014}
{Planck Collaboration}, {Ade}, P.~A.~R., {Aghanim}, N., {et~al.} 2014, \aap,
  571, A16, \dodoi{10.1051/0004-6361/201321591}

\bibitem[{{Planck Collaboration} {et~al.}(2020){Planck Collaboration},
  {Aghanim}, {Akrami}, {Ashdown}, {Aumont}, {Baccigalupi}, {Ballardini},
  {Banday}, {Barreiro}, {Bartolo}, {Basak}, {Battye}, {Benabed}, {Bernard},
  {Bersanelli}, {Bielewicz}, {Bock}, {Bond}, {Borrill}, {Bouchet}, {Boulanger},
  {Bucher}, {Burigana}, {Butler}, {Calabrese}, {Cardoso}, {Carron},
  {Challinor}, {Chiang}, {Chluba}, {Colombo}, {Combet}, {Contreras}, {Crill},
  {Cuttaia}, {de Bernardis}, {de Zotti}, {Delabrouille}, {Delouis}, {Di
  Valentino}, {Diego}, {Dor{\'e}}, {Douspis}, {Ducout}, {Dupac}, {Dusini},
  {Efstathiou}, {Elsner}, {En{\ss}lin}, {Eriksen}, {Fantaye}, {Farhang},
  {Fergusson}, {Fernandez-Cobos}, {Finelli}, {Forastieri}, {Frailis},
  {Fraisse}, {Franceschi}, {Frolov}, {Galeotta}, {Galli}, {Ganga},
  {G{\'e}nova-Santos}, {Gerbino}, {Ghosh}, {Gonz{\'a}lez-Nuevo}, {G{\'o}rski},
  {Gratton}, {Gruppuso}, {Gudmundsson}, {Hamann}, {Handley}, {Hansen},
  {Herranz}, {Hildebrandt}, {Hivon}, {Huang}, {Jaffe}, {Jones}, {Karakci},
  {Keih{\"a}nen}, {Keskitalo}, {Kiiveri}, {Kim}, {Kisner}, {Knox},
  {Krachmalnicoff}, {Kunz}, {Kurki-Suonio}, {Lagache}, {Lamarre}, {Lasenby},
  {Lattanzi}, {Lawrence}, {Le Jeune}, {Lemos}, {Lesgourgues}, {Levrier},
  {Lewis}, {Liguori}, {Lilje}, {Lilley}, {Lindholm}, {L{\'o}pez-Caniego},
  {Lubin}, {Ma}, {Mac{\'\i}as-P{\'e}rez}, {Maggio}, {Maino}, {Mandolesi},
  {Mangilli}, {Marcos-Caballero}, {Maris}, {Martin}, {Martinelli},
  {Mart{\'\i}nez-Gonz{\'a}lez}, {Matarrese}, {Mauri}, {McEwen}, {Meinhold},
  {Melchiorri}, {Mennella}, {Migliaccio}, {Millea}, {Mitra},
  {Miville-Desch{\^e}nes}, {Molinari}, {Montier}, {Morgante}, {Moss}, {Natoli},
  {N{\o}rgaard-Nielsen}, {Pagano}, {Paoletti}, {Partridge}, {Patanchon},
  {Peiris}, {Perrotta}, {Pettorino}, {Piacentini}, {Polastri}, {Polenta},
  {Puget}, {Rachen}, {Reinecke}, {Remazeilles}, {Renzi}, {Rocha}, {Rosset},
  {Roudier}, {Rubi{\~n}o-Mart{\'\i}n}, {Ruiz-Granados}, {Salvati}, {Sandri},
  {Savelainen}, {Scott}, {Shellard}, {Sirignano}, {Sirri}, {Spencer},
  {Sunyaev}, {Suur-Uski}, {Tauber}, {Tavagnacco}, {Tenti}, {Toffolatti},
  {Tomasi}, {Trombetti}, {Valenziano}, {Valiviita}, {Van Tent}, {Vibert},
  {Vielva}, {Villa}, {Vittorio}, {Wandelt}, {Wehus}, {White}, {White},
  {Zacchei}, \& {Zonca}}]{planck-collaboration:2020}
{Planck Collaboration}, {Aghanim}, N., {Akrami}, Y., {et~al.} 2020, \aap, 641,
  A6, \dodoi{10.1051/0004-6361/201833910}

\bibitem[{{Pol} {et~al.}(2019){Pol}, {Lam}, {McLaughlin}, {Lazio}, \&
  {Cordes}}]{pol:2019}
{Pol}, N., {Lam}, M.~T., {McLaughlin}, M.~A., {Lazio}, T.~J.~W., \& {Cordes},
  J.~M. 2019, \apj, 886, 135, \dodoi{10.3847/1538-4357/ab4c2f}

\bibitem[{{Prochaska} {et~al.}(2011){Prochaska}, {Weiner}, {Chen}, {Mulchaey},
  \& {Cooksey}}]{prochaska:2011}
{Prochaska}, J.~X., {Weiner}, B., {Chen}, H.~W., {Mulchaey}, J., \& {Cooksey},
  K. 2011, \apj, 740, 91, \dodoi{10.1088/0004-637X/740/2/91}

\bibitem[{{Prochaska} \& {Zheng}(2019)}]{prochaska:2019}
{Prochaska}, J.~X., \& {Zheng}, Y. 2019, \mnras, 485, 648,
  \dodoi{10.1093/mnras/stz261}

\bibitem[{{Ravi}(2019)}]{ravi:2019}
{Ravi}, V. 2019, \apj, 872, 88, \dodoi{10.3847/1538-4357/aafb30}

\bibitem[{{Ravi} {et~al.}(2019){Ravi}, {Catha}, {D'Addario}, {Djorgovski},
  {Hallinan}, {Hobbs}, {Kocz}, {Kulkarni}, {Shi}, {Vedantham}, {Weinreb}, \&
  {Woody}}]{ravi:2019a}
{Ravi}, V., {Catha}, M., {D'Addario}, L., {et~al.} 2019, \nat, 572, 352,
  \dodoi{10.1038/s41586-019-1389-7}

\bibitem[{{Shannon} {et~al.}(2018){Shannon}, {Macquart}, {Bannister}, {Ekers},
  {James}, {Os{\l}owski}, {Qiu}, {Sammons}, {Hotan}, {Voronkov}, {Beresford},
  {Brothers}, {Brown}, {Bunton}, {Chippendale}, {Haskins}, {Leach},
  {Marquarding}, {McConnell}, {Pilawa}, {Sadler}, {Troup}, {Tuthill},
  {Whiting}, {Allison}, {Anderson}, {Bell}, {Collier}, {G{\"u}rkan}, {Heald},
  \& {Riseley}}]{shannon:2018}
{Shannon}, R.~M., {Macquart}, J.~P., {Bannister}, K.~W., {et~al.} 2018, \nat,
  562, 386, \dodoi{10.1038/s41586-018-0588-y}

\bibitem[{{Sharp} {et~al.}(2006){Sharp}, {Saunders}, {Smith}, {Churilov},
  {Correll}, {Dawson}, {Farrel}, {Frost}, {Haynes}, {Heald}, {Lankshear},
  {Mayfield}, {Waller}, \& {Whittard}}]{sharp:2006}
{Sharp}, R., {Saunders}, W., {Smith}, G., {et~al.} 2006, in \procspie, Vol.
  6269, Society of Photo-Optical Instrumentation Engineers (SPIE) Conference
  Series, 62690G, \dodoi{10.1117/12.671022}

\bibitem[{{Shull} \& {Danforth}(2018)}]{shull:2018}
{Shull}, J.~M., \& {Danforth}, C.~W. 2018, \apjl, 852, L11,
  \dodoi{10.3847/2041-8213/aaa2fa}

\bibitem[{{Simha} {et~al.}(2020){Simha}, {Burchett}, {Prochaska}, {Chittidi},
  {Elek}, {Tejos}, {Jorgenson}, {Bannister}, {Bhandari}, {Day}, {Deller},
  {Forbes}, {Macquart}, {Ryder}, \& {Shannon}}]{simha:2020}
{Simha}, S., {Burchett}, J.~N., {Prochaska}, J.~X., {et~al.} 2020, \apj, 901,
  134, \dodoi{10.3847/1538-4357/abafc3}

\bibitem[{{Simha} {et~al.}(2021){Simha}, {Tejos}, {Prochaska}, {Lee}, {Ryder},
  {Cantalupo}, {Bannister}, {Bhandari}, \& {Shannon}}]{simha:2021}
{Simha}, S., {Tejos}, N., {Prochaska}, J.~X., {et~al.} 2021, arXiv e-prints,
  arXiv:2108.09881.
\newblock \doarXiv{2108.09881}

\bibitem[{{Smith} {et~al.}(2011){Smith}, {Hallman}, {Shull}, \&
  {O'Shea}}]{smith:2011}
{Smith}, B.~D., {Hallman}, E.~J., {Shull}, J.~M., \& {O'Shea}, B.~W. 2011,
  \apj, 731, 6, \dodoi{10.1088/0004-637X/731/1/6}

\bibitem[{{Spitler} {et~al.}(2016){Spitler}, {Scholz}, {Hessels}, {Bogdanov},
  {Brazier}, {Camilo}, {Chatterjee}, {Cordes}, {Crawford}, {Deneva}, {Ferdman},
  {Freire}, {Kaspi}, {Lazarus}, {Lynch}, {Madsen}, {McLaughlin}, {Patel},
  {Ransom}, {Seymour}, {Stairs}, {Stappers}, {van Leeuwen}, \&
  {Zhu}}]{spitler:2016}
{Spitler}, L.~G., {Scholz}, P., {Hessels}, J.~W.~T., {et~al.} 2016, \nat, 531,
  202, \dodoi{10.1038/nature17168}

\bibitem[{{Springel}(2005)}]{springel:2005}
{Springel}, V. 2005, \mnras, 364, 1105,
  \dodoi{10.1111/j.1365-2966.2005.09655.x}

\bibitem[{{Springel} {et~al.}(2005){Springel}, {White}, {Jenkins}, {Frenk},
  {Yoshida}, {Gao}, {Navarro}, {Thacker}, {Croton}, {Helly}, {Peacock}, {Cole},
  {Thomas}, {Couchman}, {Evrard}, {Colberg}, \& {Pearce}}]{springel:2005a}
{Springel}, V., {White}, S. D.~M., {Jenkins}, A., {et~al.} 2005, \nat, 435,
  629, \dodoi{10.1038/nature03597}

\bibitem[{{Sugai} {et~al.}(2015){Sugai}, {Tamura}, {Karoji}, {Shimono},
  {Takato}, {Kimura}, {Ohyama}, {Ueda}, {Aghazarian}, {de Arruda},
  {Barkhouser}, {Bennett}, {Bickerton}, {Bozier}, {Braun}, {Bui}, {Capocasale},
  {Carr}, {Castilho}, {Chang}, {Chen}, {Chou}, {Dawson}, {Dekany}, {Ek},
  {Ellis}, {English}, {Ferrand}, {Ferreira}, {Fisher}, {Golebiowski}, {Gunn},
  {Hart}, {Heckman}, {Ho}, {Hope}, {Hovland}, {Hsu}, {Hu}, {Huang}, {Jaquet},
  {Karr}, {Kempenaar}, {King}, {F{\`e}vre}, {Mignant}, {Ling}, {Loomis},
  {Lupton}, {Madec}, {Mao}, {Marrara}, {M{\'e}nard}, {Morantz}, {Murayama},
  {Murray}, {de Oliveira}, {de Oliveira}, {de Oliveira}, {Orndorff}, {de Paiva
  Vila{\c c}a}, {Partos}, {Pascal}, {Pegot-Ogier}, {Reiley}, {Riddle},
  {Santos}, {dos Santos}, {Schwochert}, {Seiffert}, {Smee}, {Smith},
  {Steinkraus}, {Sodr{\'e}}, {Spergel}, {Surace}, {Tresse}, {Vidal}, {Vives},
  {Wang}, {Wen}, {Wu}, {Wyse}, \& {Yan}}]{sugai:2015}
{Sugai}, H., {Tamura}, N., {Karoji}, H., {et~al.} 2015, Journal of Astronomical
  Telescopes, Instruments, and Systems, 1, 035001,
  \dodoi{10.1117/1.JATIS.1.3.035001}

\bibitem[{{Takahashi} {et~al.}(2020){Takahashi}, {Ioka}, {Mori}, \&
  {Funahashi}}]{takahashi:2020}
{Takahashi}, R., {Ioka}, K., {Mori}, A., \& {Funahashi}, K. 2020, arXiv
  e-prints, arXiv:2010.01560.
\newblock \doarXiv{2010.01560}

\bibitem[{{Tanimura} {et~al.}(2019{\natexlab{a}}){Tanimura}, {Aghanim},
  {Douspis}, {Beelen}, \& {Bonjean}}]{tanimura:2019a}
{Tanimura}, H., {Aghanim}, N., {Douspis}, M., {Beelen}, A., \& {Bonjean}, V.
  2019{\natexlab{a}}, \aap, 625, A67, \dodoi{10.1051/0004-6361/201833413}

\bibitem[{{Tanimura} {et~al.}(2020){Tanimura}, {Aghanim}, {Kolodzig},
  {Douspis}, \& {Malavasi}}]{tanimura:2020}
{Tanimura}, H., {Aghanim}, N., {Kolodzig}, A., {Douspis}, M., \& {Malavasi}, N.
  2020, \aap, 643, L2, \dodoi{10.1051/0004-6361/202038521}

\bibitem[{{Tanimura} {et~al.}(2019{\natexlab{b}}){Tanimura}, {Hinshaw},
  {McCarthy}, {Van Waerbeke}, {Aghanim}, {Ma}, {Mead}, {Hojjati}, \&
  {Tr{\"o}ster}}]{tanimura:2019}
{Tanimura}, H., {Hinshaw}, G., {McCarthy}, I.~G., {et~al.} 2019{\natexlab{b}},
  \mnras, 483, 223, \dodoi{10.1093/mnras/sty3118}

\bibitem[{{Tegmark} {et~al.}(2004){Tegmark}, {Blanton}, {Strauss}, {Hoyle},
  {Schlegel}, {Scoccimarro}, {Vogeley}, {Weinberg}, {Zehavi}, {Berlind},
  {Budavari}, {Connolly}, {Eisenstein}, {Finkbeiner}, {Frieman}, {Gunn},
  {Hamilton}, {Hui}, {Jain}, {Johnston}, {Kent}, {Lin}, {Nakajima}, {Nichol},
  {Ostriker}, {Pope}, {Scranton}, {Seljak}, {Sheth}, {Stebbins}, {Szalay},
  {Szapudi}, {Verde}, {Xu}, {Annis}, {Bahcall}, {Brinkmann}, {Burles},
  {Castander}, {Csabai}, {Loveday}, {Doi}, {Fukugita}, {Gott}, {Hennessy},
  {Hogg}, {Ivezi{\'c}}, {Knapp}, {Lamb}, {Lee}, {Lupton}, {McKay}, {Kunszt},
  {Munn}, {O'Connell}, {Peoples}, {Pier}, {Richmond}, {Rockosi}, {Schneider},
  {Stoughton}, {Tucker}, {Vanden Berk}, {Yanny}, {York}, \& {SDSS
  Collaboration}}]{tegmark:2004}
{Tegmark}, M., {Blanton}, M.~R., {Strauss}, M.~A., {et~al.} 2004, \apj, 606,
  702, \dodoi{10.1086/382125}

\bibitem[{{Tendulkar} {et~al.}(2017){Tendulkar}, {Bassa}, {Cordes}, {Bower},
  {Law}, {Chatterjee}, {Adams}, {Bogdanov}, {Burke-Spolaor}, {Butler},
  {Demorest}, {Hessels}, {Kaspi}, {Lazio}, {Maddox}, {Marcote}, {McLaughlin},
  {Paragi}, {Ransom}, {Scholz}, {Seymour}, {Spitler}, {van Langevelde}, \&
  {Wharton}}]{tendulkar:2017}
{Tendulkar}, S.~P., {Bassa}, C.~G., {Cordes}, J.~M., {et~al.} 2017, \apjl, 834,
  L7, \dodoi{10.3847/2041-8213/834/2/L7}

\bibitem[{{The CHIME/FRB Collaboration} {et~al.}(2021){The CHIME/FRB
  Collaboration}, {:}, {Amiri}, {Andersen}, {Bandura}, {Berger}, {Bhardwaj},
  {Boyce}, {Boyle}, {Brar}, {Breitman}, {Cassanelli}, {Chawla}, {Chen},
  {Cliche}, {Cook}, {Cubranic}, {Curtin}, {Deng}, {Dobbs}, {Fengqiu}, {Dong},
  {Eadie}, {Fandino}, {Fonseca}, {Gaensler}, {Giri}, {Good}, {Halpern}, {Hill},
  {Hinshaw}, {Josephy}, {Kaczmarek}, {Kader}, {Kania}, {Kaspi}, {Landecker},
  {Lang}, {Leung}, {Li}, {Lin}, {Masui}, {Mckinven}, {Mena-Parra},
  {Merryfield}, {Meyers}, {Michilli}, {Milutinovic}, {Mirhosseini},
  {M{\"u}nchmeyer}, {Naidu}, {Newburgh}, {Ng}, {Patel}, {Pen}, {Petroff},
  {Pinsonneault-Marotte}, {Pleunis}, {Rafiei-Ravandi}, {Rahman}, {Ransom},
  {Renard}, {Sanghavi}, {Scholz}, {Shaw}, {Shin}, {Siegel}, {Sikora}, {Singh},
  {Smith}, {Stairs}, {Tan}, {Tendulkar}, {Vanderlinde}, {Wang}, {Wulf}, \&
  {Zwaniga}}]{the-chime/frb-collaboration:2021}
{The CHIME/FRB Collaboration}, {:}, {Amiri}, M., {et~al.} 2021, arXiv e-prints,
  arXiv:2106.04352.
\newblock \doarXiv{2106.04352}

\bibitem[{Virtanen {et~al.}(2020)Virtanen, Gommers, Oliphant, Haberland, Reddy,
  Cournapeau, Burovski, Peterson, Weckesser, Bright, {van der Walt}, Brett,
  Wilson, Millman, Mayorov, Nelson, Jones, Kern, Larson, Carey, Polat, Feng,
  Moore, {VanderPlas}, Laxalde, Perktold, Cimrman, Henriksen, Quintero, Harris,
  Archibald, Ribeiro, Pedregosa, {van Mulbregt}, \& {SciPy 1.0
  Contributors}}]{virtanen:2020}
Virtanen, P., Gommers, R., Oliphant, T.~E., {et~al.} 2020, Nature Methods, 17,
  261, \dodoi{10.1038/s41592-019-0686-2}

\bibitem[{{Walters} {et~al.}(2019){Walters}, {Ma}, {Sievers}, \&
  {Weltman}}]{walters:2019}
{Walters}, A., {Ma}, Y.-Z., {Sievers}, J., \& {Weltman}, A. 2019, \prd, 100,
  103519, \dodoi{10.1103/PhysRevD.100.103519}

\bibitem[{{Wechsler} \& {Tinker}(2018)}]{wechsler:2018}
{Wechsler}, R.~H., \& {Tinker}, J.~L. 2018, \araa, 56, 435,
  \dodoi{10.1146/annurev-astro-081817-051756}

\bibitem[{{Werk} {et~al.}(2014){Werk}, {Prochaska}, {Tumlinson}, {Peeples},
  {Tripp}, {Fox}, {Lehner}, {Thom}, {O'Meara}, {Ford}, {Bordoloi}, {Katz},
  {Tejos}, {Oppenheimer}, {Dav{\'e}}, \& {Weinberg}}]{werk:2014}
{Werk}, J.~K., {Prochaska}, J.~X., {Tumlinson}, J., {et~al.} 2014, \apj, 792,
  8, \dodoi{10.1088/0004-637X/792/1/8}

\bibitem[{{Zhang} {et~al.}(2021){Zhang}, {Yan}, {Li}, {Zhang}, \&
  {Wang}}]{zhang:2021}
{Zhang}, Z.~J., {Yan}, K., {Li}, C.~M., {Zhang}, G.~Q., \& {Wang}, F.~Y. 2021,
  \apj, 906, 49, \dodoi{10.3847/1538-4357/abceb9}

\bibitem[{{Zhu} \& {Feng}(2020)}]{zhu:2020}
{Zhu}, W., \& {Feng}, L.-L. 2020, arXiv e-prints, arXiv:2011.08519.
\newblock \doarXiv{2011.08519}

\end{thebibliography}
\bibliographystyle{aasjournal}


\end{CJK*}
\end{document}